\documentclass[acmlarge]{acmart}

\AtBeginDocument{%
  \providecommand\BibTeX{{%
    \normalfont B\kern-0.5em{\scshape i\kern-0.25em b}\kern-0.8em\TeX}}}

\usepackage{graphicx}
\usepackage{caption}
\usepackage{subcaption}
\usepackage{url}
\usepackage{xcolor}
\usepackage[capitalise,noabbrev]{cleveref}

\begin{document}

\title{Efficacy of the Confinement Policies on the COVID-19 Spread Dynamics in the Early Period of the Pandemic}

\author{Mehedi Hassan}
\email{mehedi.hassan1@louisiana.edu}
\affiliation{%
  \institution{University of Louisiana at Lafayette}
  \city{Lafayette}
  \state{LA}
  \postcode{70504}
}
\author{Md Enamul Haque}
\email{enamulh@stanford.edu}
\affiliation{%
  \institution{Stanford University}
  \city{Stanford}
  \state{CA}
  \postcode{94305}
}

\author{Mehmet Engin Tozal}
\email{metozal@louisiana.edu}
\affiliation{%
  \institution{University of Louisiana at Lafayette}
  \city{Lafayette}
  \state{LA}
  \postcode{70504}
}

\renewcommand{\shortauthors}{Hassan, et al.}

\begin{abstract}
Spread dynamics and the confinement policies of COVID-19 exhibit different patterns for different countries.
Numerous factors affect such patterns within each country.
Examining these factors, and analyzing the confinement practices allow government authorities to implement effective policies in the future.
In addition, they help the authorities to distribute healthcare resources optimally without overwhelming their systems.
In this empirical study, we use a clustering-based approach, Hierarchical Cluster Analysis (HCA) on time-series data to capture the spread patterns at various countries.
We particularly investigate the confinement policies adopted by different countries and their impact on the spread patterns of COVID-19 .
We limit our investigation to the early period of the pandemic, because many governments tried to respond rapidly and aggressively in the beginning.
Moreover, these governments adopted diverse confinement policies based on trial-and-error in the beginning of the pandemic.
We found that implementations of the same confinement policies may exhibit different results in different countries.
Specifically, lockdowns become less effective in densely populated regions, because of the reluctance to comply with social distancing measures.
Lack of testing, contact tracing, and social awareness in some countries forestall people from self-isolation and maintaining social distance.
Large labor camps with unhealthy living conditions also aid in high community transmissions in countries depending on foreign labor.
Distrust in government policies and fake news instigate the spread in both developed and under-developed countries.
Large social gatherings play a vital role in causing rapid outbreaks almost everywhere.
An early and rapid response at the early period of the pandemic is necessary to contain the spread, yet it is not always sufficient.
\end{abstract}




\maketitle

\section{Introduction}
The novel coronavirus disease (COVID-19) was first detected in Wuhan, China on December 31, 2019, following the reports of a cluster of cases of a typical pneumonia~\cite{paules2020coronavirus}. 
On January 30, 2020, the World Health Organization (WHO) declared the outbreak of COVID-19 as a Public Health Emergency of International Concern (PHEIC)~\cite{wang2020novel}. 
WHO advised for early detection, isolation, and treatment of COVID-19 cases as well as social distancing measures. 
Despite the initially infected geographic regions being isolated and locked down by the Chinese government, the COVID-19 soon spread into other parts of the world.
Due to the rapid increase in the number of detected cases and deaths, WHO declared the outbreak of COVID-19 as a pandemic on March 11, 2020~\cite{Who_pandemic}. 
There are more than 248 million cases reported in over 180 countries to date.
To control the spread of COVID-19 and not to overwhelm the medical facilities, different governments have decided to take various confinement policies or preventive measures in the early period of the pandemic.

Multiple studies have employed different mathematical models~\cite{fanelli2020analysis, rocklov2020covid, mohamed2020exploring, elmousalami2020day, deb2020time, jewell2020predictive} to predict the transmission of the COVID-19 disease.
However, these predictive models do not account for individual social activity or the effect of confinement policies.
As a result, these studies often failed to predict or explain the sudden surge of new cases.
Other studies have focused on the impact of different confinement policies adopted by different authorities~\cite{gatto2020spread, wells2020impact, courtemanche2020strong}.
However, these studies mostly focus on one country,~\emph{e.g.,} the provinces of China, Italy, or USA~\cite{castorina2020data, wells2020impact, gatto2020spread, deb2020time} or a set of selected countries~\cite{hamzah2020coronatracker, dey2020analyzing, ruiz2020covid}. 
Yet, the spread characteristics differ from one country to another depending on their confinement policies.
Moreover, implementations of the same confinement policies may exhibit different results in different countries.
Therefore, examining the spread patterns, and analyzing the confinement policies over multiple countries are needed to determine effective preventive measures.
Understanding the effect of these policies can help countries to manage and distribute resources efficiently, and fight against COVID-19 or similar outbreaks in the future.

In the early period of the pandemic, countries have adopted various confinement policies to reduce or contain the COVID-19 spread.
Lockdowns, social distancing, isolation of infected patients, large gathering bans, travel or mobility restrictions, contact tracing, testing, or curfews have been the most common forms of confinement policies or preventive measures.
Implementations of strict policies after the orders by governments or local authorities have also been challenging.
Many countries succeeded to contain or flatten the spread through imposing strict lockdowns, travel restrictions, and social distancing while others,~\emph{e.g.,} South Korea, Netherlands, and Denmark were successful without any strict confinement policies.
In some countries, the spread increased despite strict measures due to the lack of social awareness, not maintaining social distancing, large gatherings, or distrust to comply with government policies.
As a result, medical facilities were easily overwhelmed and people did not access adequate and proper healthcare, which may cause panic among the population.
The introduction and distribution of several COVID-19 vaccines (\emph{e.g.,} Pfizer, Moderna, and Johnson \& Johnson) have helped controlling spread in some countries recently.
Even after the vaccination, countries such as India, Netherlands, and Canada are still unable to contain the spread due to either the lack of adequate doses or the distrust/misinformation of the effect of vaccines.
Vaccination as a confinement policy is out of the scope of our study, because we only focus on the early period of the pandemic before the vaccines were developed.

In this study, we employ Hierarchical Cluster Analysis (HCA)~\cite{kaufman2009finding} with Dynamic-Time Warping (DTW) as a distance metric~\cite{sakoe1978dynamic} to capture the patterns and analyze the time series data of different COVID-19 affected countries.
We use the global confirmed cases data from the GitHub repository~\cite{CovidDataset} maintained by the Johns Hopkins University, Center for Systems Science, and Engineering (JHU CSSE).
We select the countries having more than 10,000 confirmed cases and at least 90 days of data in the early period of the pandemic, which spans an over-four-months period from January 22 to May 31 2020 for this study.
We investigate the early period, because (i) most governments have tried to respond rapidly and aggressively in the beginning of the pandemic; (ii) the confinement policies in the initial period have been more diverse; (iii) lack of immediate, widespread and accurate information caused diverse public reactions in the beginning; and (iv) the information gathered later allows us to meticulously assess the successful and unsuccessful policies in the initial period.
We divide the time-series data of each country into three stages: early, middle, and post stages.
Each stage contains 30 days of time series data, not necessarily starting on the same date.
We apply hierarchical clustering on the early and post stages to analyze the spread dynamics in different countries at two different time intervals.
Then, we use the middle-stage time-series data to explain the variation between the early and post-stage periods.
Our main contributions in this study are listed as follows:
\begin{itemize}
	\item We demonstrate that Hierarchical Cluster Analysis (HCA) with Dynamic Time Warping (DTW) as a distance metric to capture the time-series based COVID-19 spread patterns has been effective.
	\item We include a wide range of countries to compare and contrast the COVID-19 spread dynamics as well as the counter-measures implemented by those countries.
	\item We propose segmenting the time series data into early and post-stage periods and use the middle-stage period to understand the spread dynamics that led countries to move from an early-stage cluster to another post-stage cluster.
	\item We analyze the efficacy of different confinement policies or preventive measures on the spread dynamics of different countries in detail.
\end{itemize}

Our findings suggest that some countries were successful to reduce the spread, while others were unsuccessful, despite imposing lockdowns or other strict measures.
Through early and rapid response (Romania, Israel, Portugal, Austria, and China), contact tracing (Israel, Philippines, Germany, and China), improved testing capacity (Spain, Germany, Austria, and China), or social distancing (Italy and Belgium), many countries have relatively reduced or flattened the spread.
On the other hand, slow or delayed response (USA, U.K., Mexico, and Brazil), lack of testing (USA, Afghanistan, India, Spain, and Mexico), lack of social awareness (Afghanistan, India, Italy, and Russia), large gatherings during festivals (Iran and France), failure to maintain social distancing policies (Iran, India, Qatar, and Saudi Arabia), migrant workers in large and dense labor camps (Kuwait, U.A.E., Qatar, Oman, Saudi Arabia, and Singapore), fake news (Dominican Republic), distrust in government-issued policies (Dominican Republic and Russia) or poor healthcare facilities (Egypt) are the main factors influencing the increase in the spread.
Moreover, lockdowns have become ineffective in densely populated areas (India).
Some countries such as Denmark, Netherlands, Switzerland, Japan, South Korea, Indonesia, Belarus, and Sweden opted for different and more relaxed confinement policies, as they did not impose any lockdowns.
Among these countries, Denmark, Netherlands, Switzerland, and South Korea were successful to flatten the curve.
Rigorous testing capacity, compliance with social distancing measures, and self-discipline or self-responsibility helped these countries to control the spread.
However, the confinement policies adopted by Belarus, Sweden, Indonesia, and Japan have failed to contain the spread due to lack of testing, not imposing strict lockdowns or social distancing measures, or large gatherings during festivals.

The rest of the paper is organized as follows. 
Section~\ref{related_work} presents the related work. 
Section~\ref{Methodology} introduces the technical approaches used in this study. 
Section~\ref{results} describes empirical results. 
Section~\ref{disucssion} provides detailed discussions on the analyses of the clusters. 
Finally, Section~\ref{conclusion} concludes our study.

\section{Related Work}\label{related_work}

COVID-19 data analysis has leveraged different kinds of methodologies.
Kumar used a hierarchical clustering technique with euclidean distance as a distance metric to perform the data analysis on 27 Indian states~\cite{kumarmonitoring}.
Confirmed, deaths and recovered cases were used to cluster the states into six groups.
The author suggested optimizing the monitoring techniques such as screenings, lockdowns, curfews, and improving medical facilities which provide valuable insight into the seriousness of the disease spread to the government, police, and healthcare authorities.
Castorina~\emph{et al.}~\cite{castorina2020data} analyzed the COVID-19 data using generalized Gompertz law.
The authors used the data of four countries, including China, Italy, South Korea, and Singapore, and evaluated the saturation points of the above-mentioned countries.
The study concluded that strong confinement policy in China has helped to saturate the spread, the new growth rate in Singapore may point to a strong spreading, and the next data on South Korea and Italy need to be evaluated to determine if the spread in these two countries will reach a saturation point or not.

More recent studies discussed the impact of social distancing and other preventive measures against the spread of COVID-19.
Wells~\emph{et al.}~\cite{wells2020impact} analyzed the impact of travel restrictions on the growth of COVID-19 spread.
The authors utilized Monte Carlo simulations and estimated that the travel restrictions in Hubei reduced the rate of COVID-19 disease exportation by 81\%.
Courtemanche~\emph{et al.}~\cite{courtemanche2020strong} discussed the impact of social distancing measures in the USA.
The authors evaluated four types of measures such as large event ban, closure of schools, closure of entertainment venues, and shelter-in-place-orders (SIPOs).
The study suggested that without the social distancing measures, the number of confirmed cases would have risen to 35 times by 27 April.

Many researchers have proposed mathematical prediction models based on the daily incidence data.
Fanelli and Piazza~\cite{fanelli2020analysis} proposed a susceptible-infected-recovered-deaths (SIRD) model to predict the outbreak in China, Italy, and France.
Data from 22 January to 15 March were analyzed.
They observed that the recovery rate is the same but death rates were different for these countries.
Kucharski~\emph{et al.}~\cite{kucharski2020early} proposed a susceptible-exposed-infectious-removed (SEIR) based prediction model, using a combination of four datasets.
The model estimated the transmission dynamics in Wuhan.
Rocklov~\emph{et al.}~\cite{rocklov2020covid} analyzed the confirmed cases data of the Diamond Ship Cruise and estimated the reproduction number using the SEIR model.
The authors recommended evacuating immediately after an outbreak is confirmed.
Naji~\cite{mohamed2020exploring} proposed a susceptible-exposed-infected-recovered-death (SEIRD) model to analyze the confirmed, recovery, and death cases in Morocco.
The proposed model estimated the time of the peak and the minimum number of infection rates.
Fang~\emph{et al.}~\cite{fang2020transmission} proposed an SEIR based prediction model to analyze the COVID-19 spread data in China.
The authors discussed the impact of preventive measures taken by the Chinese government and recommended other affected countries to follow the same measures.
Gatto~\emph{et al.}~\cite{gatto2020spread} also leveraged an SEIR based prediction model.
However, this study focused only on Italy.
The results indicated that strict social distancing measures have reduced the spread by 45\%.
Liu~\emph{et al.}~\cite{liu2020machine} proposed a machine learning based prediction model, Augmented ARGOnet.
The proposed system provides a short-term forecast of COVID-19 spread in the provinces of China.
The proposed model outperformed a persistence model, which is used as a baseline, in 27 out of 32 provinces of China.

Elmousalami and Hassanien~\cite{elmousalami2020day} analyzed the daily incidence data using a mathematical model.
The model observed that without the preventive measures, the incidence rate grows at more than 25\%.
Deb and Majumdar~\cite{deb2020time} proposed a different mathematical model to analyze the daily incidence rate in the provinces of China and five other countries including Italy, USA, Iran, South Korea, and India.
The proposed model provides insights into the lockdown effect and estimates the significant change of growth in spread patterns and reproduction numbers.

In these recent studies, data analysis is performed in some selected countries.
Most of these studies are based on a prediction model that estimates the peak infection rates and reproduction rates with a small amount of data.
However, in this study, we use a more generalized approach by selecting a wide range of countries all over the world.
We included the countries having more than 10,000 cases and at least 90 days of data to have a sufficient amount of data for analysis.
Our goal in this study is to shed light on the COVID-19 spread dynamics in different regions in the world by analyzing time-series data belonging to three consecutive time periods.

\section{Methods}\label{Methodology}

 In this section, we explain the methodology used in this study.
 We first standardize the time-series data belonging to different countries.
 Next, we divide the time-series data into three periods: early-stage, middle-stage, and post-stage.
 Then we use Hierarchical Cluster Analysis (HCA) with Dynamic Time Warping (DTW) to cluster early and post-stage time-series.
 Finally, we use the middle-stage to explain the discrepancies between the early-stage and post-stage clusters.
 We perform Hierarchical Cluster Analysis (HCA) with Dynamic Time Warping (DTW) as a distance metric and complete linkage as a linkage method to compare and contrast the spread dynamics in various countries.

\subsection{Hierarchical Cluster Analysis (HCA)}
Cluster analysis groups the data that share meaningful or useful characteristics. 
Performing cluster analysis allows us to discover patterns or characteristics that are previously not noticeable. 
There are two popular techniques for cluster analysis: hierarchical and non-hierarchical~\cite{yim2015hierarchical}.
In hierarchical clustering techniques, data points are merged or dissected to form homogeneous clusters.
In non-hierarchical clustering techniques, an initial set of cluster centroids are first established. 
Then, each data point is assigned to the nearest cluster and cluster centroids are updated. 
This process continues until the centroids of the clusters are not updated anymore~\cite{morissette2013k}.

The goal of hierarchical clustering is to find sub-groups of data points that are homogeneous to each other.
This technique is especially suitable when the ideal number of clusters in the data is not anticipated before-hand. 
A dendrogram, tree-like visual representation, is generated to visually inspect each possible number of clusters from 1 to $k$, where $k$ is the total number of data points. 
There are two types of hierarchical clustering: agglomerative and divisive~\cite{yim2015hierarchical}. 
In the agglomerative or bottom-up approach, each data point is assigned to its cluster.
Then two closest clusters are merged to form a cluster and this process continues until there is only one cluster containing all the data points.
In the divisive or top-down approach, all the data points are assigned to one cluster. 
Next, the cluster is split iteratively until all the data points are assigned to singleton clusters.
We use agglomerative hierarchical clustering, as our goal is to cluster the countries exhibiting the same spreading patterns.

To perform a hierarchical clustering analysis, a proximity matrix between each pair of data points is required.
The proximity matrix is updated every time two clusters are merged.
The proximity matrix is computed using a distance metric.
The most widely used distance metrics are Euclidean distance, Manhattan distance, Mahalanobis distance, or Cosine similarity. 
The closeness between two clusters is measured by a linkage method. The most popular linkage methods are complete, average, single, and centroid. 
In this study, we use Dynamic Time Warping (DTW) as a distance metric (described in section~\ref{DTW}) and complete linkage as a linkage method (described in section~\ref{linkage}).

In unsupervised learning, hierarchical clustering in the time-series analysis is desirable over other clustering methods in three ways.
First, HCA does not need any prior information of the data or the structure of the data for grouping the homogeneous data points~\cite{fong2012using}.
Second, HCA does not need to predefine the desired number of clusters~\cite{fong2012using, ozkocc2020clustering}.
Third, HCA provides a graphical representation (\emph{i.e.,} dendrogram) of the homogeneous groups in the data~\cite{ozkocc2020clustering}.
Dendrogram allows us to have an idea about the data and how the data points are clustered together at different distance levels.
Hierarchical clustering enables us to determine the spreading patterns of various countries.
Therefore, it is suitable to use HCA on time-series data of various COVID-19 affected countries.

\subsection{Dynamic Time Warping (DTW)}\label{DTW}
Dynamic time warping (DTW) is a distance metric that is used to find patterns in time-series data. 
DTW is first used in speech recognition~\cite{sakoe1978dynamic}. 
It has been applied to other domains such as robotics~\cite{schmill1999learned}, data mining~\cite{rakthanmanon2012searching}, handwriting recognition~\cite{rath2003word}, and gesture recognition~\cite{ten2007multi}. 
DTW calculates the similarity between two time-series by building one-to-many and many-to-one matches to minimize the cumulative distance between them. 
Generally, DTW calculates the distance between all pairs of indexes in two time-series which is computationally costly,~\emph{i.e.,} the time complexity is $O$($n^2$), and the space complexity is $O$($n^2$). 
However, it can be optimized using a warping window~\cite{salvador2007toward} with time complexity $O$($n$) and space complexity $O$($n$) by reducing the distance computations.
Let $X$ = $\left \{ x_1, x_2, ... , x_m \right \}$ and $Y$ = $\left \{ y_1, y_2, ... , y_n \right \}$ be two time series of lengths $m$ and $n$, respectively. DTW distance is calculated between $X$ and $Y$ with a window size $w$ by using Equation~\ref{dtw}.

\begin{equation}
	D\left ( i,j \right ) = d\left ( x_i, y_j \right ) + \min \begin{cases}D\left ( i-1,j \right )\\ D\left ( i,j-1 \right )\\ D\left ( i-1, j-1 \right ) \end{cases}
	\label{dtw}
\end{equation}
where $0 \leq  i <  m$ and $\left ( i-\left \lfloor w/2 \right \rfloor \right ) \leq  j \leq  \left ( i+\left \lfloor w/2 \right \rfloor \right )$.
DTW measures the similarity in terms of shape or pattern with shifting or scaling at any period between two time-series data~\cite{fong2012using}.
The correlation coefficient also captures shape or pattern.
However, it is not convenient to apply to time series with shifting and variable amplitude~\cite{li2012grid}.
Therefore, DTW is a preferable distance metric in our study to capture the spread patterns among various countries.

\subsection{Linkage Method: Complete}\label{linkage}
The linkage method is used to compute the similarity/dissimilarity between clusters when multiple data points are residing inside them. 
The objective is to merge a cluster with the nearest cluster. 
In this study, we use complete linkage which is also referred to as maximum linkage. 
Complete linkage considers the maximum distance between one data point in the first cluster and another data point in the second cluster. 
Sequentially, 
two clusters with the minimum of the maximum distance values considered to be the closest to each other and merged.
After merging two clusters their distance values are updated in the proximity matrix.

Let $C$ be the clusters present in the data at any arbitrary moment, $X$ and $Y$ be two clusters in $C$,~\emph{i.e.,} $X, Y \in C$. 
The distance between two clusters for all pairs of clusters ($X$, $Y$) in $C$ is calculated using Equation~\ref{linkage_eq}. 
Two clusters with the minimum value are merged and the proximity matrix is updated according to Equation~\ref{distance}:

\begin{equation}
	\min\limits_{X,Y \in C} D\left ( X,Y \right ) = \max\limits_{x\in X, y\in Y} d\left ( x,y \right ) 
	\label{linkage_eq}
\end{equation}

\begin{equation}
	D\left ( X,Y \right ) = \max\limits_{x\in X, y\in Y} d\left ( x,y \right ) 
	\label{distance}
\end{equation}
where $d$($x$,$y$) is the distance between two time series $x$ and $y$ at any arbitrary point, and $D$($X$,$Y$) is the distance between two clusters.

Hierarchical clustering with the single linkage method tends to produce more loose clusters~\cite{yim2015hierarchical}.
In centroid linkage, the smaller clusters are more similar to the new larger cluster than to their individual clusters. 
Thus, it may cause an inversion in the dendrogram.
The average and complete linkage both produce compact clusters.
However, complete linkage ensures the grouping of most correlated data points or clusters~\cite{camiz2007comparison}.
Therefore, the complete linkage is more applicable to our dataset.


\subsection{Data Preparation}
\begin{figure}[p]
	\centering
	\includegraphics[width=\linewidth]{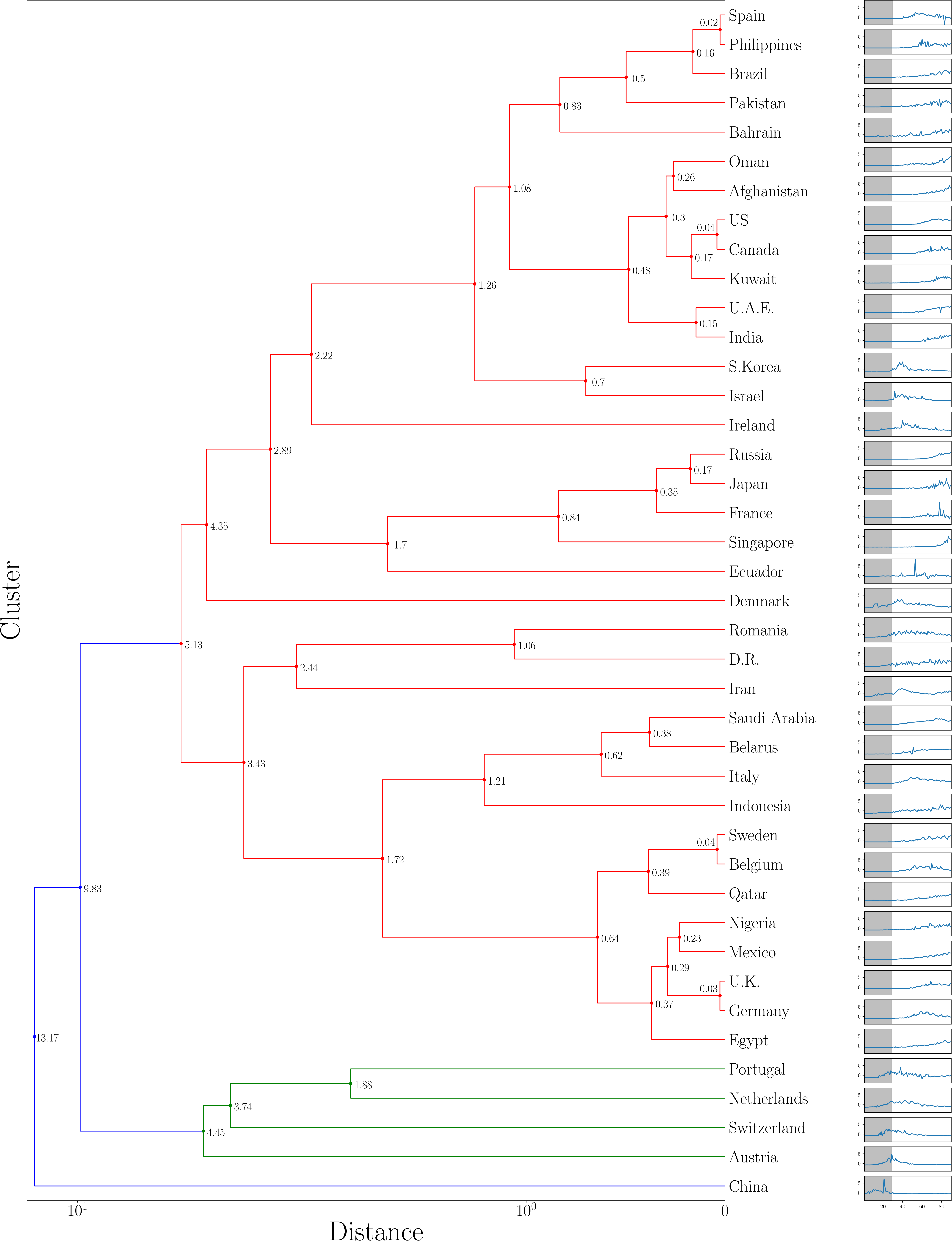}
	\caption{Dendrogram of 41 countries in the early-stage.}
	\label{fig:dendo_early}
\end{figure}

\begin{figure}[p]
	\centering
	\includegraphics[width=\linewidth]{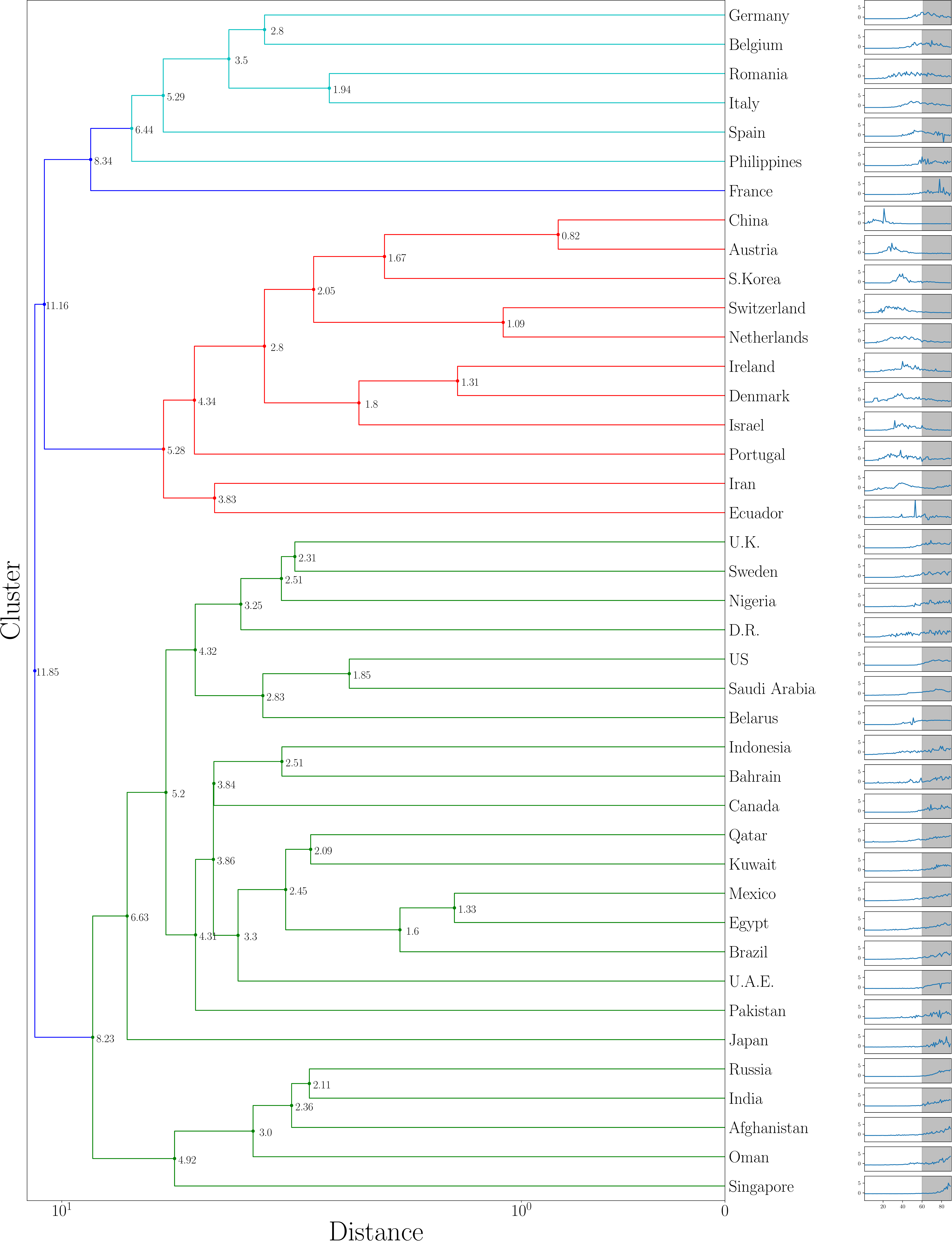}
	\caption{Dendrogram of 41 countries in the post-stage.}
	\label{fig:dendo_post}
\end{figure}

We use the publicly available COVID-19 time-series dataset published and maintained by the Johns Hopkins University, Center for Systems Science and Engineering (JHU CSSE). 
The dataset is updated daily on their GitHub repository~\cite{CovidDataset}. 
Confirmed, recovered and death cases are available for both country and state/county levels (for some countries such as Australia, Canada, China, France, Netherlands, U.K., and USA) in different files. 
The dataset consists of the cumulative number of confirmed cases from 22 January 2020 to date. 
Our study focuses on the daily incidence rate at the country level. 
Therefore, we merged all state/province level data into one record to represent their corresponding countries, removed the latitude and longitude information; and converted the cumulative data to daily case numbers.
Next, we transformed the data using standardization~\cite{wang2012use}. 
A standard score is defined as the difference of a sample in population and mean of the population, divided by the standard deviation (SD) of the population. 
Standardization is needed to convert all data into a similar scale so that all countries have the same weight in clustering.
We applied the following equation to transform the data into their z-scores:
\begin{equation}
	z_{ij}= \frac{x_{ij}-\mu_{j} }{\sigma_{j} }
\end{equation}
where $z_{ij}$ is the standardized $i$-th case number of the $j$-th country, $x_{ij}$ is the $i$-th case number of the $j$-th country,  $\mu_{j}$ is the mean of the daily case numbers of the $j$-th country and $\sigma_{j}$ is the standard deviation of the daily case numbers of $j$-th country.

Next, we apply hierarchical clustering with complete linkage and DTW as the distance metric with a window size of seven.
The daily new confirmed cases linearly increase or decrease throughout a week in most of the countries.
Due to weekly holidays, Saturdays and Sundays have fewer new cases whereas Mondays and Tuesdays have more new cases comparatively.
Therefore, we set the window size to seven to precisely capture the weekly spread pattern among different countries.
To ensure the reproducibility of our results, we used python's sci-kit-learn package ~\cite{scikit-learn} for agglomerative hierarchical clustering, scipy package ~\cite{scipy} for dendrograms, and dtaidistance package~\cite{DTW} for DTW.

\section{Results}\label{results}

\begin{figure}[t!]
	\centering
	\includegraphics[width=0.65\linewidth]{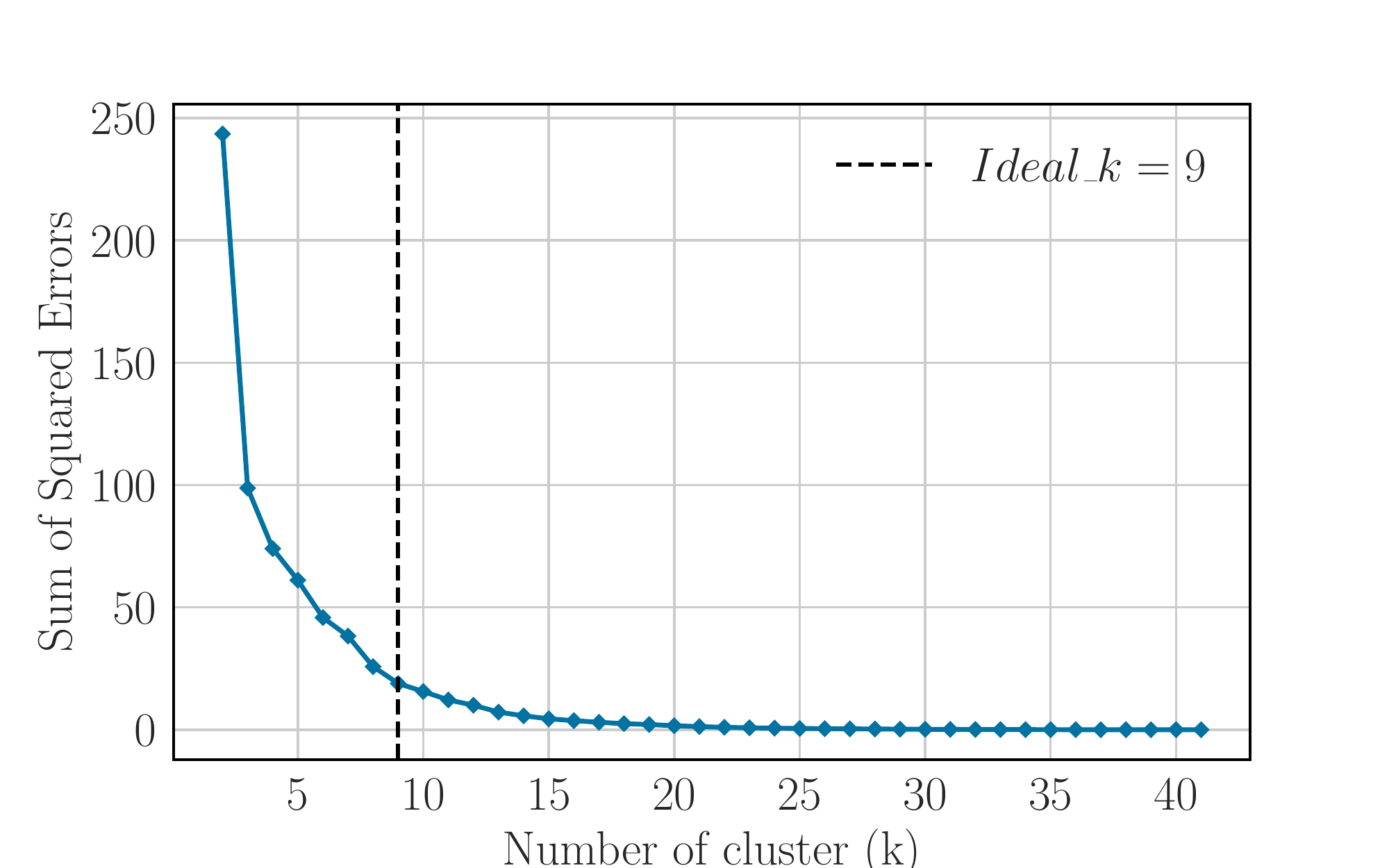}
	\caption{Ideal number of clusters using Elbow Method.}
	\label{fig:train_performance}
\end{figure}

We generate dendrograms for both early and post stages.
Dendrogram of 41 countries with at least 10,000 cases along with their respective time-series for early and post stages are shown in Figure~\ref{fig:dendo_early} and Figure~\ref{fig:dendo_post}, respectively.
The formations of clusters and their symmetrical log distances are shown on the left-hand side of the dendrograms.
The time series of each country is shown on the right-hand side of the dendrograms, highlighting the early-stage in Figure~\ref{fig:dendo_early} and the post-stage in Figure~\ref{fig:dendo_post}.
By visually inspecting the dendrogram of early-stage, we deduced that the ideal number of clusters is nine. 
We validated our decision by using the elbow method as shown in Figure~\ref{fig:train_performance}. 
We computed the within-cluster sum of squared errors for 1 to $k$ clusters, where $k$ is the maximum number of countries in the dataset. 
Thus, the value of $k$ is 41 which is the total number of countries present in the dataset after preprocessing. 
We also set the ideal number of clusters in the post-stage to 9, same as the early-stage, to be able to track the changes between the early and post stages.

\begin{table*}[t!]
	\centering
	\caption{Country-wise type and level of measures taken by the government.}
	\label{tab:measures}
	\renewcommand{\arraystretch}{0.85}
	\resizebox{1\linewidth}{!}{%
		\begin{tabular}{|c|c|c|c|c|c|}
			\hline
			\textbf{Countries} &
			\textbf{\begin{tabular}[c]{@{}c@{}}First Case \\ Confirmed\end{tabular}} &
			\textbf{Type of Measure} &
			\textbf{Level} &
			\textbf{Start Date} &
			\textbf{End Date} \\ \hline \hline
			Afghanistan  & 24 Feb     & Lockdown            & State   & 28 March (Day 33) & 21 May (Day 87)   \\ \hline
			Austria      & 25 Feb     & Lockdown            & Country & 16 March (Day 20) & 13 April (Day 48) \\ \hline
			Bahrain &
			24 Feb &
			Strict restrictions &
			Country &
			18 March (Day 23) &
			7 May (Day 63) \\ \hline
			Belarus      & 28 Feb     & -                   & -       & No Lockdown       & No Lockdown       \\ \hline
			Belgium      & 4 Feb      & Lockdown            & Country & 18 March (Day 43) & 4 May (Day 90)    \\ \hline
			Brazil       & 26 Feb     & Lockdown            & State   & 17 March (Day 20) & 7 April (Day 41)  \\ \hline
			Canada       & 26 Jan     & Strict restrictions & State   & 18 March (Day 52) & 11 May            \\ \hline
			China        & 31 Dec     & Lockdown            & State   & 23 Jan (Day 1)    & 8 April (Day 76)  \\ \hline
			\begin{tabular}[c]{@{}c@{}}Dominican \\ Republic\end{tabular} &
			1 March &
			Strict restrictions &
			Country &
			19 March (Day 18) &
			Not Lifted \\ \hline
			Denmark      & 27 Feb     & Lockdown            & Country & 11 March (Day 13) & 13 April (Day 46) \\ \hline
			Ecuador      & 1 March    & Lockdown            & Country & 15 March (Day 14) & Not Lifted        \\ \hline
			Egypt        & 14 Feb     & Lockdown            & Country & 19 March (Day 34) & Not Lifted        \\ \hline
			France       & 24 Jan     & Lockdown            & Country & 17 March (Day 53) & 11 May            \\ \hline
			Germany      & 27 Jan     & Lockdown            & Country & 23 March (Day 56) & 10 May            \\ \hline
			India        & 30 Jan     & Lockdown            & Country & 25 March (Day 55) & Not Lifted        \\ \hline
			Indonesia    & 2 March    & Strict restrictions & Country & 24 April (Day 53) & 1 June            \\ \hline
			Iran         & 19 Feb     & Lockdown            & Country & 14 March (Day 24) & 20 April (Day 61) \\ \hline
			Ireland      & 29 Feb     & Lockdown            & Country & 12 March (Day 12) & 18 May (Day 78)   \\ \hline
			Israel       & 21 Feb     & Lockdown            & State   & 2 April (Day 41)  & Not Lifted        \\ \hline
			Italy        & 31 Jan     & Lockdown            & Country & 8 March (Day 37)  & 18 May            \\ \hline
			Japan        & 16 Jan     & -                   & -       & No Lockdown       & No Lockdown       \\ \hline
			Kuwait       & 24 Feb     & Strict restrictions & Country & 14 March (Day 19) & 31 May            \\ \hline
			Mexico       & 28 Feb     & Lockdown            & Country & 23 March (Day 24) & 1 June            \\ \hline
			Netherlands  & 27 Feb     & -                   & -       & No Lockdown       & No Lockdown       \\ \hline
			Nigeria      & 28 Feb     & Lockdown            & State   & 30 March (Day 31) & 4 May (Day 66)    \\ \hline
			Oman         & 24 Feb     & Lockdown            & State   & 10 April (Day 46) & 29 May            \\ \hline
			Pakistan     & 26 Feb     & Lockdown            & Country & 22 March (Day 25) & 9 May (Day 73)    \\ \hline
			Philippines  & 30 Jan     & Lockdown            & State   & 15 March (Day 45) & 31 May            \\ \hline
			Portugal     & 2 March    & Lockdown            & Country & 19 March (Day 17) & 2 May (Day 61)    \\ \hline
			Qatar        & 29 Feb     & Lockdown            & State   & 11 March (Day 12) & Not Lifted        \\ \hline
			Romania      & 26 Feb     & Lockdown            & Country & 25 March (Day 28) & 15 May (Day 78)   \\ \hline
			Russia       & 31 Jan     & Lockdown            & Country & 30 March (Day 59) & 12 May            \\ \hline
			South Korea  & 20 Jan     & -                   & -       & No Lockdown       & No Lockdown       \\ \hline
			Saudi Arabia & 2 March    & Lockdown            & State   & 23 March (Day 21) & Not Lifted        \\ \hline
			Singapore    & 23 January & Lockdown            & Country & 7 April (Day 75)  & Not Lifted        \\ \hline
			Spain        & 1 Feb      & Lockdown            & Country & 14 March (Day 42) & Not lifted        \\ \hline
			Sweden       & 31 Jan     & -                   & -       & No Lockdown       & No Lockdown       \\ \hline
			Switzerland &
			25 Feb &
			Strict restrictions &
			Country &
			16 March (Day 20) &
			11 May (Day 76) \\ \hline
			U.A.E.       & 29 Jan     & Lockdown            & Country & 5 April (Day 67)  & 28 May            \\ \hline
			U.K.         & 31 Jan     & Lockdown            & Country & 23 March (Day 52) & 1 June            \\ \hline
			US           & 20 Jan     & Lockdown            & State   & 16 March (Day 54) & 27 April-30 May   \\ \hline
		\end{tabular}%
	}
\end{table*}

\begin{figure*}[t!]
	\centering
	\begin{subfigure}[]{0.33\linewidth}
		\includegraphics[width=\linewidth]{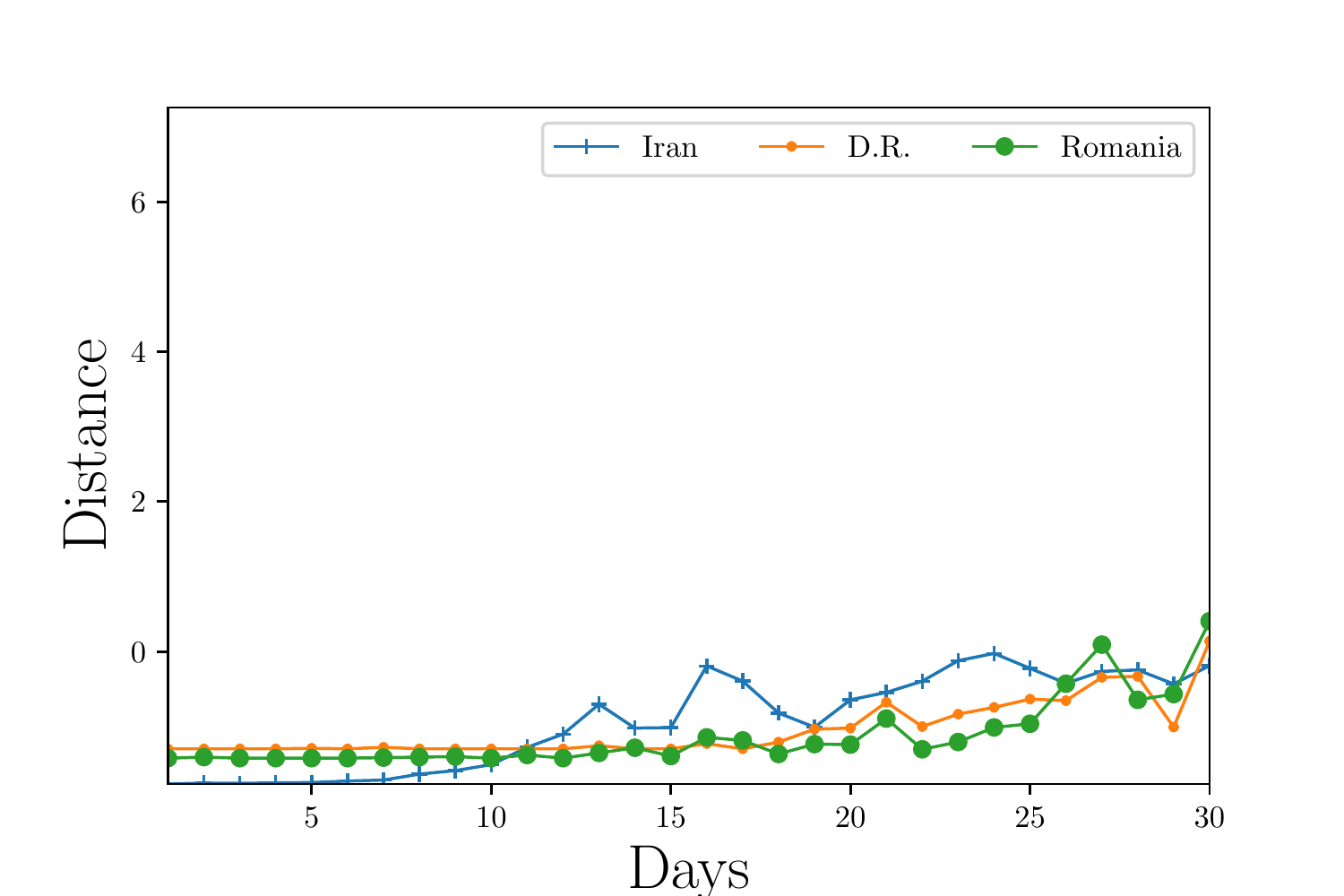}
		\subcaption{Cluster-1}
		\label{fig:Early_stage_cluster_1}
	\end{subfigure}\hfill
	\begin{subfigure}[]{0.33\linewidth}
		\includegraphics[width=\linewidth]{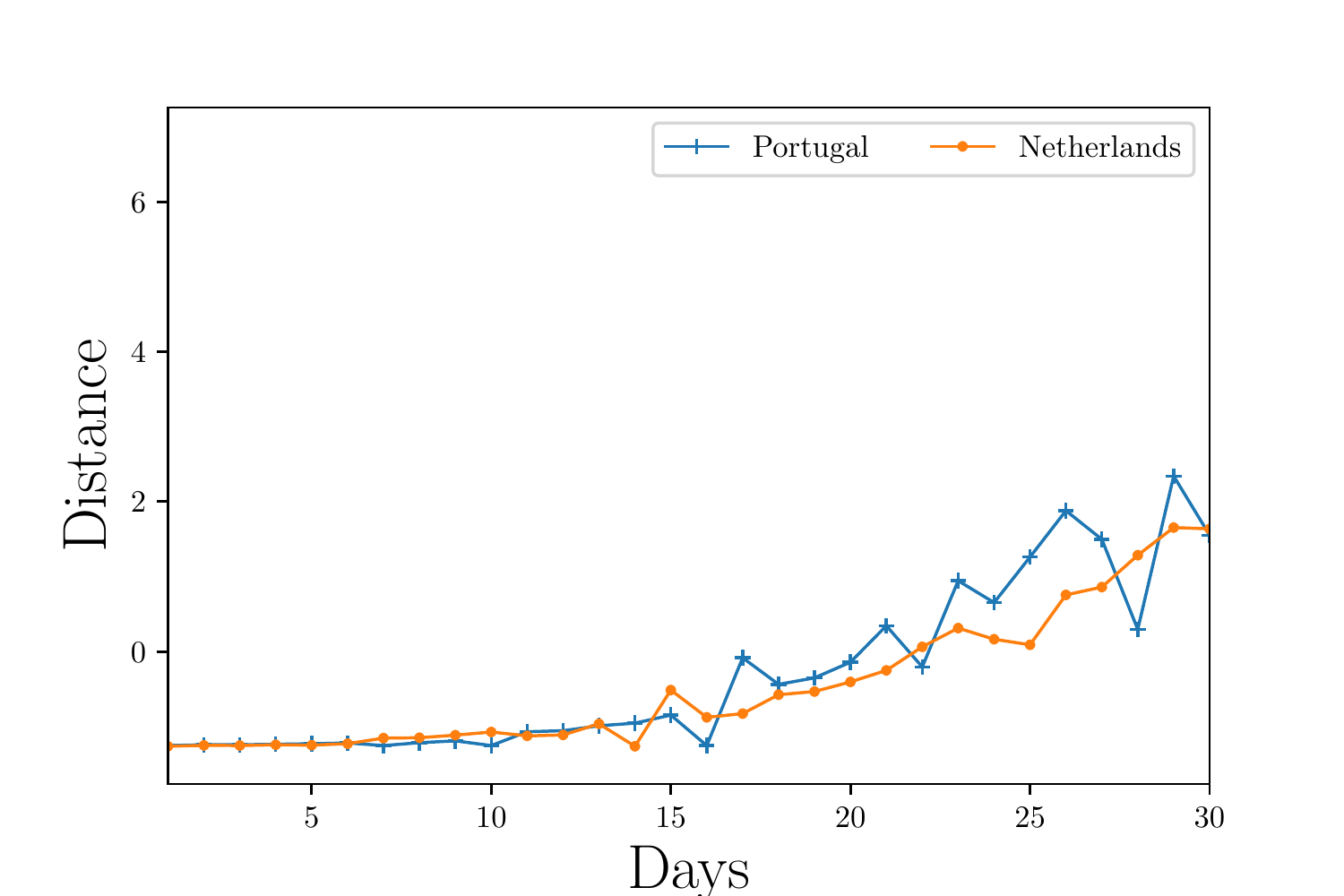}
		\subcaption{Cluster-2}
		\label{fig:Early_stage_cluster_2}
	\end{subfigure}\hfill
	\begin{subfigure}[]{0.33\linewidth}
		\includegraphics[width=\linewidth]{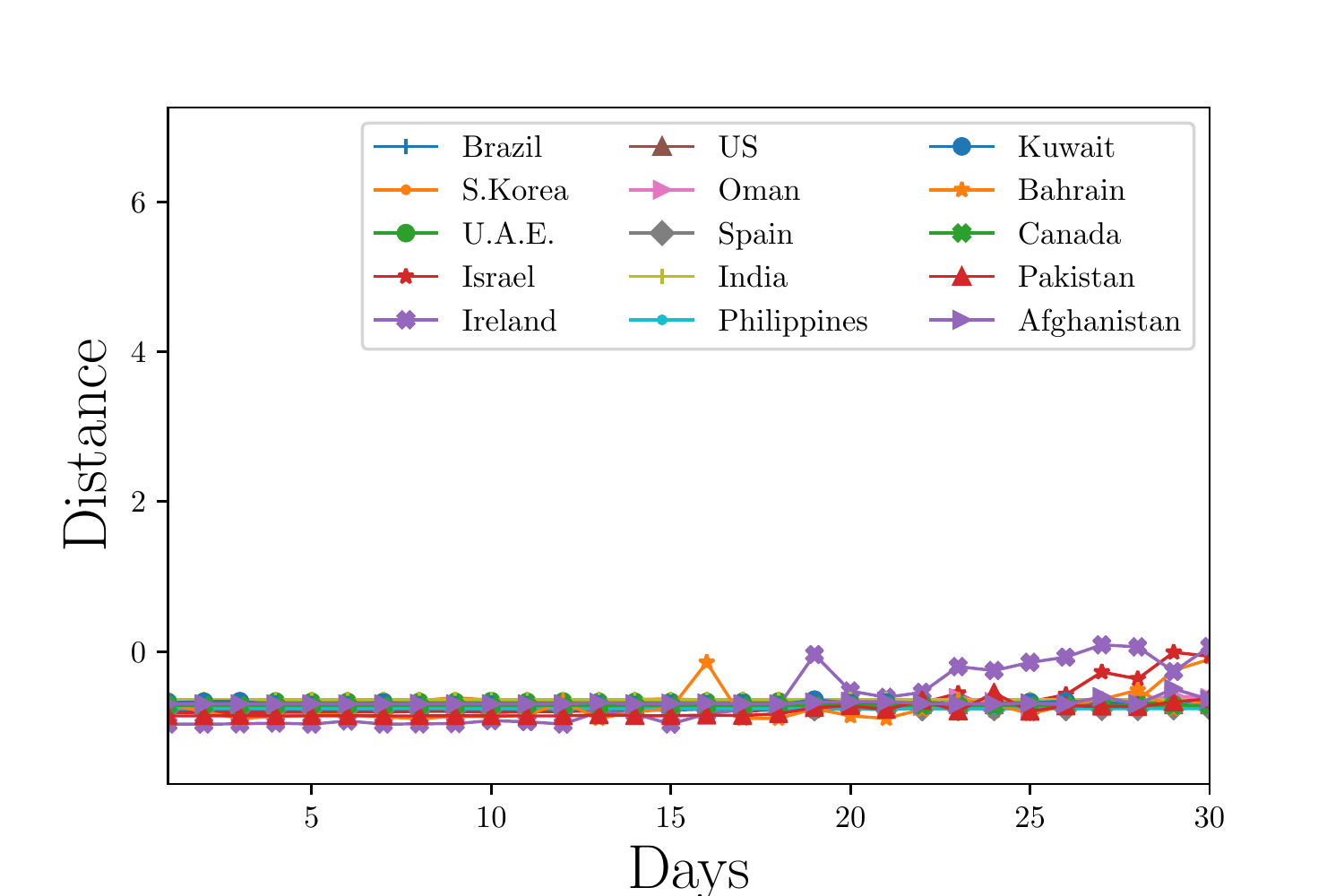}
		\subcaption{Cluster-3}
		\label{fig:Early_stage_cluster_3}
	\end{subfigure}\hfill
	\begin{subfigure}[]{0.33\linewidth}
		\includegraphics[width=\linewidth]{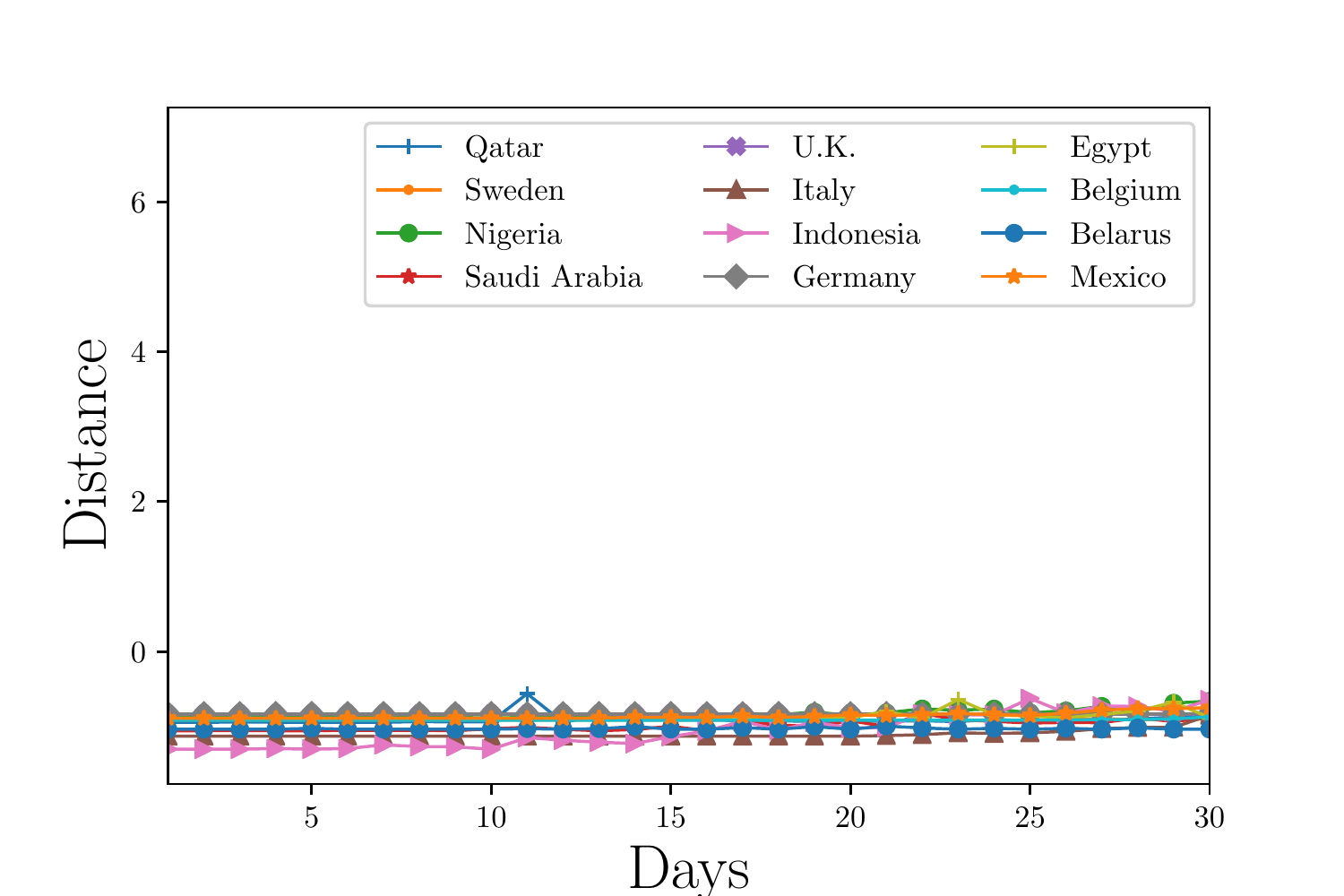}
		\subcaption{Cluster-4}
		\label{fig:Early_stage_cluster_4}
	\end{subfigure}\hfill
	\begin{subfigure}[]{0.33\linewidth}
		\includegraphics[width=\linewidth]{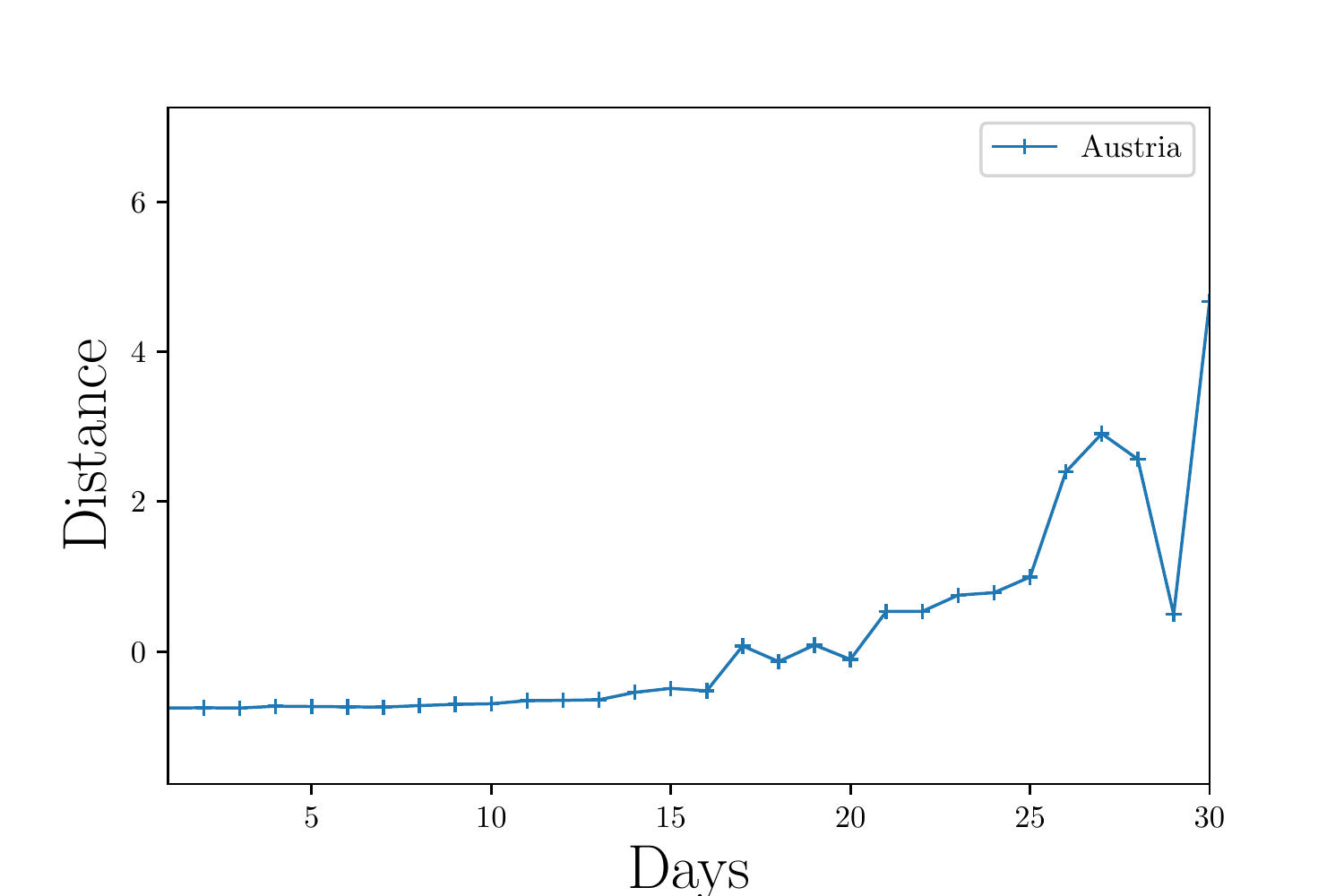}
		\subcaption{Cluster-5}
		\label{fig:Early_stage_cluster_5}
	\end{subfigure}\hfill
	\begin{subfigure}[]{0.33\linewidth}
		\includegraphics[width=\linewidth]{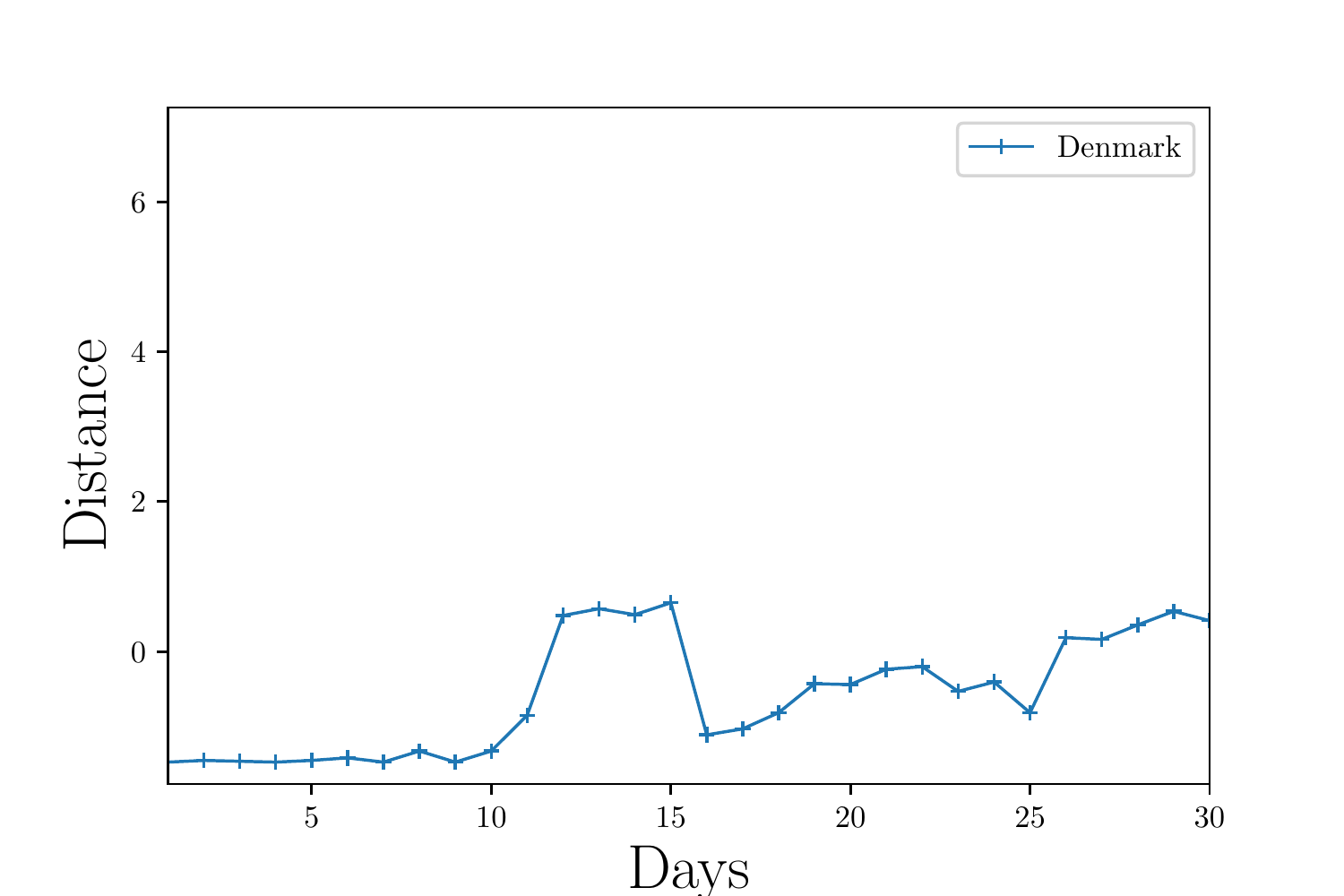}
		\subcaption{Cluster-6}
		\label{fig:Early_stage_cluster_6}
	\end{subfigure}\hfill
	\begin{subfigure}[]{0.33\linewidth}
		\includegraphics[width=\linewidth]{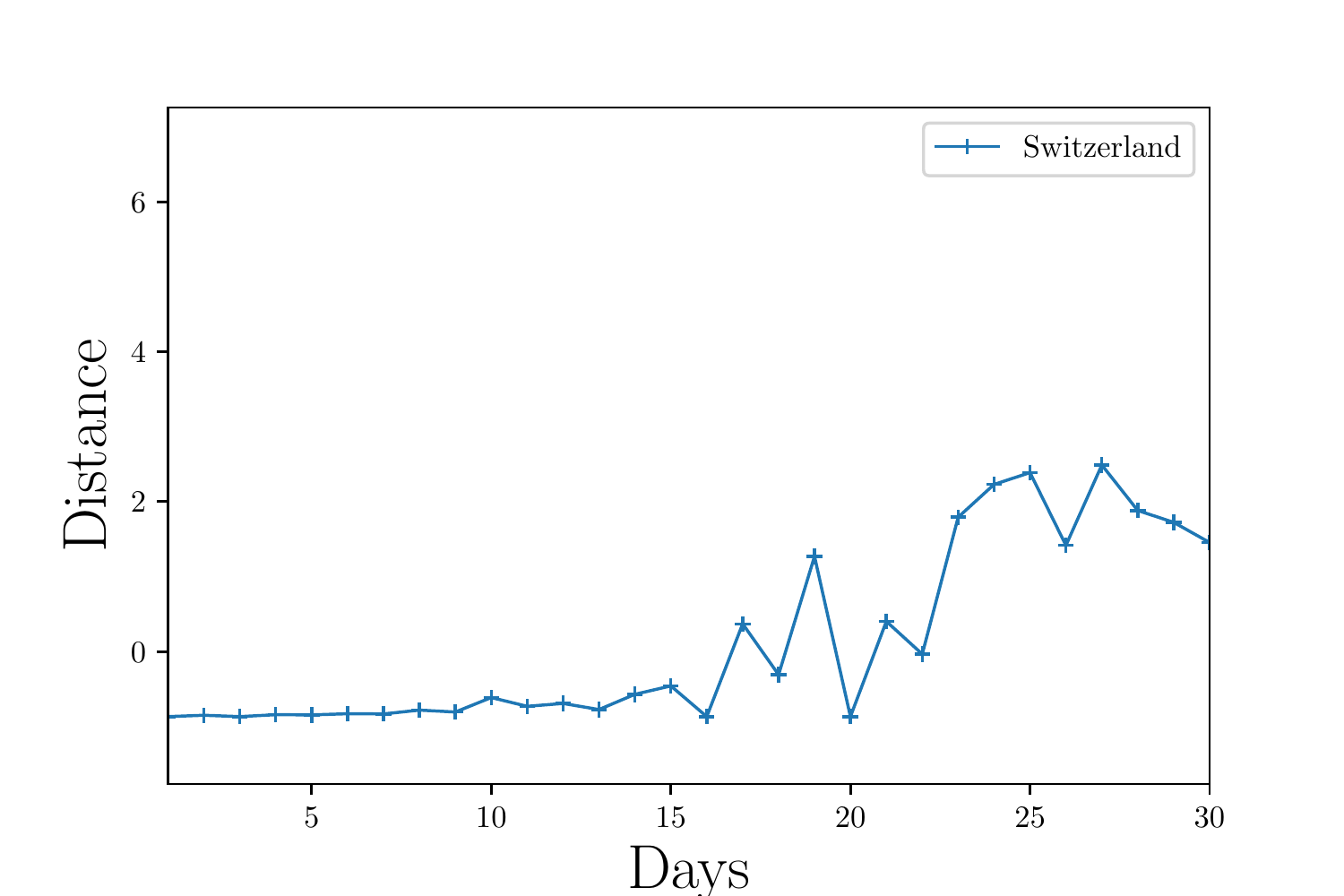}
		\subcaption{Cluster-7}
		\label{fig:Early_stage_cluster_7}
	\end{subfigure}\hfill
	\begin{subfigure}[]{0.33\linewidth}
		\includegraphics[width=\linewidth]{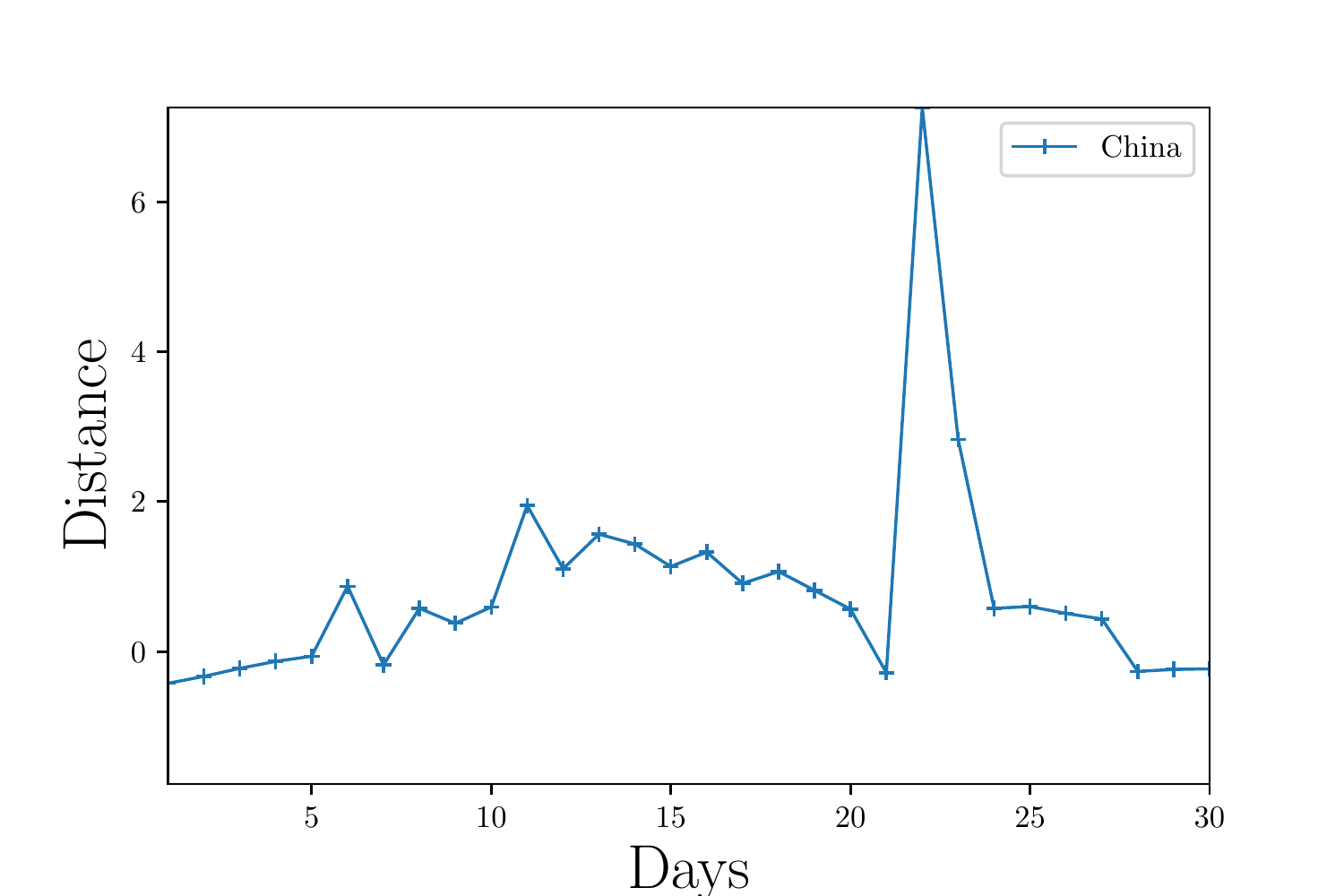}
		\subcaption{Cluster-8}
		\label{fig:Early_stage_cluster_8}
	\end{subfigure}\hfill
	\begin{subfigure}[]{0.33\linewidth}
		\includegraphics[width=\linewidth]{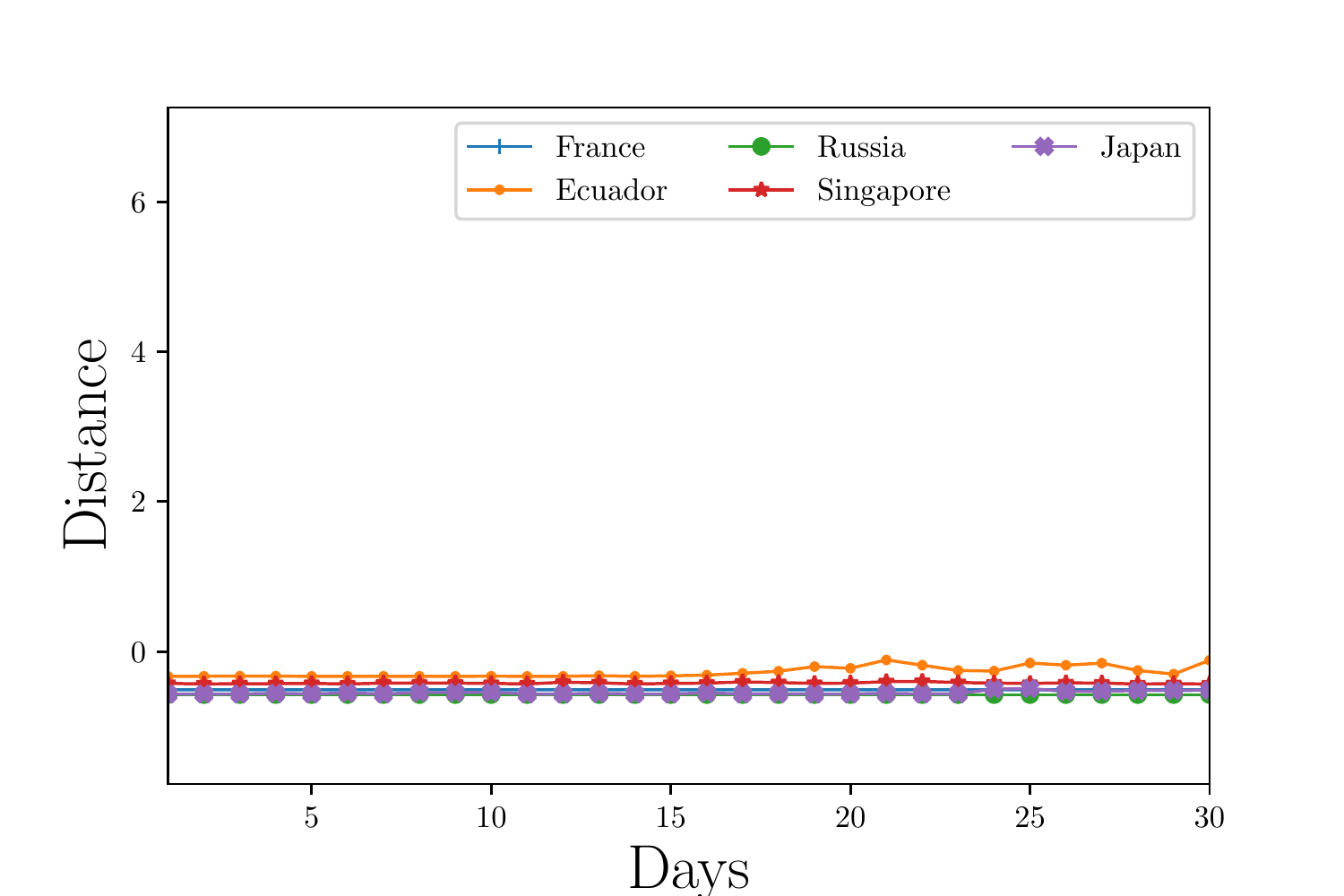}
		\subcaption{Cluster-9}
		\label{fig:Early_stage_cluster_9}
	\end{subfigure}\hfill
	\caption{Different clusters in the early-stage after applying Agglomerative Hierarchical Clustering.}
	\label{fig:early_stage_clusters}
\end{figure*}

\begin{figure*}[t!]
	\centering
	\begin{subfigure}[]{0.33\linewidth}
		\includegraphics[width=\linewidth]{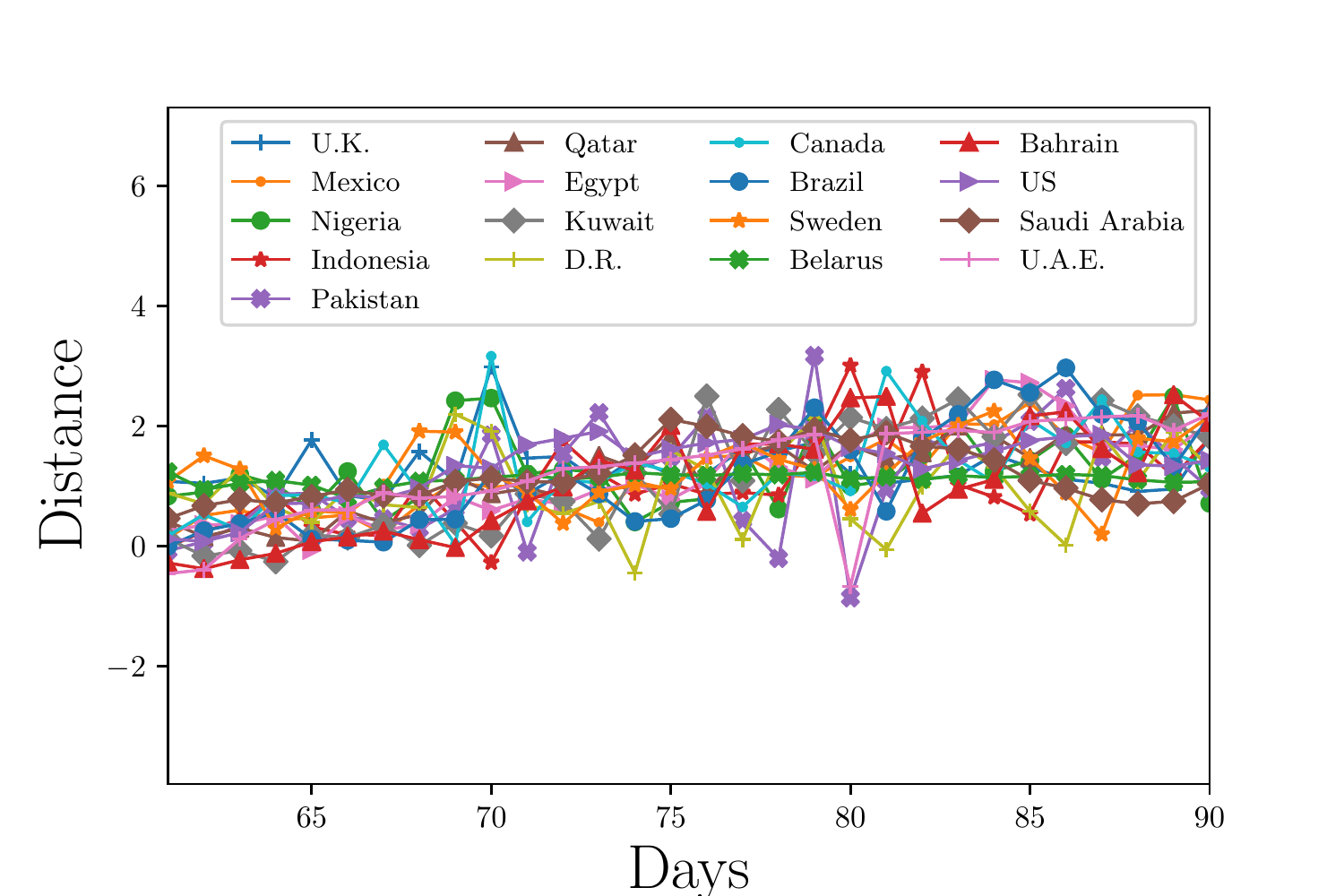}
		\subcaption{Cluster-1}
		\label{fig:Late_stage_cluster_1}
	\end{subfigure}\hfill
	\begin{subfigure}[]{0.33\linewidth}
		\includegraphics[width=\linewidth]{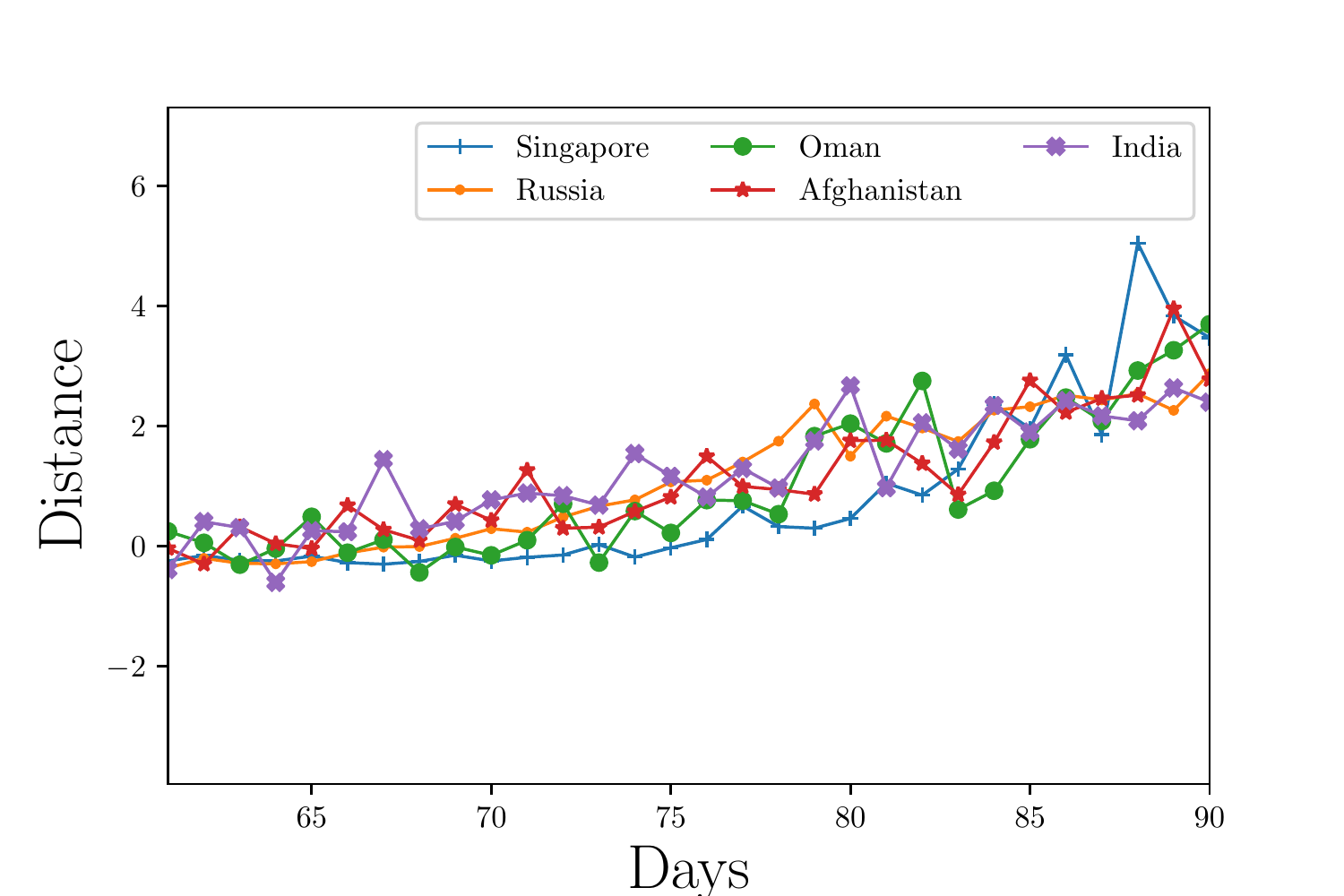}
		\subcaption{Cluster-2}
		\label{fig:Late_stage_cluster_2}
	\end{subfigure}\hfill
	\begin{subfigure}[]{0.33\linewidth}
		\includegraphics[width=\linewidth]{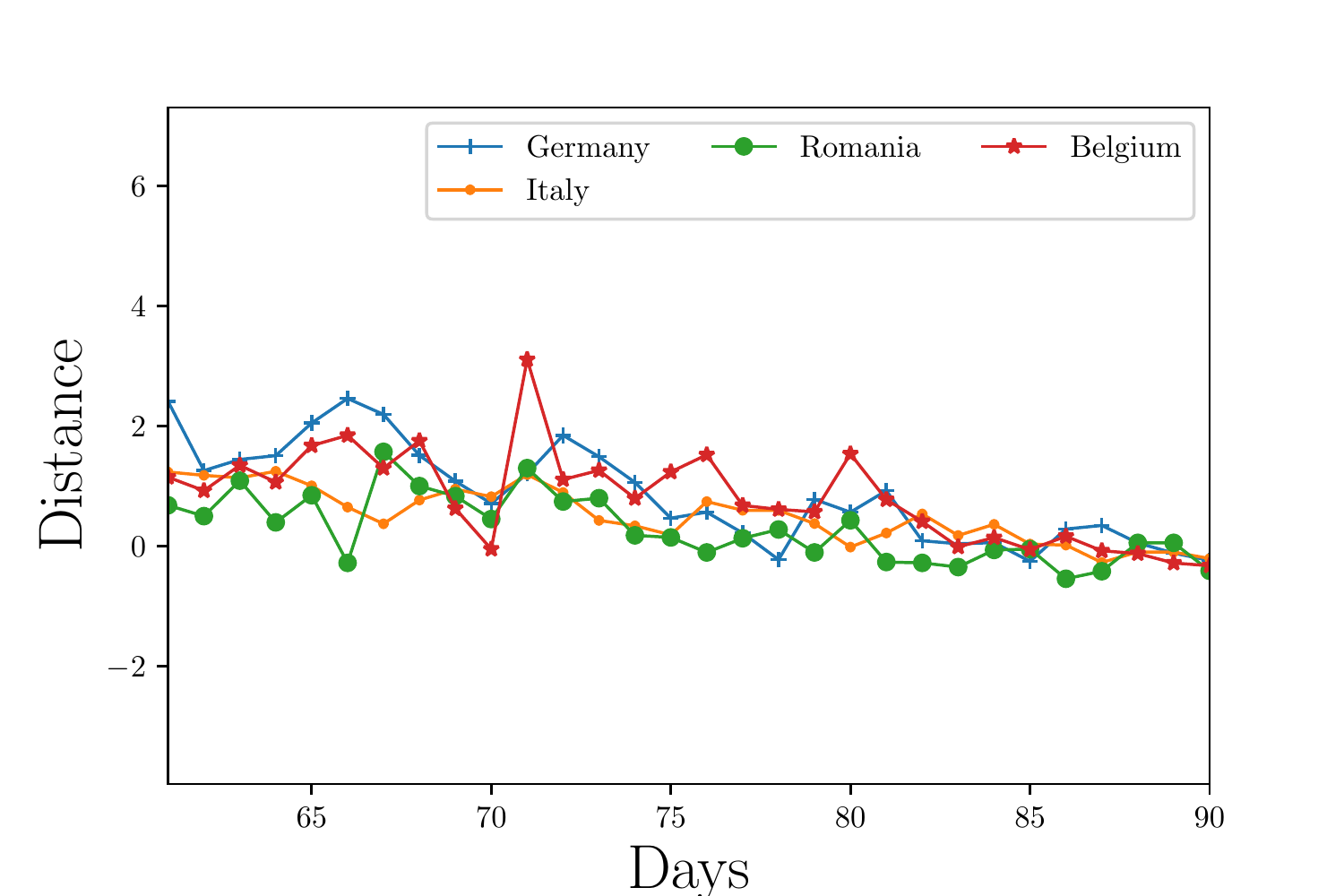}
		\subcaption{Cluster-3}
		\label{fig:Late_stage_cluster_3}
	\end{subfigure}\hfill
	\begin{subfigure}[]{0.33\linewidth}
		\includegraphics[width=\linewidth]{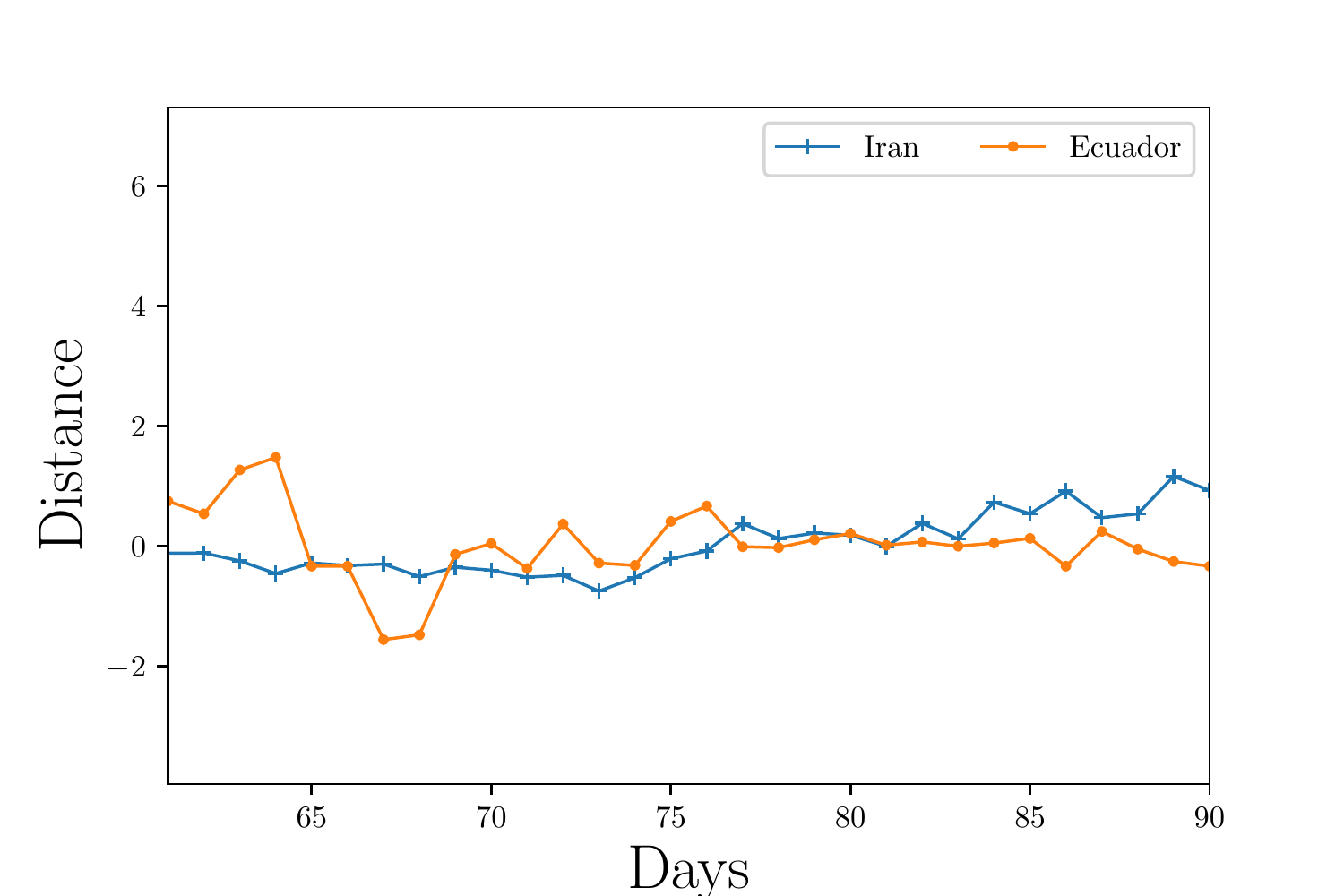}
		\subcaption{Cluster-4}
		\label{fig:Late_stage_cluster_4}
	\end{subfigure}\hfill
	\begin{subfigure}[]{0.33\linewidth}
		\includegraphics[width=\linewidth]{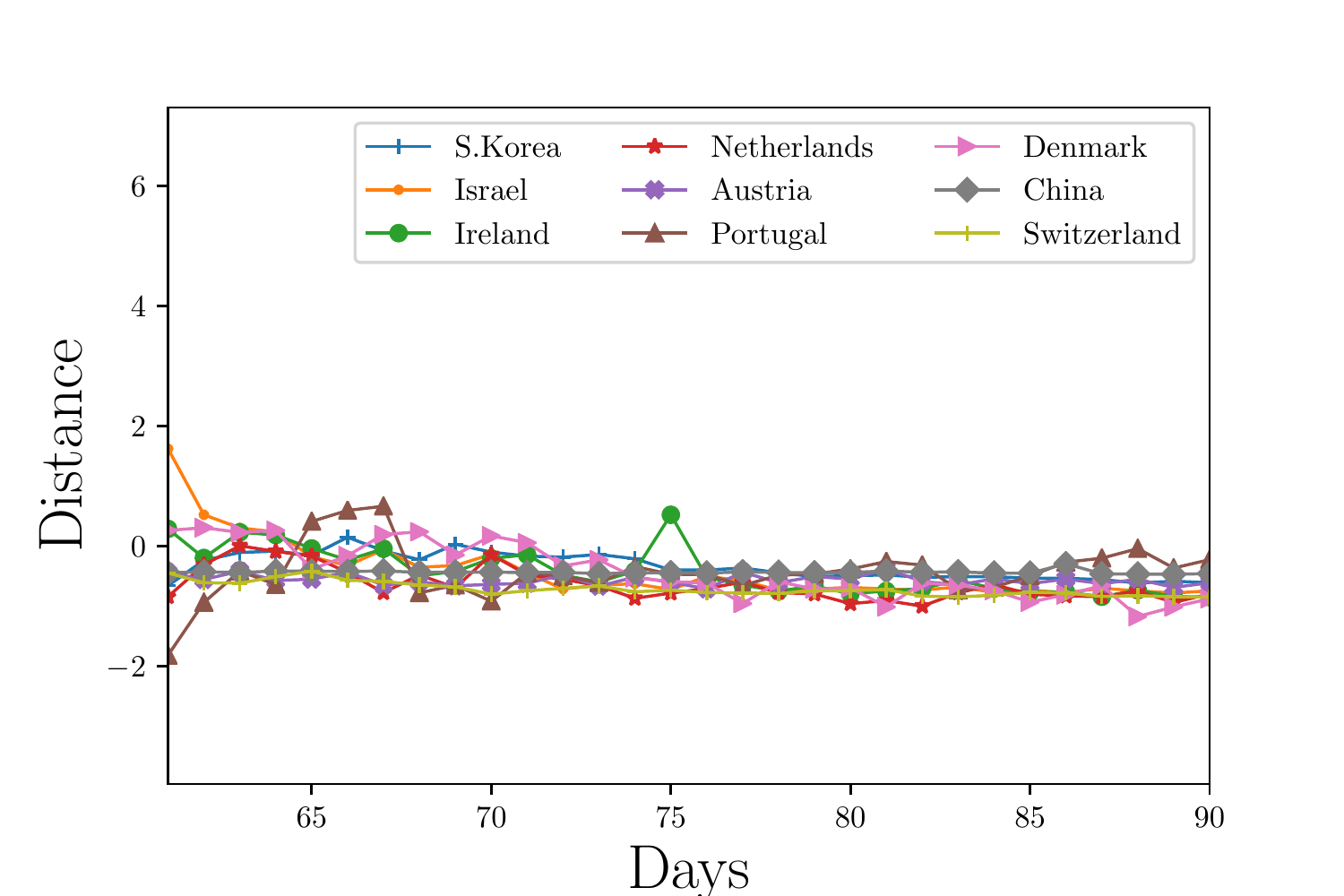}
		\subcaption{Cluster-5}
		\label{fig:Late_stage_cluster_5}
	\end{subfigure}\hfill
	\begin{subfigure}[]{0.33\linewidth}
		\includegraphics[width=\linewidth]{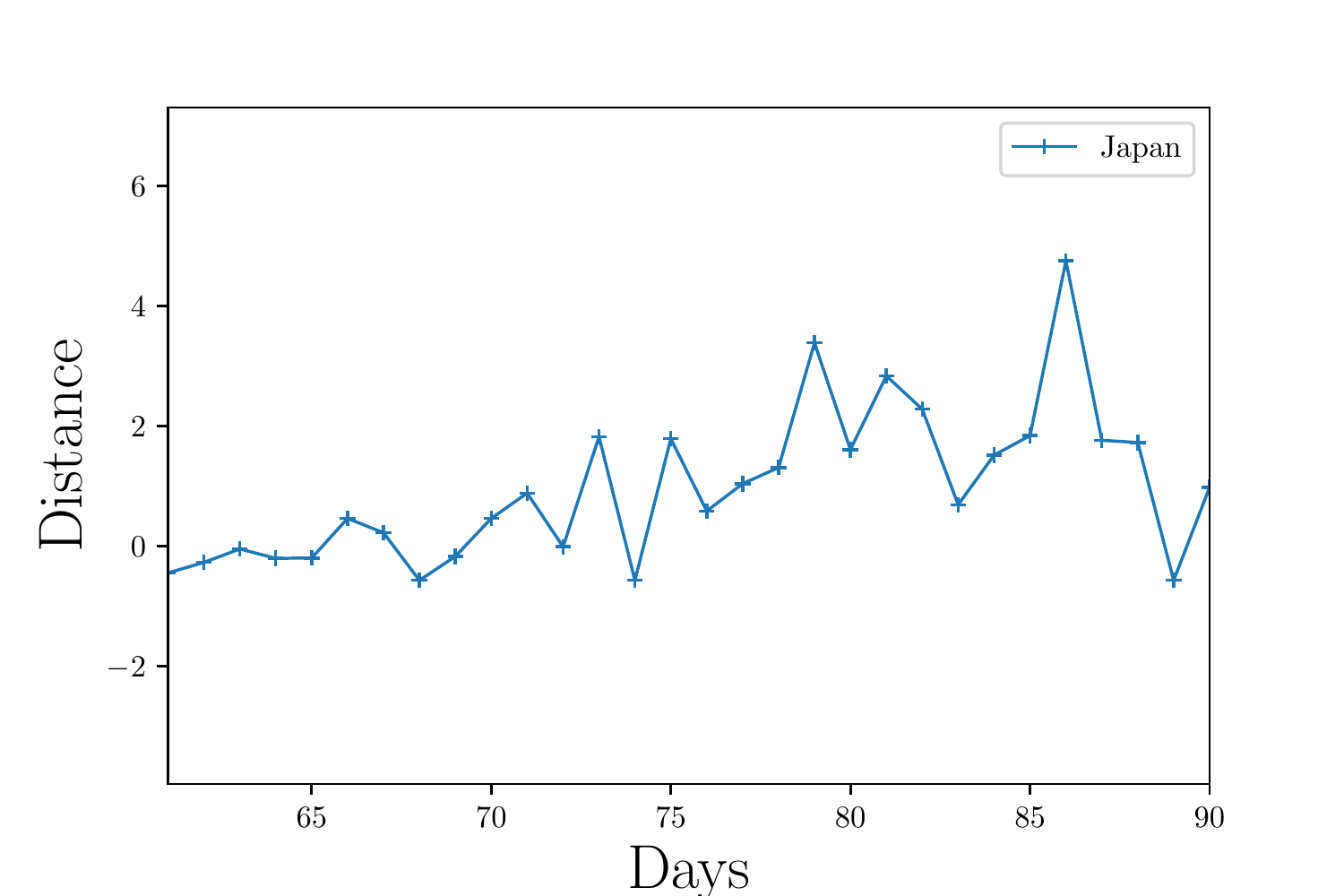}
		\subcaption{Cluster-6}
		\label{fig:Late_stage_cluster_6}
	\end{subfigure}\hfill
	\begin{subfigure}[]{0.33\linewidth}
		\includegraphics[width=\linewidth]{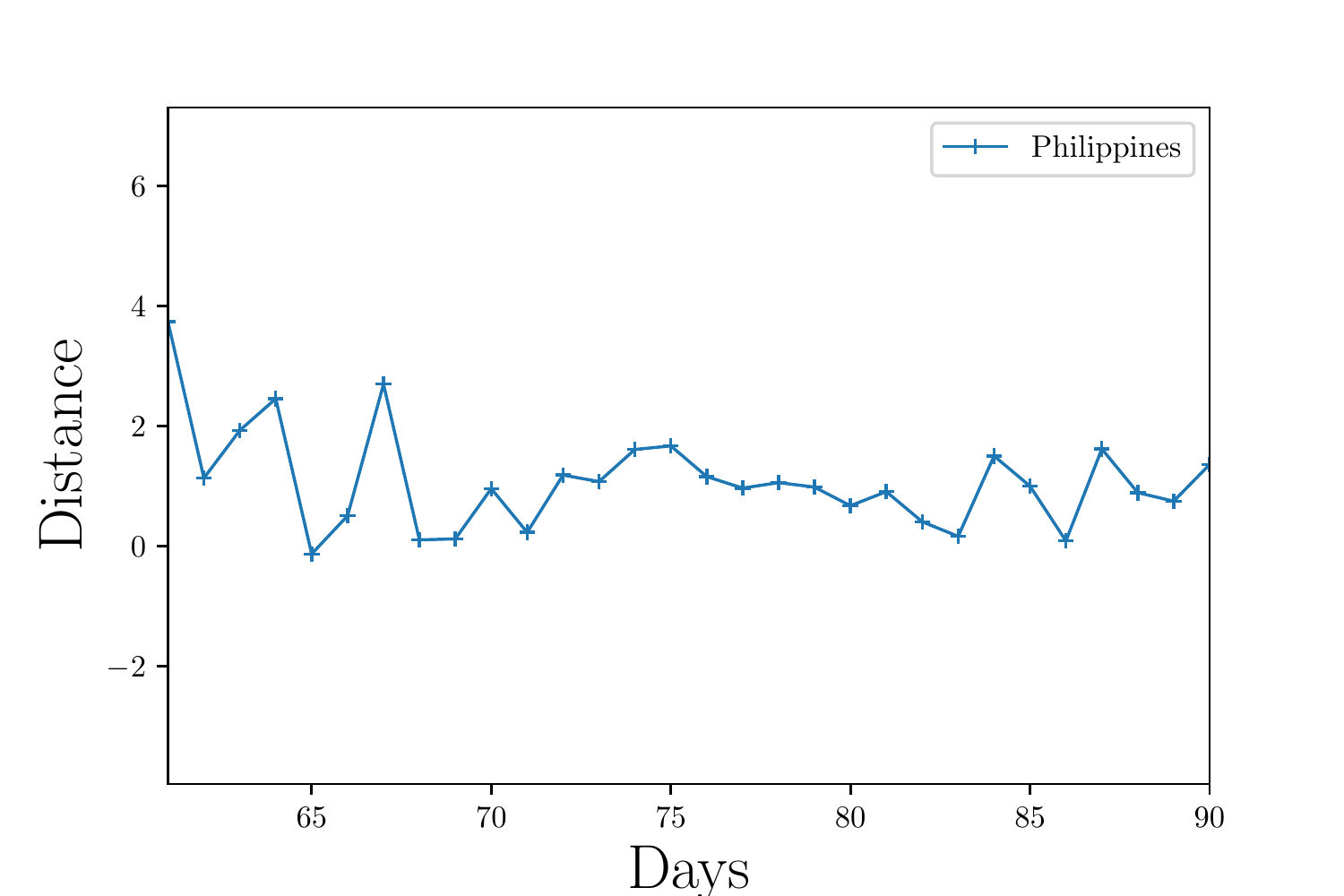}
		\subcaption{Cluster-7}
		\label{fig:Late_stage_cluster_7}
	\end{subfigure}\hfill
	\begin{subfigure}[]{0.33\linewidth}
		\includegraphics[width=\linewidth]{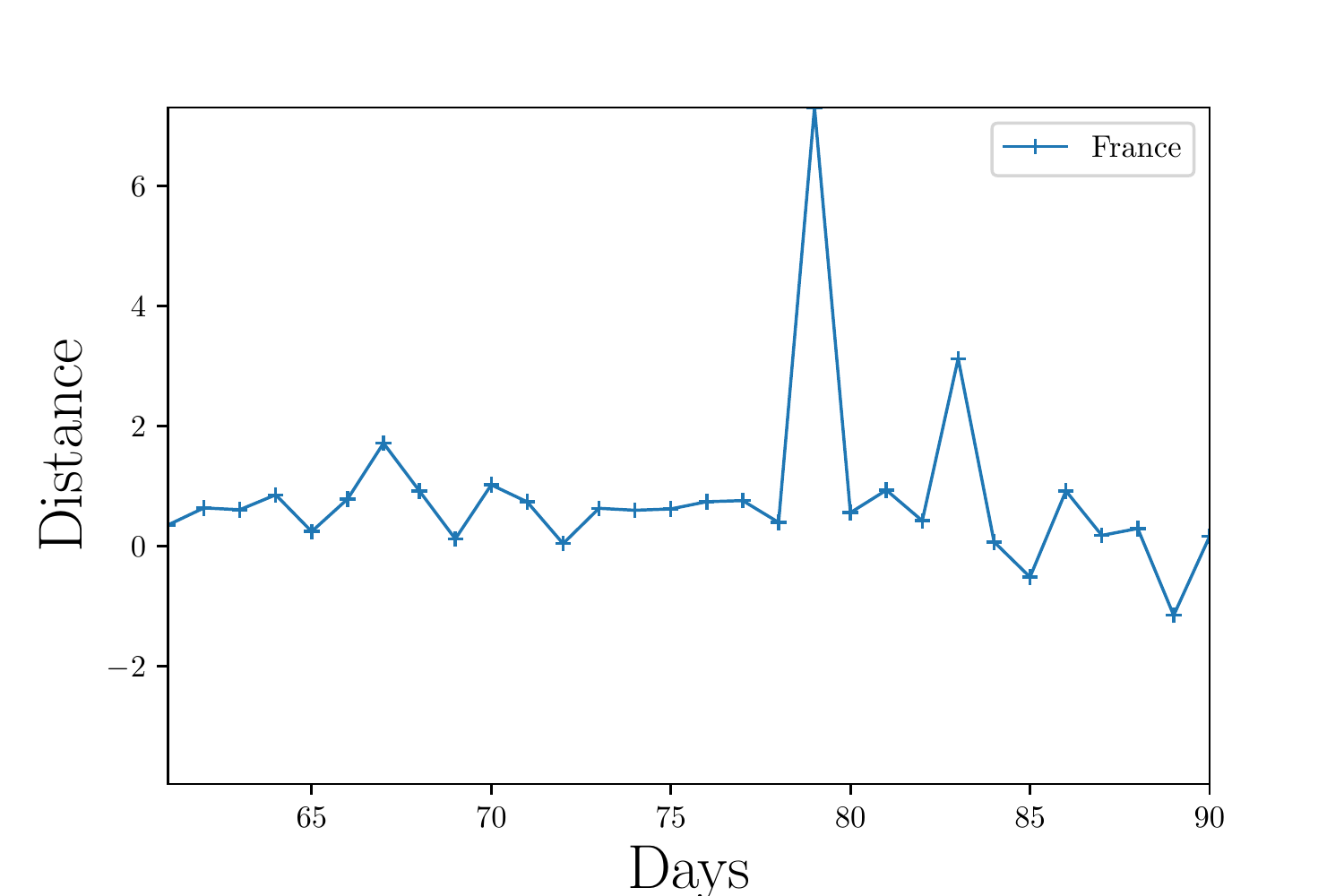}
		\subcaption{Cluster-8}
		\label{fig:Late_stage_cluster_8}
	\end{subfigure}\hfill
	\begin{subfigure}[]{0.33\linewidth}
		\includegraphics[width=\linewidth]{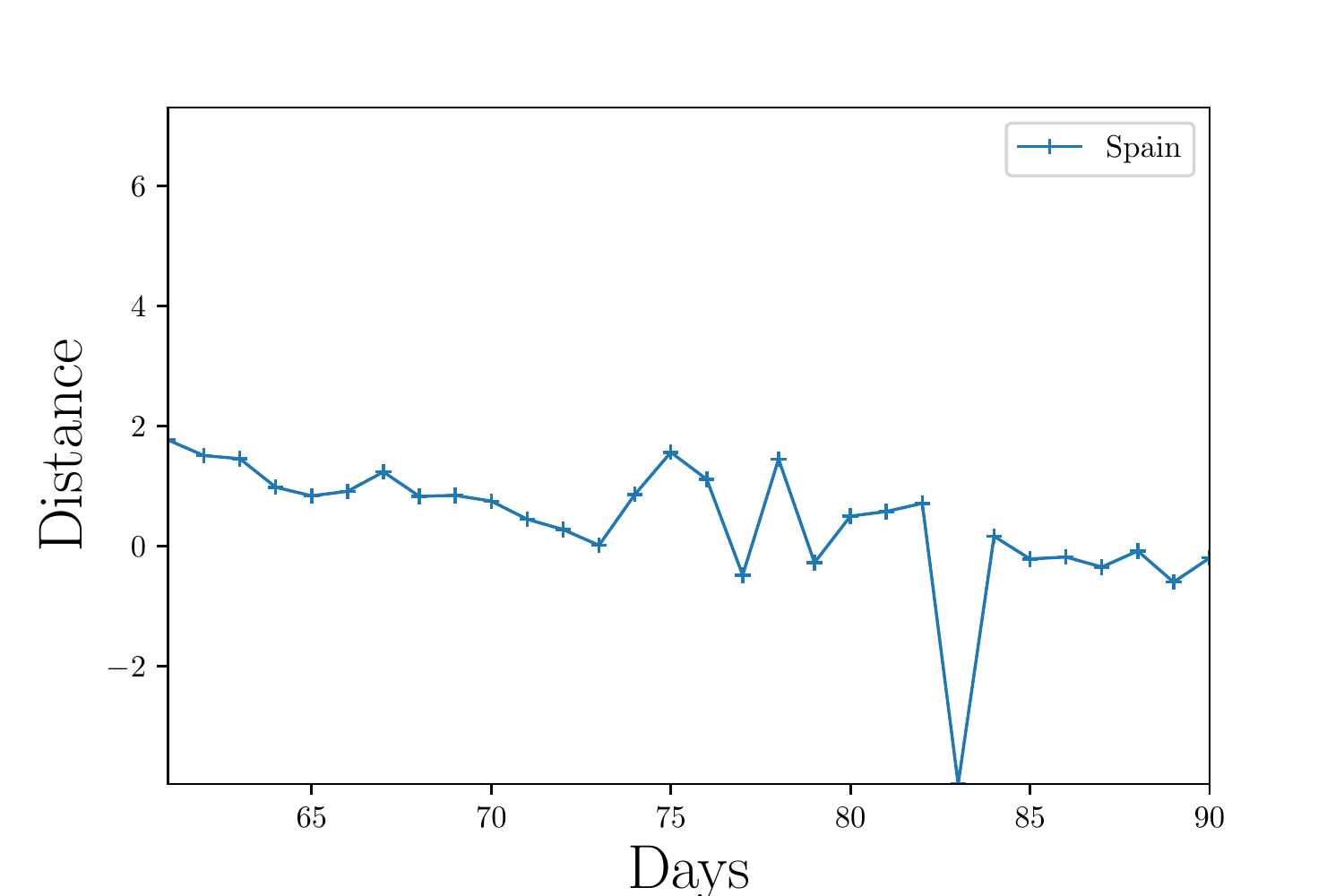}
		\subcaption{Cluster-9}
		\label{fig:Late_stage_cluster_9}
	\end{subfigure}\hfill
	\caption{Different clusters in the post-stage after applying Agglomerative Hierarchical Clustering.}
	\label{fig:Late_stage_clusters}
\end{figure*}

\cref{fig:early_stage_clusters,fig:Late_stage_clusters} show the time-series data of different countries residing in different clusters of the early and post stages, respectively.
In the early-stage, the largest cluster is Cluster-3 with 15 countries.
Cluster-4 is the second-largest one containing 12 countries.
Cluster 1, 2, and 9 have 3, 2, and 5 countries, respectively.
The rest of the clusters \emph{i.e.,} Cluster 5, 6, 7, and 8, are singleton clusters.
In the post-stage, Cluster-1 is the largest cluster with 17 countries.
The second-largest cluster is Cluster-5 containing 9 countries. 
Cluster 2, 3, and 4 have 5, 4, and 2 countries, respectively.
Similar to the early-stage, there are four singleton clusters in the post-stage,~\emph{i.e.,} Cluster 6, 7, 8, and 9.
The cluster labels obtained from performing agglomerative hierarchical clustering is shown in Appendix~\ref{appendix_cluster_labels}, Table~\ref{tab:cluster_labels}.
Please note that the cluster labels do not carry any additional information except that the countries clustered together exhibit similar behavior.
To emphasize this fact and demonstrate the countries clustered together better, we show the clusters on a world map in~\ref{appendix_cluster}, Figure~\ref{fig:world_map_clusters}.

Additionally, we investigated the confinement policies or preventive measures adopted by each country.
We collected the preventive-measures data from various credible news sources~\cite{afg_lock, afg_lift, austria_denmark_lock_lift, bahrain_lock, bahrain_lock_lift, belarus, Belgium_lock, Belgium_lock_lift, brazil_lock, canada_lock, canada_lock_lift, china_lock, china_lock_lift, dr_lock, ecu_lock, ecu_lock_lift, egypt_lock, france_lock, france_lock_lift, germany_lock, Germany_lock_lift, India_lock, India_lock_lift, Indonesia_lock, Iran_lock, Iran_lock_lift, Italy_lock, Italy_lock_lift, Japan_no_lock, Mexico_lock, Netherlands_no_lock, Nigeria_lock, Oman_lock_lift, Oman_lock, Pakistan_lock,  Pakistan_lock_lift, Philippines_lock, Portugal_lock, Portugal_lock_lift, Romania_lock, Romania_lock_lift, Russia_lock_lift, Qatar_lock, Korea_no_lock, algaissi2020preparedness, Singapore_lock, Spain_lock, Sweden_no_lock, Switzerland_lock_lift, Switzerland_lock, UAE_lock_lift, UAE_lock, UK_lock, USA_lock, USA_lock_lift}   
and summarised them in Table~\ref{tab:measures}.
Although we were not able to contact the health authorities of the countries presented in Table~\ref{tab:measures}, we assume that the information provided by the news outlets were correct and/or dependable.

\section{Discussion}\label{disucssion}

The only way to fight against the COVID-19 disease is to reduce the virus spread as much as possible, as no cure or vaccine had been developed in the early period of the pandemic.
To this end, various confinement policies or preventive measures have been taken by different countries around the world.
The travel bans, mobility restrictions, bans on large gatherings, social distancing, quarantine or isolation of infected patients, and lockdowns of infected cities, states or countries, are the most widely adopted the preventive measures.
Researchers have studied the effects of the preventive measures on the spread of COVID-19~\cite{chinazzi2020effect}.
In the early period of the pandemic, it is essential to restrict mobility within a city, state, or country as well as traveling outside of the infected areas~\cite{chinazzi2020effect, fang2020transmission}.
The government of China issued travel restrictions on January 22, 2020.
Chinazzi~\emph{et al.}~\cite{chinazzi2020effect} estimated that a 77\% reduction in cases exported from China to the outside world due to the travel restrictions.
Air travel restrictions have severely reduced the cases in some European countries~\cite{linka2020outbreak}.
However, the preventive measures can become less effective if a country already has a high infection rate~\cite{fang2020transmission}.
Therefore, the data of each country is needed to be analyzed individually to recognize the factors that affect the spread.

We utilize the Google COVID-19 Community Mobility Reports~\cite{mob_Dataset} to investigate the effect of lockdowns or other preventive measures on the mobility of the people.
The mobility data at any period is compared with a baseline.
The baseline is calculated using a 5-week period from January 3 to February 6, 2020.
Positive values mean increased mobility, whereas negative values indicate reduced mobility.
The mobility data is available for all countries except China and Iran.

\subsection{Early-Stage: Cluster-1}

\begin{figure*}[t!]
	\centering
	\begin{subfigure}[b]{0.48\linewidth}
		\includegraphics[width=\linewidth]{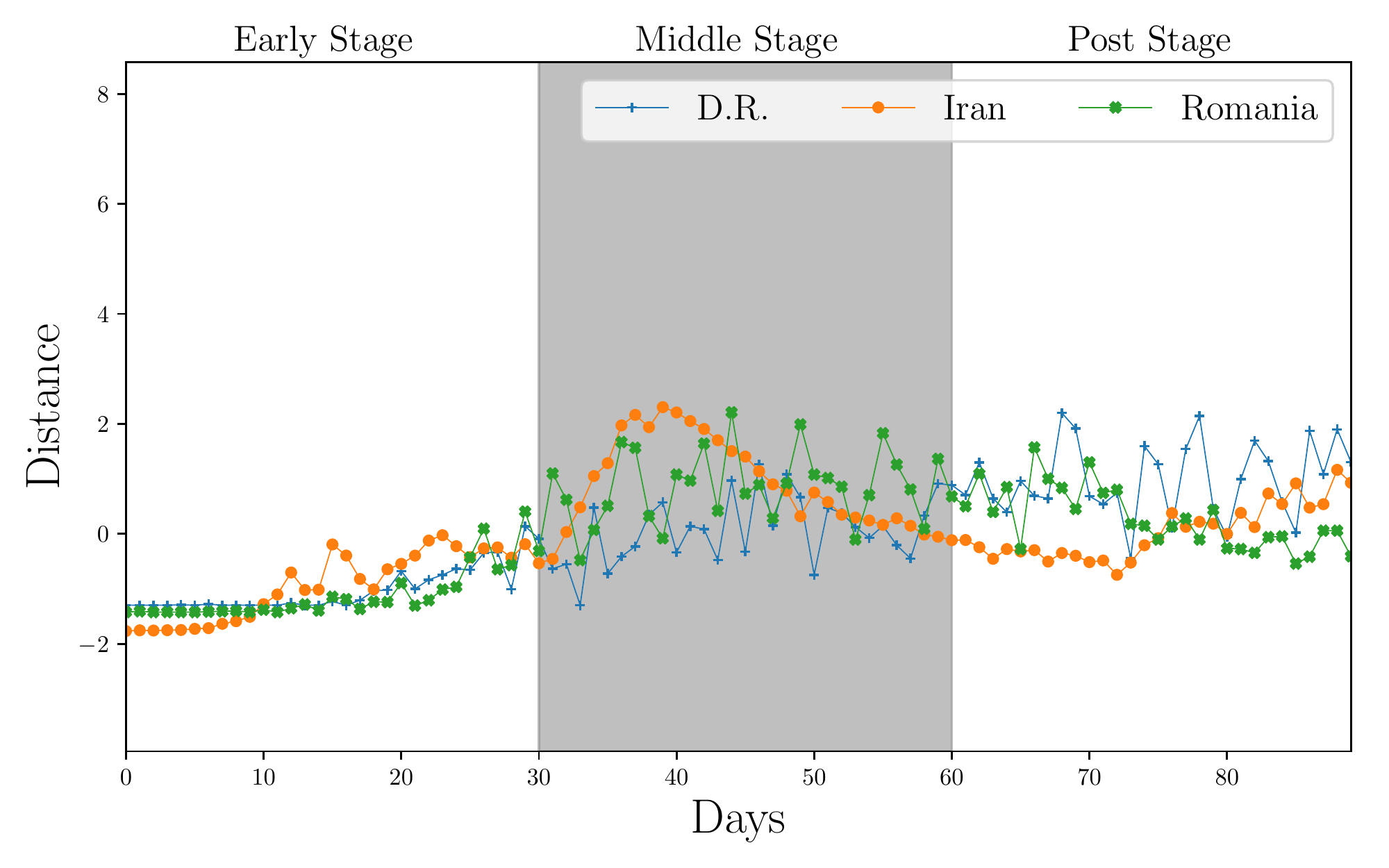}
		\subcaption{Spread Pattern}
		\label{fig:Dr_Iran_Romania}
	\end{subfigure}\hfill
	\begin{subfigure}[b]{0.48\linewidth}
		\includegraphics[width=\linewidth]{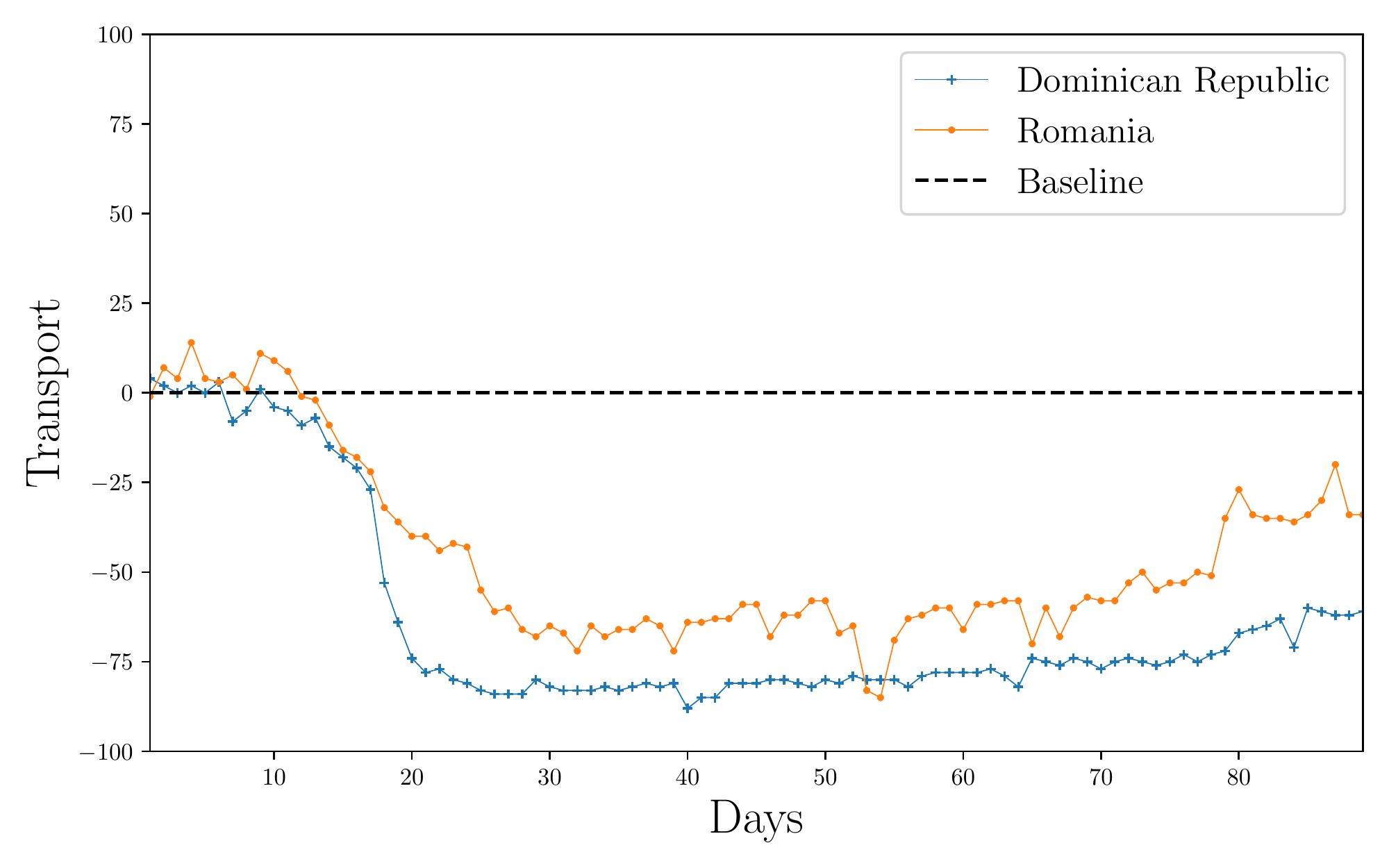}
		\subcaption{Mobility Pattern.}
		\label{fig:Dr_Iran_Romania_mobility}
	\end{subfigure}\hfill
	\caption{Early, middle and post stage spread patterns of Dominican Republic, Iran, Romania and their mobility during the same period.}
\end{figure*}

Early-stage clustering in Figure~\ref{fig:Early_stage_cluster_1} shows that the Dominican Republic (D.R.), Iran, and Romania are initially clustered together. 
According to Table~\ref{tab:measures}, Dominican Republic imposed a lockdown on March 19 (day 18 in the time series data)~\cite{dr_lock}, Iran on March 14 (day 24)~\cite{Iran_lock}, and Romania on March 25 (day 28)~\cite{Romania_lock}.
However, all three countries are in different clusters in the post-stage due to different spread patterns in their middle stages.

In Figure~\ref{fig:Dr_Iran_Romania}, Iran has a very high spread compared to the Dominican Republic and Romania in the middle-stage due to the Nowruz festival~\cite{Iran_nowruz}.
Iran contains the spread from day 40 according to its time-series data in Figure~\ref{fig:Dr_Iran_Romania}.
In the post-stage, Iran has a lower spread pattern than the other two countries. 
As a result, Iran lifted the lockdown on day 61~\cite{Iran_lock_lift}.
On the other hand, Iran has suffered from lifting the restrictions as the daily new cases increased after day 70.
It is reported that not wearing masks and not maintaining social distancing measures after the reopening are the main reasons for a surge of new cases~\cite{Why_iran}.
Figure~\ref{fig:Dr_Iran_Romania} also shows that the Dominican Republic and Romania have dissimilar peaks.
The Dominican Republic did not lift the strict restrictions in the post-stage.
Figure~\ref{fig:Dr_Iran_Romania_mobility} shows that mobility in the Dominican Republic is greatly reduced after day 18.
Despite the preventive measures taken by the government, the spread has been gradually increasing in the post-stage. 
Tapia states that the reasons behind the failure to control the spread are fake news regarding COVID-19 treatment and the public distrust in the government~\cite{tapia2020covid}.
In Romania, the spread has started to rise after day 10 according to its time-series data in Figure~\ref{fig:Dr_Iran_Romania}.
According to Table~\ref{tab:measures}, Romania imposed lockdown on March 25 (day 28)~\cite{Romania_lock}.
In Figure~\ref{fig:Dr_Iran_Romania_mobility}, the effect of the lockdown can be seen in the mobility data of Romania. 
The mobility has started to reduce around day 15, but it reduced significantly when the lockdown was imposed on day 28.
Dascalu states that Romania is successful in containing the spread in the post-stage by implementing preventive measures and addressing the issues of the healthcare system early and fast~\cite{dascalu2020successes} as the daily cases started to decrease after day 75.

Despite being in lockdowns, large gatherings, failure to maintain the social distancing protocols, fake news, or distrust in government actions of preventive measures have increased the spread in the post stages for the countries in cluster-1. 
On the other hand, early and fast actions have enabled these countries to avoid massive outbreaks.

\begin{figure*}[t!]
	\centering
	\begin{subfigure}[b]{0.48\linewidth}
		\includegraphics[width=\linewidth]{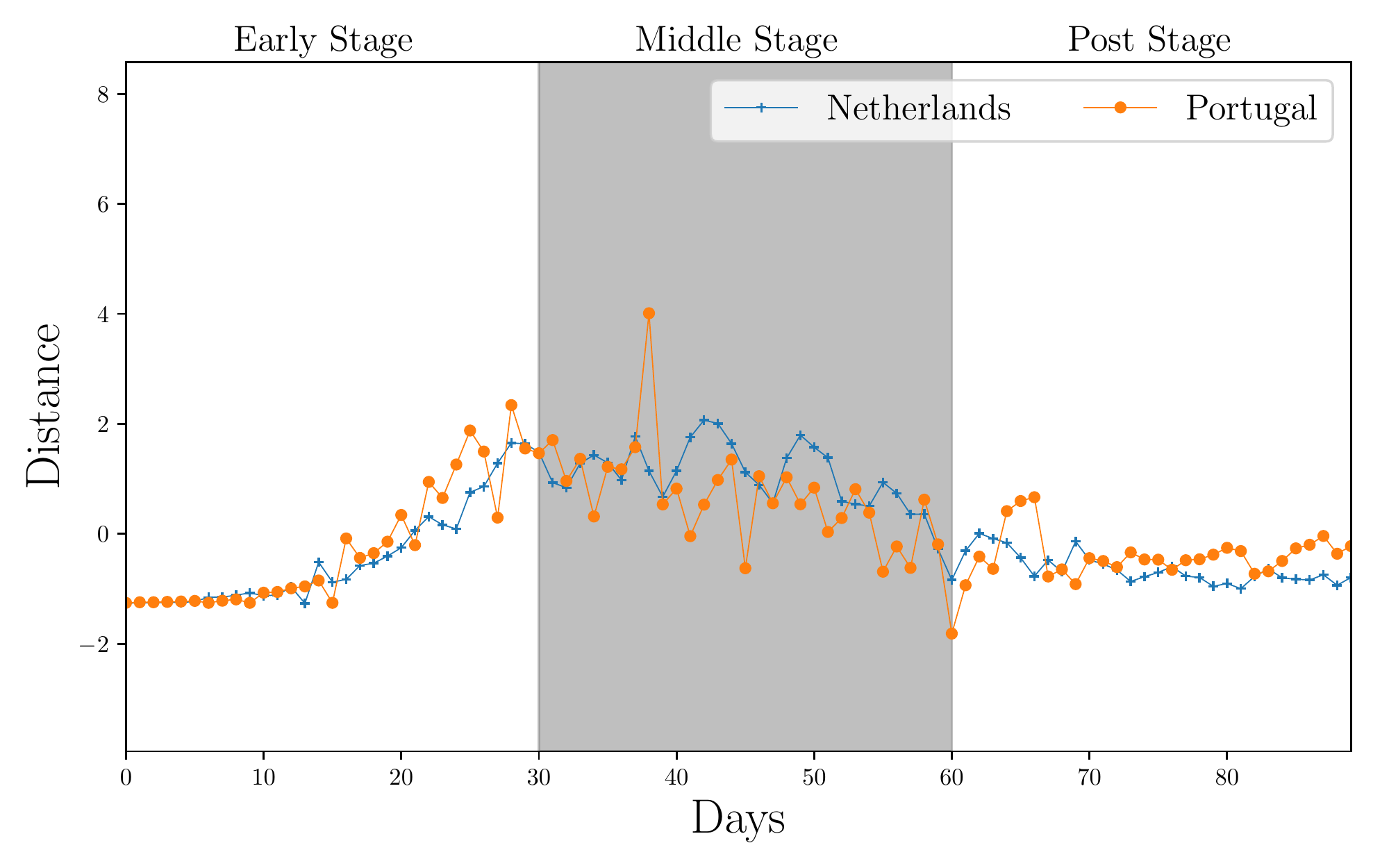}
		\subcaption{Spread Pattern}
		\label{fig:Net_Por}
	\end{subfigure}\hfill
	\begin{subfigure}[b]{0.48\linewidth}
		\includegraphics[width=\linewidth]{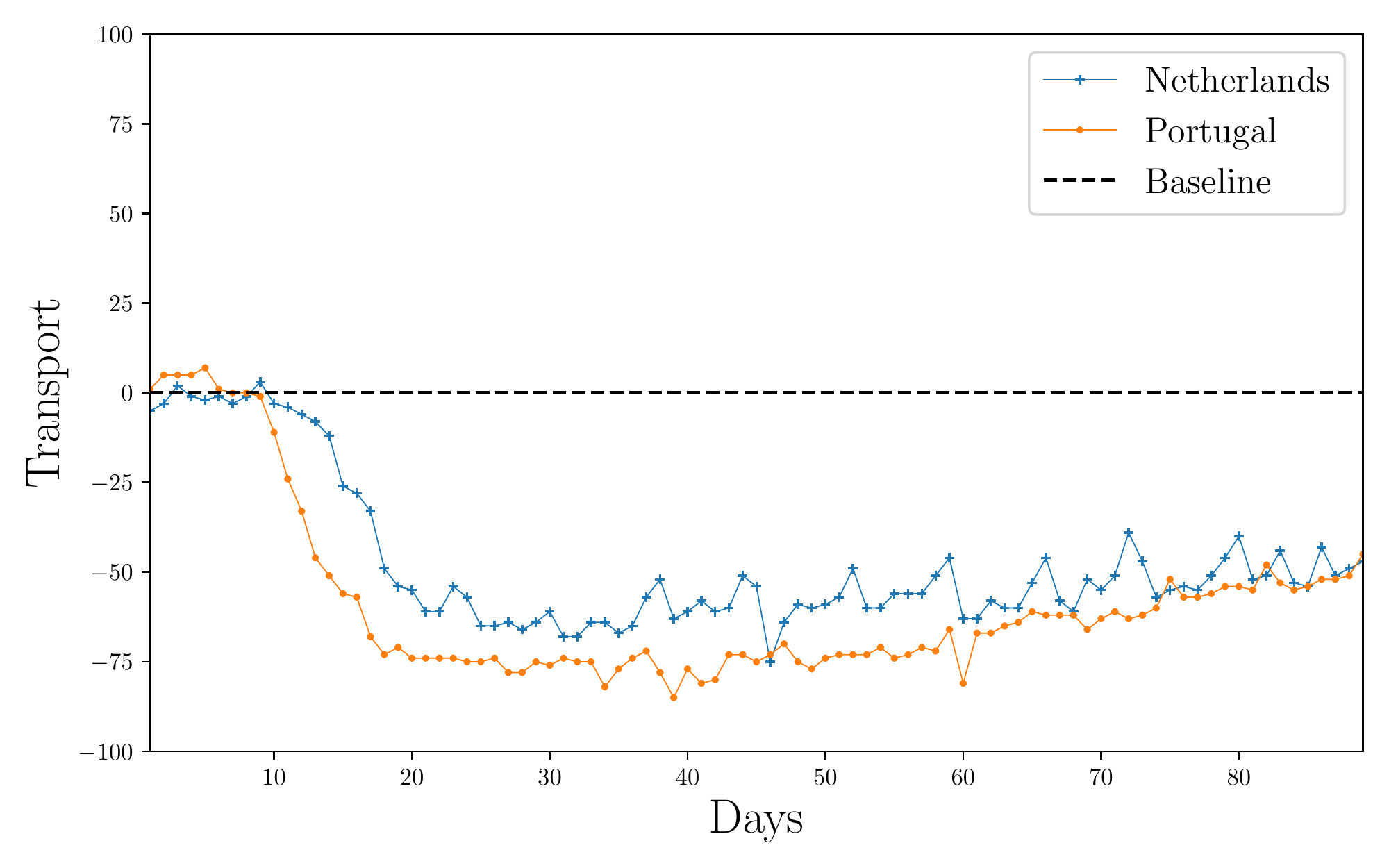}
		\subcaption{Mobility Pattern.}
		\label{fig:Net_Por_mobility}
	\end{subfigure}\hfill
	\caption{Early, middle and post stage spread patterns of Netherlands and Portugal, and their mobility during the same period.}
\end{figure*}

\subsection{Early-Stage: Cluster-2}

Early-stage clustering in Figure~\ref{fig:Early_stage_cluster_2} shows that the Netherlands and Portugal are in the same cluster. 
Two countries take different approaches to prevent the COVID-19 spread.
The Netherlands did not impose traditional lockdowns like other countries. 
Rather, the government implemented an "Intelligent Lockdown" which includes work from home and an urge to avoid leaving homes as much as possible~\cite{de2020covid}. 
On the other hand, Portugal imposed a total lockdown on March 19 (day 17)~\cite{Portugal_lock} and lifted it on May 2 (day 61)~\cite{Portugal_lock_lift}.
Figure~\ref{fig:Net_Por} depicts the spread pattern of the Netherlands and Portugal.
The spread increased for both countries at the end of the early-stage. 
Figure~\ref{fig:Net_Por_mobility} suggests that the mobility in Portugal reduced by almost 75\% around day 20.
Euro News and Politico reported that early implementation of the preventive measures along with the compliance of the population with the government orders played key roles in controlling the spread~\cite{why_portugal_1, why_portugal_2}.
On the other hand, Figure~\ref{fig:Net_Por_mobility} presents that the intelligent lockdown drastically reduced the mobility in the Netherlands.
Haas~\emph{et al.}~\cite{de2020covid} stated that self-discipline and self-responsibility of the Dutch made it possible to carry out the government orders.

Both countries in Cluster-2 adopted different preventive measures, but were successful in the containment of the spread.
Self-discipline, self-responsibility, and compliance to follow and maintain the preventive measures have helped these countries to control the spread.

\begin{figure*}[t!]
	\centering
	\begin{subfigure}[b]{0.48\linewidth}
		\includegraphics[width=\linewidth]{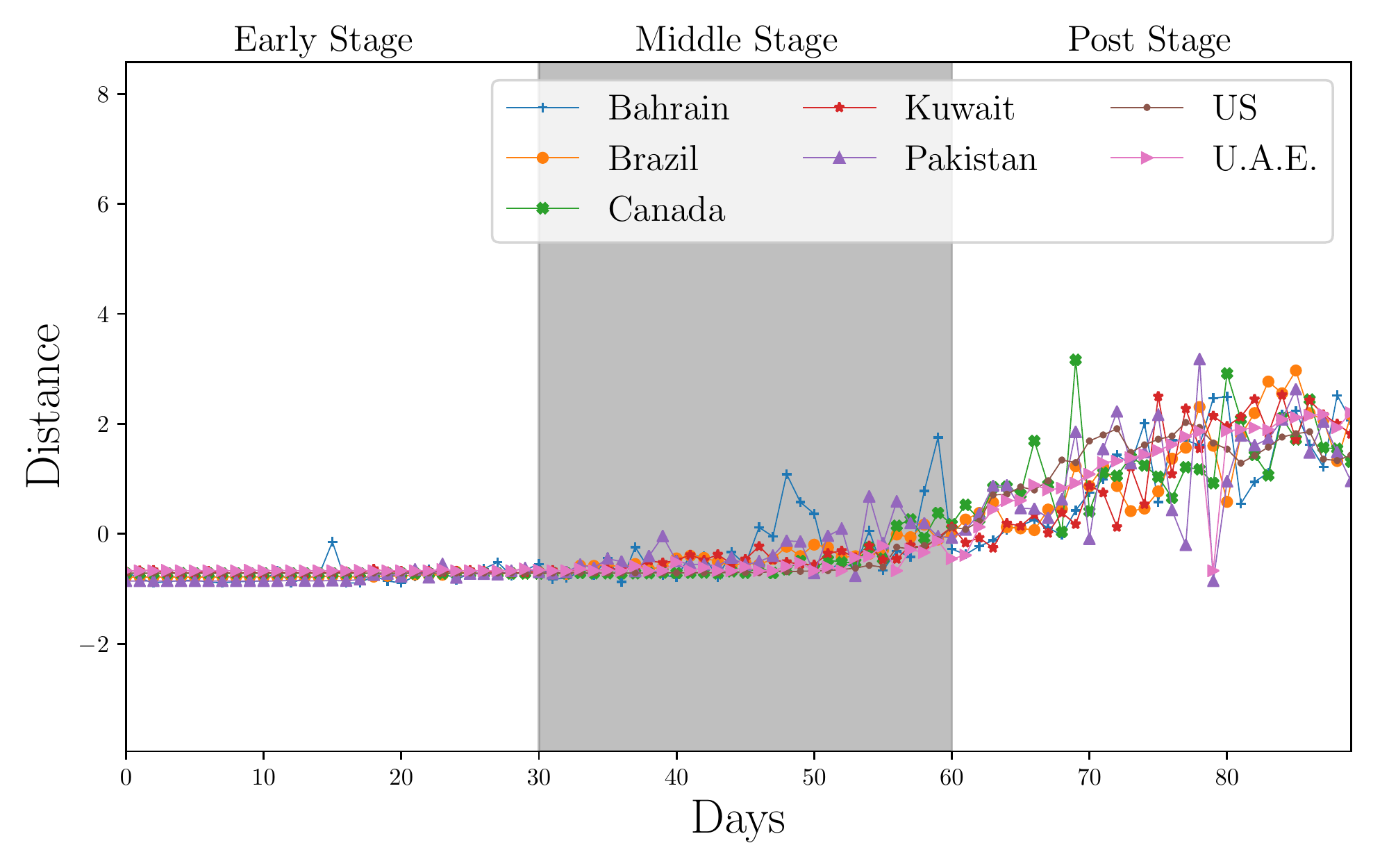}
		\subcaption{Spread Pattern}
		\label{fig:early_cluster_3_1}
	\end{subfigure}\hfill
	\begin{subfigure}[b]{0.48\linewidth}
		\includegraphics[width=\linewidth]{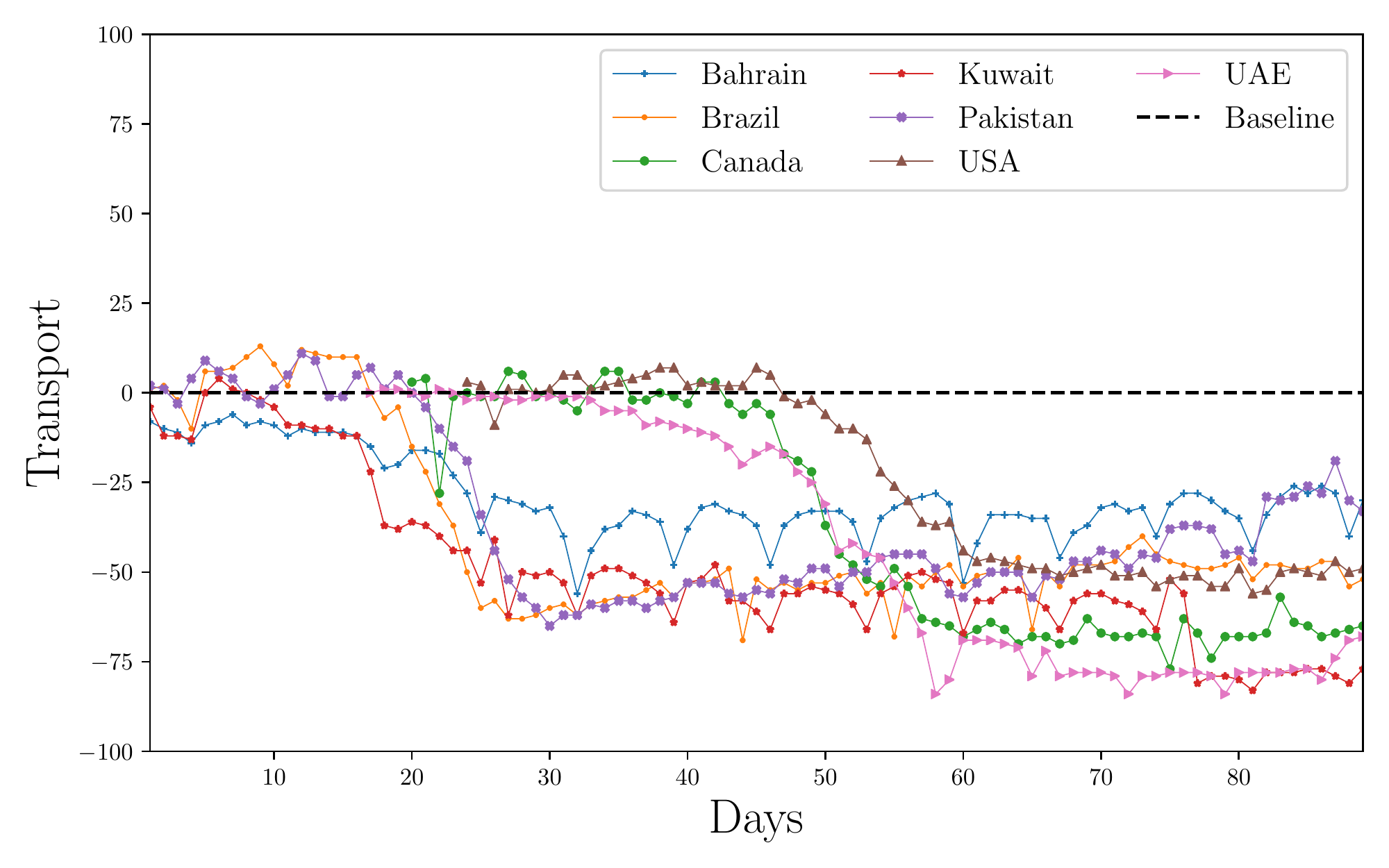}
		\subcaption{Mobility Pattern.}
		\label{fig:early_cluster_3_1_mobility}
	\end{subfigure}\hfill
	\caption{Early, middle and post stage spread patterns of Bahrain, Brazil, Canada, Kuwait, Pakistan, USA and U.A.E., and their mobility during the same period.}
\end{figure*}

\subsection{Early-Stage: Cluster-3}

Cluster-3 of the early-stage is the largest cluster consisting of 15 countries.
Among the countries in this cluster, Bahrain, Brazil, Canada, Kuwait, Pakistan, USA, and U.A.E. remain clustered together in Cluster-1; India, Afghanistan, and Oman in Cluster-2; Ireland, Israel, and South Korea in Cluster-5; the Philippines in Cluster-7 and Spain in Cluster-9 in the post stages according to Appendix~\ref{appendix_cluster_labels}, Table~\ref{tab:cluster_labels}.


Bahrain, Brazil, Canada, Kuwait, Pakistan, USA, and U.A.E. have a common spread pattern.
These countries initially have a low spread in the early-stage.
But their spread rise in the post-stage.
Early, middle, and post-stage spread patterns of these seven countries are shown in Figure~\ref{fig:early_cluster_3_1} and their respective mobility during the first 90 days of the pandemic are shown in Figure~\ref{fig:early_cluster_3_1_mobility}.
As the spread was low in the early-stage, Bahrain only imposed restrictions on large gatherings and ordered to maintain social distancing on March 18 (day 23)~\cite{bahrain_lock}.
According to Figure~\ref{fig:early_cluster_3_1_mobility}, such measures reduced the mobility from 25\% to 50\%. 
Bahrain lifted the restrictions on May 7 (day 63)~\cite{bahrain_lock_lift}.
The impact of lifting the restrictions can be seen in Figure~\ref{fig:early_cluster_3_1} as the spread started to rise in the post-stage.
Brazil imposed state-wide social distancing measures on March 17 (day 20)~\cite{brazil_lock}.
Their mobility data also shows that more than 60\% reduction from day 25 (see Figure~\ref{fig:early_cluster_3_1_mobility}).
Brazil lifted the restrictions on April 21 (day 55).
As a consequence, the spread started to rise in the post-stage around day 70.
Canada imposed a state-wide lockdown starting from March 18 (day 52).
The country restricted traveling and closed the borders~\cite{canada_lock}.
Figure~\ref{fig:early_cluster_3_1_mobility} shows that mobility is decreased by 70\% after the declaration of the lockdown.
The spread continued to increase in the post-stage. 
Canada started to lift restrictions on May 11~\cite{canada_lock_lift} which is after the first 90 days since the first case.
Kuwait imposed a country-wide curfew and restricted air and border travel beginning March 22 (day 19)~\cite{Kuwait_lock}.
The mobility data in Figure~\ref{fig:early_cluster_3_1_mobility} shows that the curfew reduced the internal movement.
However, the continued increase in spread (see Figure~\ref{fig:early_cluster_3_1}) indicates that a complete lockdown was needed to be enforced instead of a curfew.
Pakistan imposed a nationwide lockdown on March 22 (day 25)~\cite{Pakistan_lock}.
The impact of the lockdown can be seen in their mobility data in Figure~\ref{fig:early_cluster_3_1_mobility}.
Surprisingly, Pakistan lifted the lockdown on May 9 (day 73)~\cite{Pakistan_lock_lift} while the spread continued to increase.
Such action only worsened the spread in Pakistan.
To this date, the USA has the highest number of cumulative cases in the world.
The USA imposed state-wide lockdowns and restricted air travel starting from March 16~\cite{USA_lock}.
Figure~\ref{fig:early_cluster_3_1_mobility} shows that lockdowns were only able to reduce mobility by 50\% which is less than most of the countries.
The USA saw a peak in the spread around day 70 (see Figure~\ref{fig:early_cluster_3_1}).

\begin{figure*}[t!]
	\centering
	\begin{subfigure}[b]{0.48\linewidth}
		\centering
		\includegraphics[width=\linewidth]{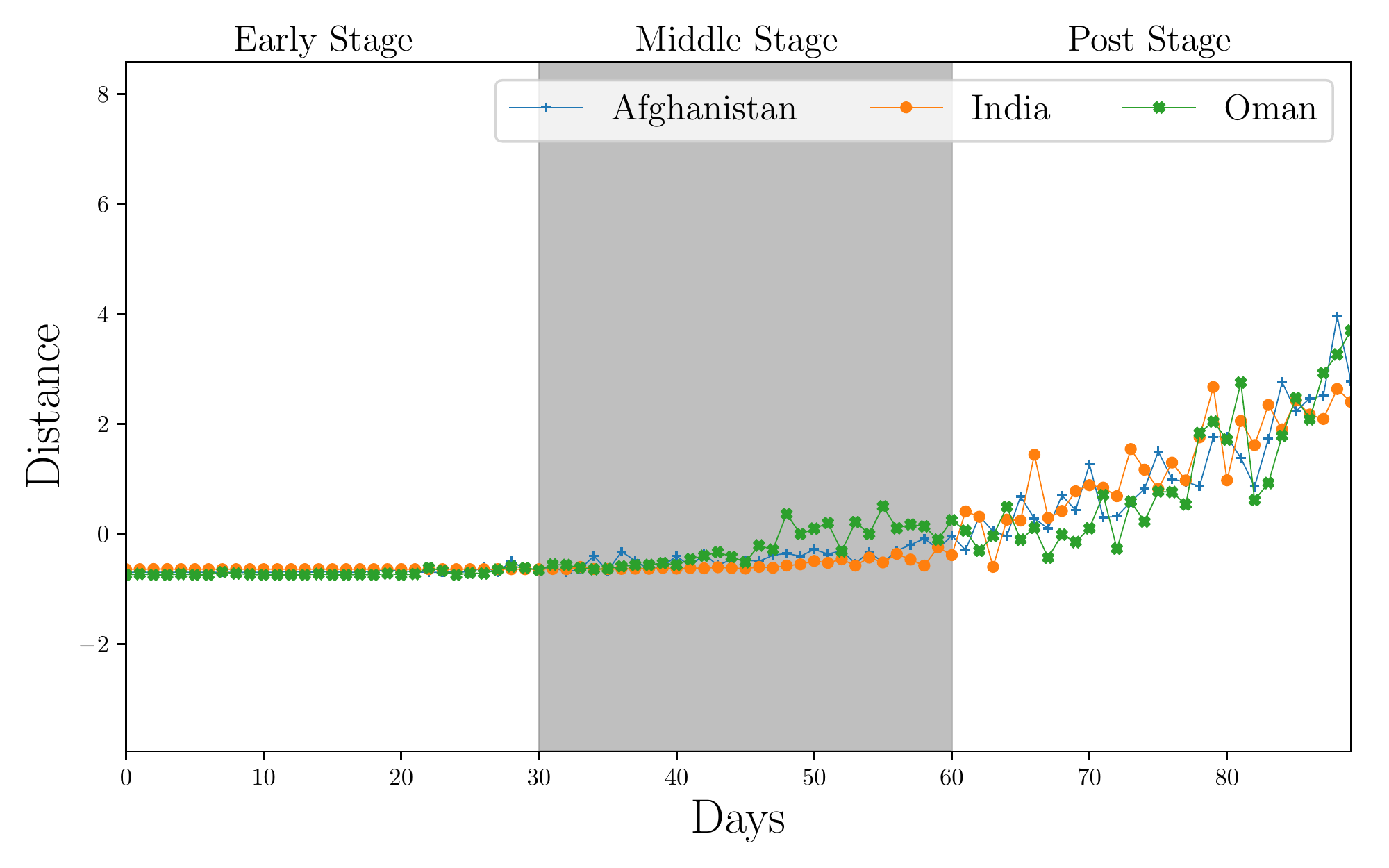}
		\subcaption{Spread Pattern}
		\label{fig:early_cluster_3_2}
	\end{subfigure}\hfill
	\begin{subfigure}[b]{0.48\linewidth}
		\centering
		\includegraphics[width=\linewidth]{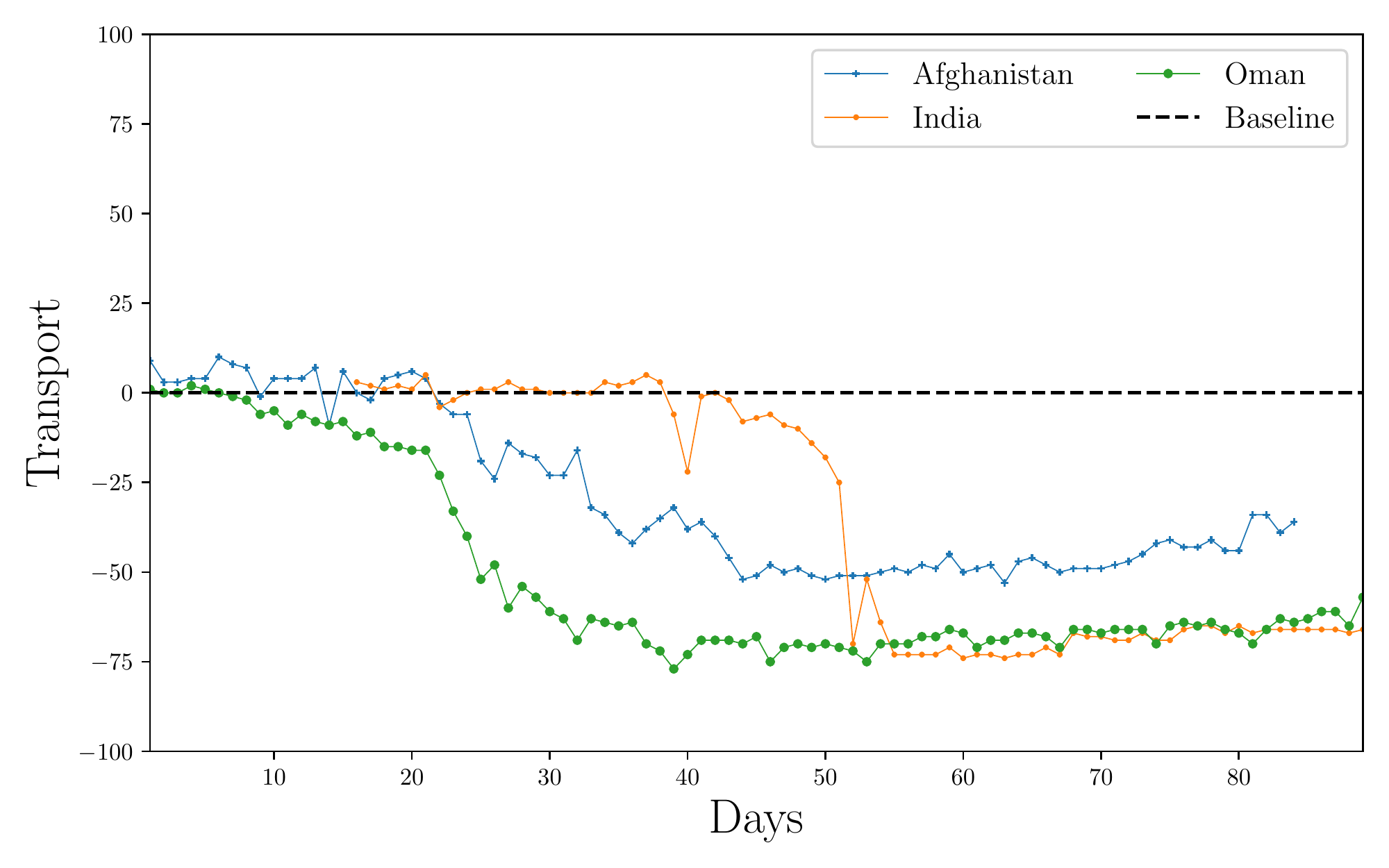}
		\subcaption{Mobility Pattern.}
		\label{fig:early_cluster_3_2_mobility}
	\end{subfigure}\hfill
	\caption{Early, middle and post stage spread patterns of Afghanistan, India, and Oman, and their mobility during the same period.}
\end{figure*}

Slow response and delayed testing in the early stages of the pandemic contributed to the spread in the USA~\cite{schneider2020failing, Why_USA}.
The states imposed lockdowns independently.
The country as a whole did not implement a unified lockdown which made it difficult to control the spread.
The spread in U.A.E starts to increase around day 60 according to Figure~\ref{fig:early_cluster_3_1}.
The government imposed a lockdown on April 5 (day 67)~\cite{UAE_lock}.
Enforcing a lockdown reduced the mobility according to Figure~\ref{fig:early_cluster_3_1_mobility}.
Unfortunately, failure to comply with social distancing measures and lack of social awareness are the factors behind the increasing spread in the country.

In summary, these seven countries suffered high spread in the post-stage due to different factors.
Slow response, lack of testing, lack of contact tracing, state-wide lockdowns or curfews instead of unified lockdowns, and not maintaining social distancing measures properly are the worsening factors.

According to Figure~\ref{fig:early_cluster_3_2}, the spread in Afghanistan, India, and Oman started to rise from the mid-point of the middle-stage.
Afghanistan imposed a state-wide lockdown from March 28 (day 33)~\cite{afg_lock} while India and Oman imposed nationwide lockdown from March 25 (day 55)~\cite{India_lock} and April 10 (day 46)~\cite{Oman_lock}, respectively.
Figure~\ref{fig:early_cluster_3_2_mobility} shows that the mobility in India has reduced significantly after the order of the lockdown.
The mobility in Oman has reduced even before the lockdown, because the government closed schools, offices, places of worship and ordered to maintain social distancing measures.
Similar to Wuhan, the Muscat state of Oman has suffered from the spread significantly compared to other states.
Lack of social awareness, social gatherings, and migrant workers living and working in close proximity contributed to the rising spread in Oman~\cite{why_Oman_1, why_Oman_2, why_Oman_3}.
Despite the lockdown, the spread continues to increase in Afghanistan due to the lack of testing, social awareness, social gatherings, and case importations from foreign countries, especially from Iran~\cite{why_afg, why_afg_2, why_afg_3}.
India fails to maintain social distancing measures because of the lack of social awareness in the highly populated and dense country~\cite{why_India}. 
Hence, India has also suffered from a high spread in the post-stage.

The spread in these countries continues to rise in the post-stage despite adopting lockdowns and social distancing measures.
Several factors including lack of social awareness in the densely populated regions, failure to maintain social distancing measures, and lack of rigorous testing supplement the spread.

\begin{figure*}[t!]
	\centering
	\begin{subfigure}[b]{0.48\linewidth}
		\includegraphics[width=\linewidth]{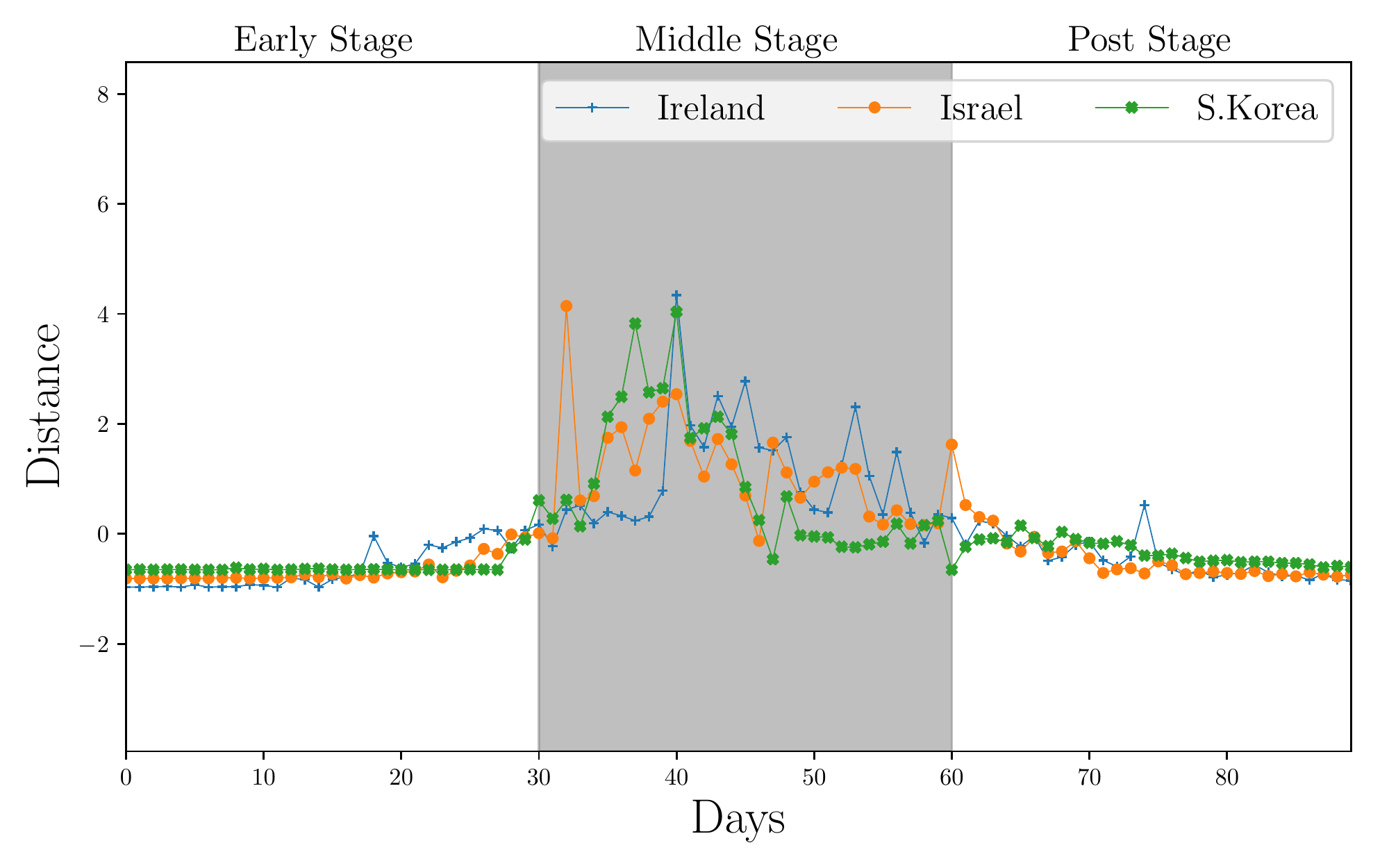}
		\subcaption{Spread Pattern}
		\label{fig:early_cluster_3_3}
	\end{subfigure}\hfill
	\begin{subfigure}[b]{0.48\linewidth}
		\includegraphics[width=\linewidth]{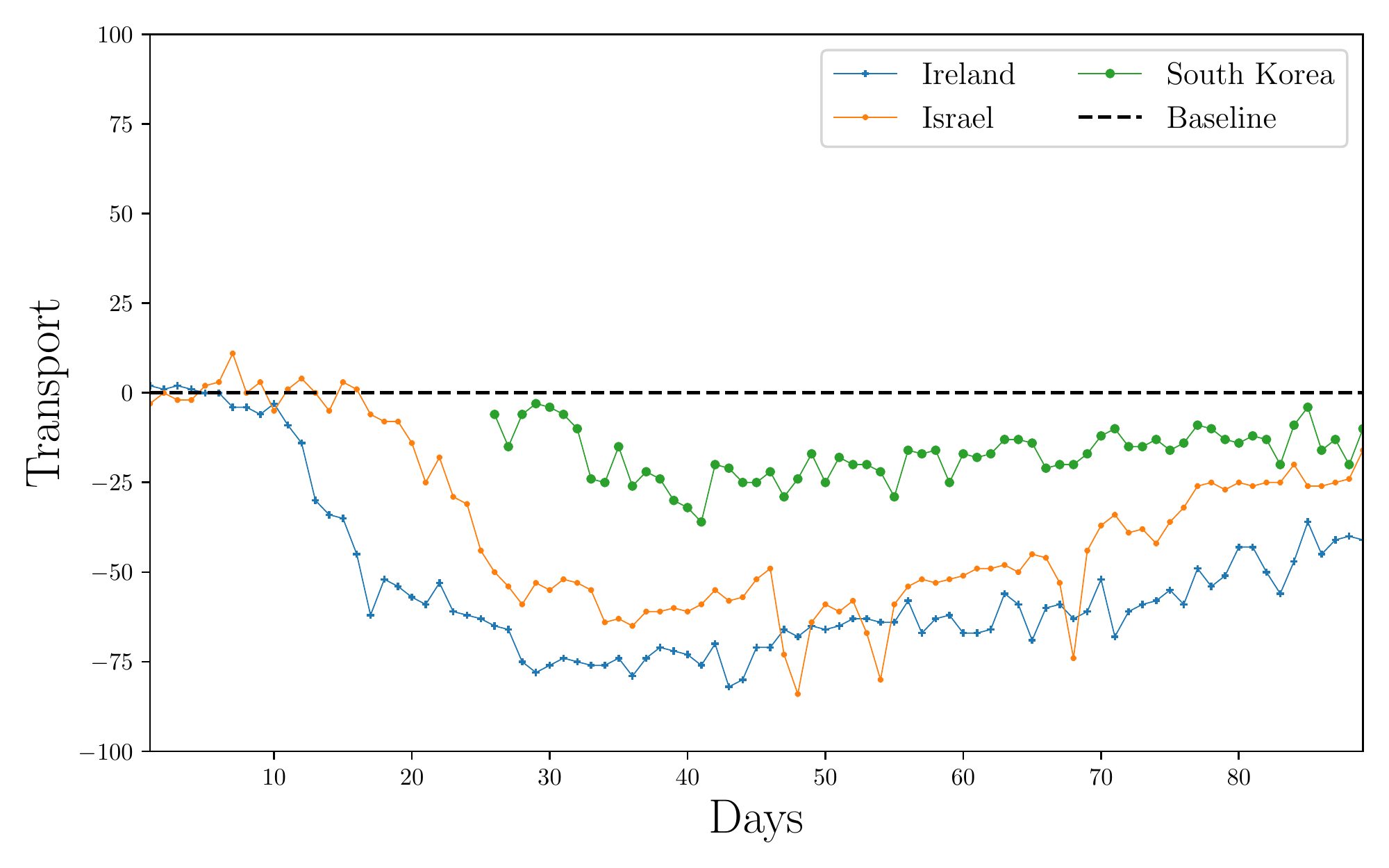}
		\subcaption{Mobility Pattern.}
		\label{fig:early_cluster_3_3_mobility}
	\end{subfigure}\hfill
	\caption{Early, middle and post stage spread patterns of Ireland, Israel, and South Korea, and their mobility during the same period.}
\end{figure*}

Ireland, Israel, and South Korea exhibit low spread patterns in the post stages according to Figure~\ref{fig:early_cluster_3_3}.
Ireland was quick to respond and imposed a lockdown on March 12 (day 12)~\cite{Ireland_lock}.
Their mobility was also reduced as soon as the lockdown was ordered (see Figure~\ref{fig:early_cluster_3_3_mobility}).
Ireland saw a peak in the spread in the middle-stage.
The government lifted the restrictions on May 18 (day 78)~\cite{Ireland_lock_lift}.
By successfully enforcing a lockdown and supplying adequate medical equipment, Ireland was able to control the spread in the post-stage~\cite{Why_Ireland}.
Israel imposed a state-wide lockdown on March 25 (day 33)~\cite{waitzberg2020israel}.
Like Ireland, Israel had its peak in the early part of the middle-stage as shown in Figure~\ref{fig:early_cluster_3_3}.
Rapid response and contact tracing helped Israel to slow down the spread in the post-stage~\cite{waitzberg2020israel}.
South Korea did not impose any lockdowns.
Rather, they enforced isolation and quarantine for the infected patients, rigorous contact tracing, and increased testing capacity to contain the spread~\cite{oh2020national}.

\begin{figure*}[t!]
	\centering
	\begin{subfigure}[b]{0.48\linewidth}
		\includegraphics[width=\linewidth]{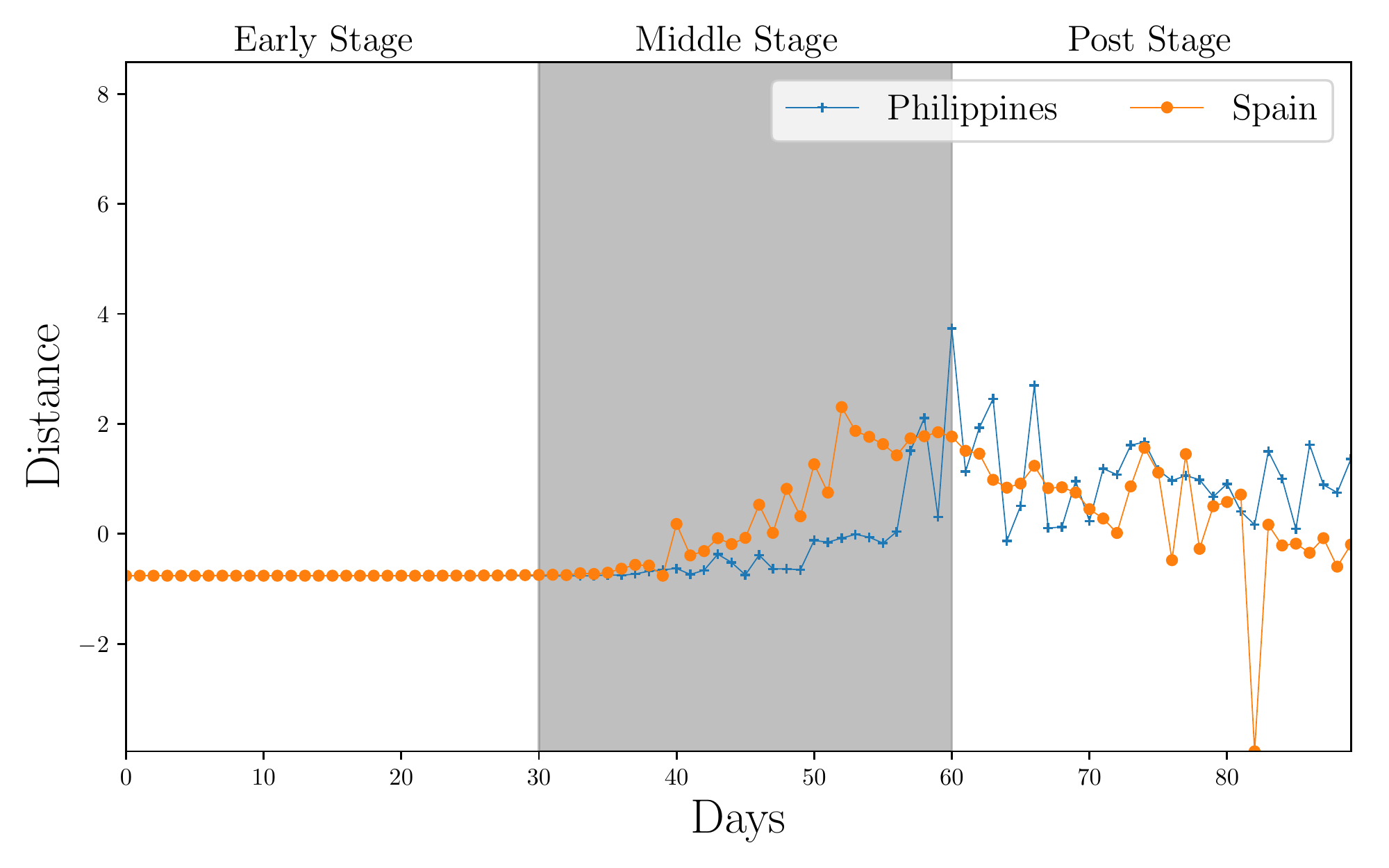}
		\subcaption{Spread Pattern}
		\label{fig:early_cluster_3_4}
	\end{subfigure}\hfill
	\begin{subfigure}[b]{0.48\linewidth}
		\includegraphics[width=\linewidth]{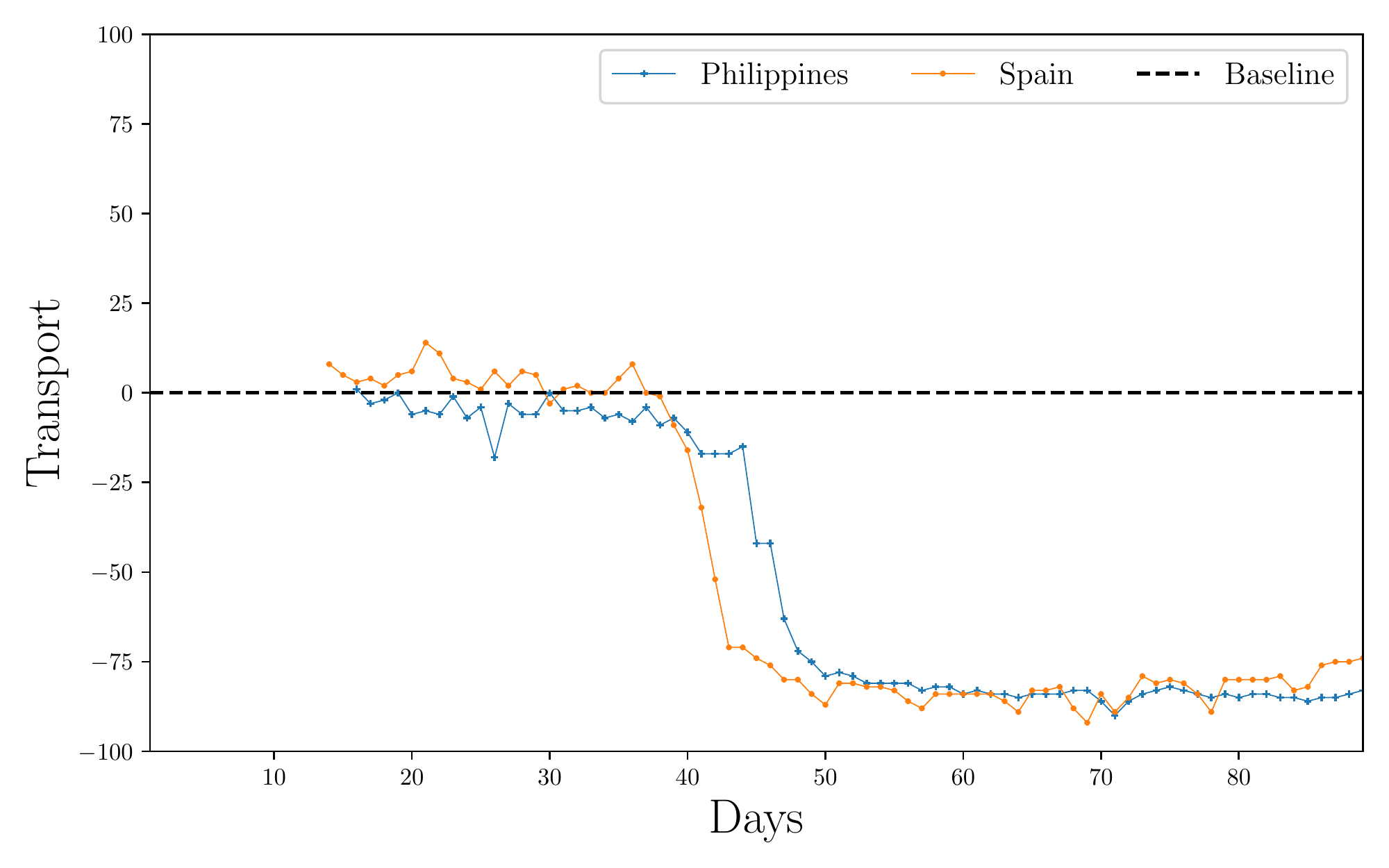}
		\subcaption{Mobility Pattern.}
		\label{fig:early_cluster_3_4_mobility}
	\end{subfigure}\hfill
	\caption{Early, middle and post stage spread patterns of Philippines, and Spain, and their mobility during the same period.}
\end{figure*}

These countries exhibit high spread patterns in the middle-stage.
Through early and rapid response, contact tracing, and meticulous testing facilities, these countries reduced the spread significantly in the post-stage.
Among these countries, the response of South Korea has been exceptional.

Early-stage clustering in Figure~\ref{fig:Early_stage_cluster_3} shows that the Philippines and Spain remain in the same cluster.
\cref{fig:Late_stage_cluster_7,fig:Late_stage_cluster_9} show that they reside in singleton clusters in the post stages due to different spread patterns in their middle stages.
Figure~\ref{fig:early_cluster_3_4} shows that the spread in the Philippines has started to increase in the middle-stage.
The country is under lockdown since March 15 (day 45)~\cite{Philippines_lock}.
Figure~\ref{fig:early_cluster_3_4_mobility} shows that the mobility in the Philippines has reduced to almost 85\% after the enforcement of the lockdown.
The peak in the spread can be seen between days 60 to 67.
Lack of testing and contact tracing are the reasons behind the increasing spread in the Philippines~\cite{why_Philippines}.
Lockdown helped the country to decrease the spread in the latter part of the post-stage.
Spain imposed a countrywide lockdown on March 14 (day 42)~\cite{Spain_lock} as the spread started to increase.
The mobility data in Figure~\ref{fig:early_cluster_3_4_mobility} indicates that Spain implemented and maintained strict measures.
Spain had a peak in the latter part of the middle-stage.
It has been reported that the lack of testing at the beginning of the pandemic supplement the spread~\cite{Why_Spain}.
However, by implementing strict preventive measures and improving the testing capacity, Spain started to contain the spread in the latter part of the post-stage according to Figure~\ref{fig:early_cluster_3_4}.

Lack of testing and contact tracing prevents people from self-isolation.
As a consequence, it aids in high community transmission in these countries.

\subsection{Early-Stage: Cluster-4}

Early-stage clustering in Figure~\ref{fig:Early_stage_cluster_4} shows that 12 countries in Cluster-4 exhibit a similar spread pattern. 
The countries in this cluster move to two different clusters in the post-stage.
Germany, Italy, and Belgium remain clustered together (Cluster-3 in the post-stage) as they are in the early-stage.
The remaining nine countries move to Cluster-1 in the post-stage as shown in Appendix~\ref{appendix_cluster_labels}, Table~\ref{tab:cluster_labels}.

\begin{figure*}[t!]
	\centering
	\begin{subfigure}[b]{0.48\linewidth}
		\includegraphics[width=\linewidth]{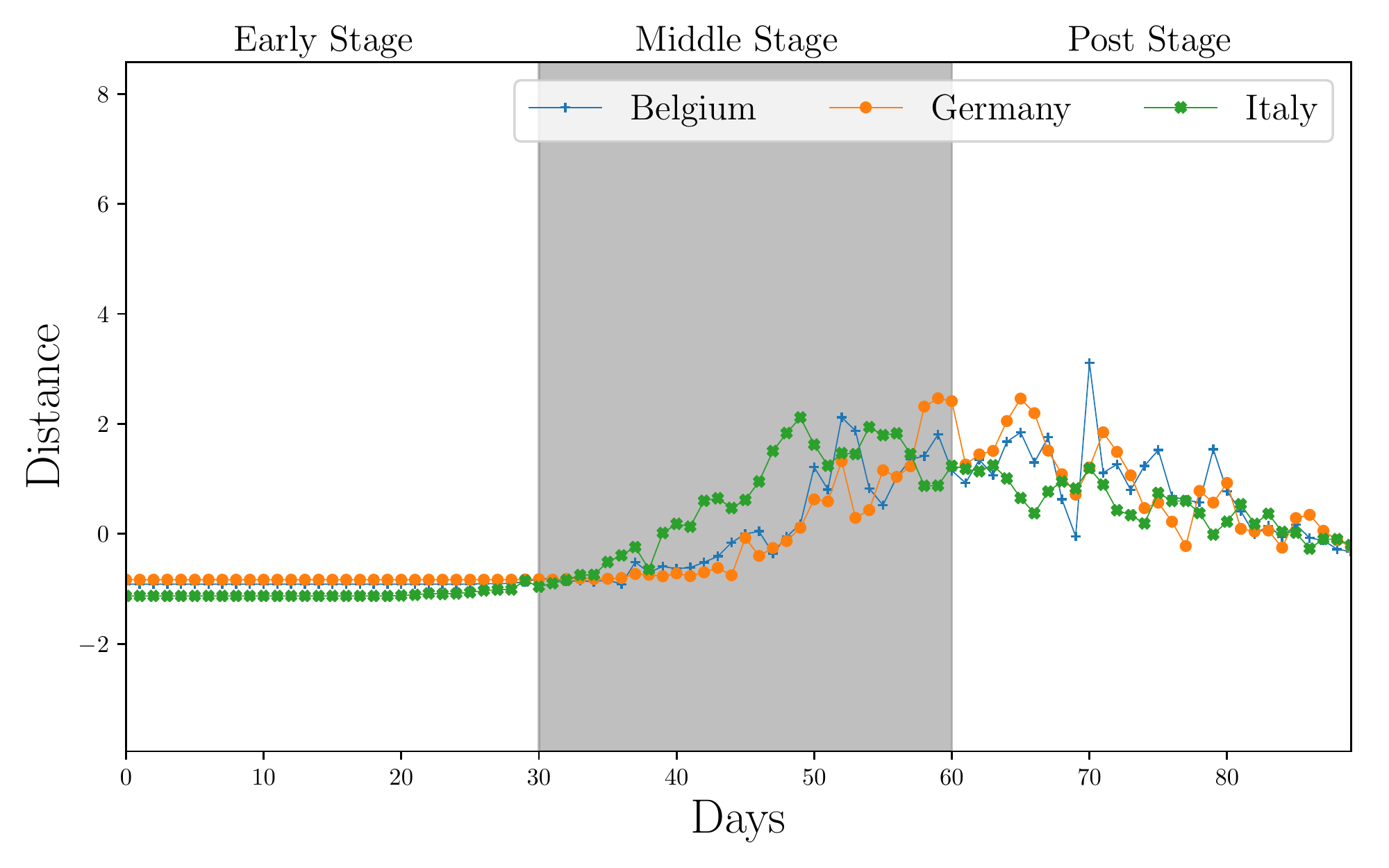}
		\subcaption{Spread Pattern}
		\label{fig:early_cluster_4_1}
	\end{subfigure}\hfill
	\begin{subfigure}[b]{0.48\linewidth}
		\includegraphics[width=\linewidth]{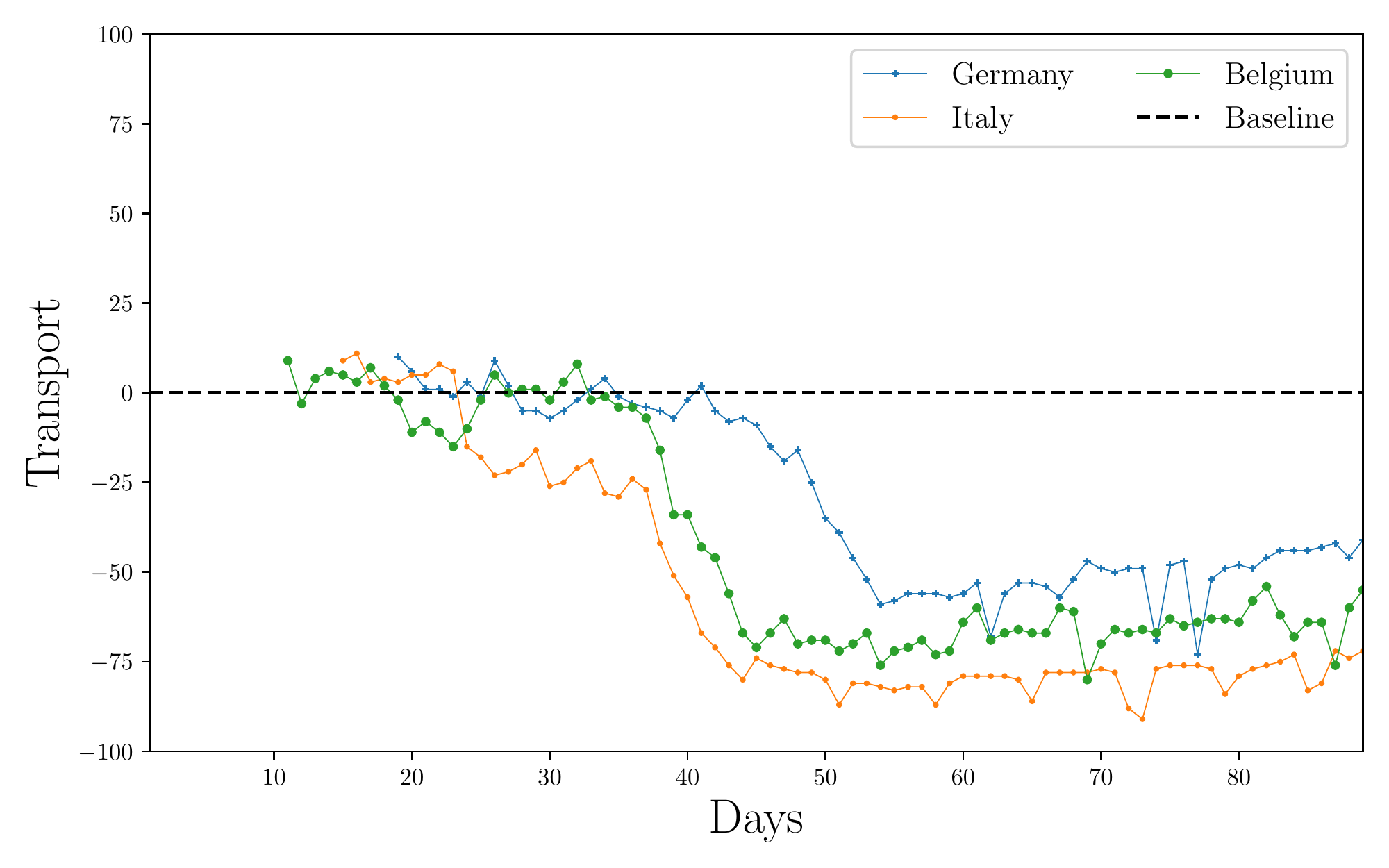}
		\subcaption{Mobility Pattern.}
		\label{fig:early_cluster_4_1_mobility}
	\end{subfigure}\hfill
	\caption{Early, middle and post stage spread patterns of Germany, Italy, and Belgium, and their mobility during the same period.}
\end{figure*}

In Figure~\ref{fig:early_cluster_4_1}, Germany, Italy, and Belgium had a very low spread in the early-stage.
The spread started to increase in the middle-stage in these three countries.
According to Table~\ref{tab:measures}, the lockdowns were imposed on March 23 (day 56)~\cite{germany_lock}, March 8 (day 37)~\cite{Italy_lock}, and March 18 (day 43)~\cite{Belgium_lock} in Germany, Italy, and Belgium, respectively.
Germany was in a complete lockdown starting on March 23 (day 56)~\cite{germany_lock}.
Figure~\ref{fig:early_cluster_4_1_mobility} shows the reduction in mobility as the lockdown was enforced.
According to Figure~\ref{fig:early_cluster_4_1}, Germany started to contain the spread in the last part of the post-stage by implementing and maintaining the lockdown measures, contact tracing, and widespread testing~\cite{why_Germany, why_Germany_2}.
A sudden spike of new cases forces Italy to be in a complete lockdown starting on March 8 (day 37)~\cite{Italy_lock}.
However, the people moved to different cities before the orders were in effect causing high local transmission of the spread~\cite{Italy_lock}.
Figure~\ref{fig:early_cluster_4_1_mobility} supports this fact as the highest reduction in mobility is six days after the lockdown was ordered.
Lack of social awareness and a high number of case importations from China in the early-stage, and densely populated cities in the Northern region contributed to the rise of spread in Italy~\cite{Why_Italy_1, Why_Italy_2}.
In Figure~\ref{fig:early_cluster_4_1}, the decreasing spread pattern in the post-stage suggests that Italy started to contain the spread by imposing the complete lockdown~\cite{Why_Italy_3}.
Belgium closed its borders and imposed strict social distancing measures on March 18 (day 43)~\cite{Belgium_lock}.
Figure~\ref{fig:early_cluster_4_1_mobility} suggests that mobility in Belgium reduced significantly as soon as the preventive measures were taken.
In Figure~\ref{fig:early_cluster_4_1}, the effect of the lockdown can be seen from the data as the spread started to decrease for Belgium in the post-stage.
Hence, the government lifted the restrictions on May 4 (day 90)~\cite{Belgium_lock_lift}.

\begin{figure*}[t!]
	\centering
	\begin{subfigure}[b]{0.48\linewidth}
		\includegraphics[width=\linewidth]{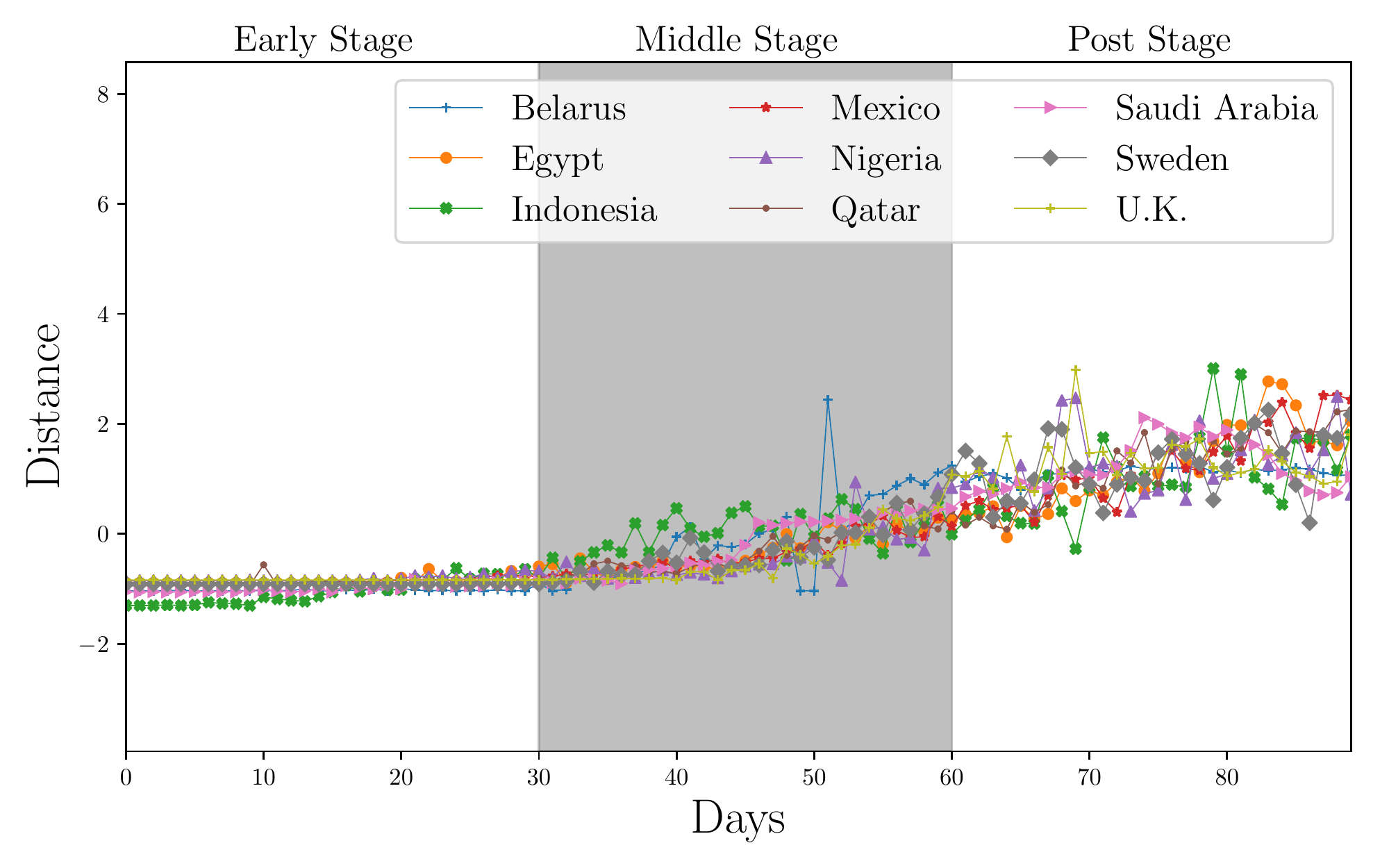}
		\subcaption{Spread Pattern}
		\label{fig:early_cluster_4_2}
	\end{subfigure}\hfill
	\begin{subfigure}[b]{0.48\linewidth}
		\includegraphics[width=\linewidth]{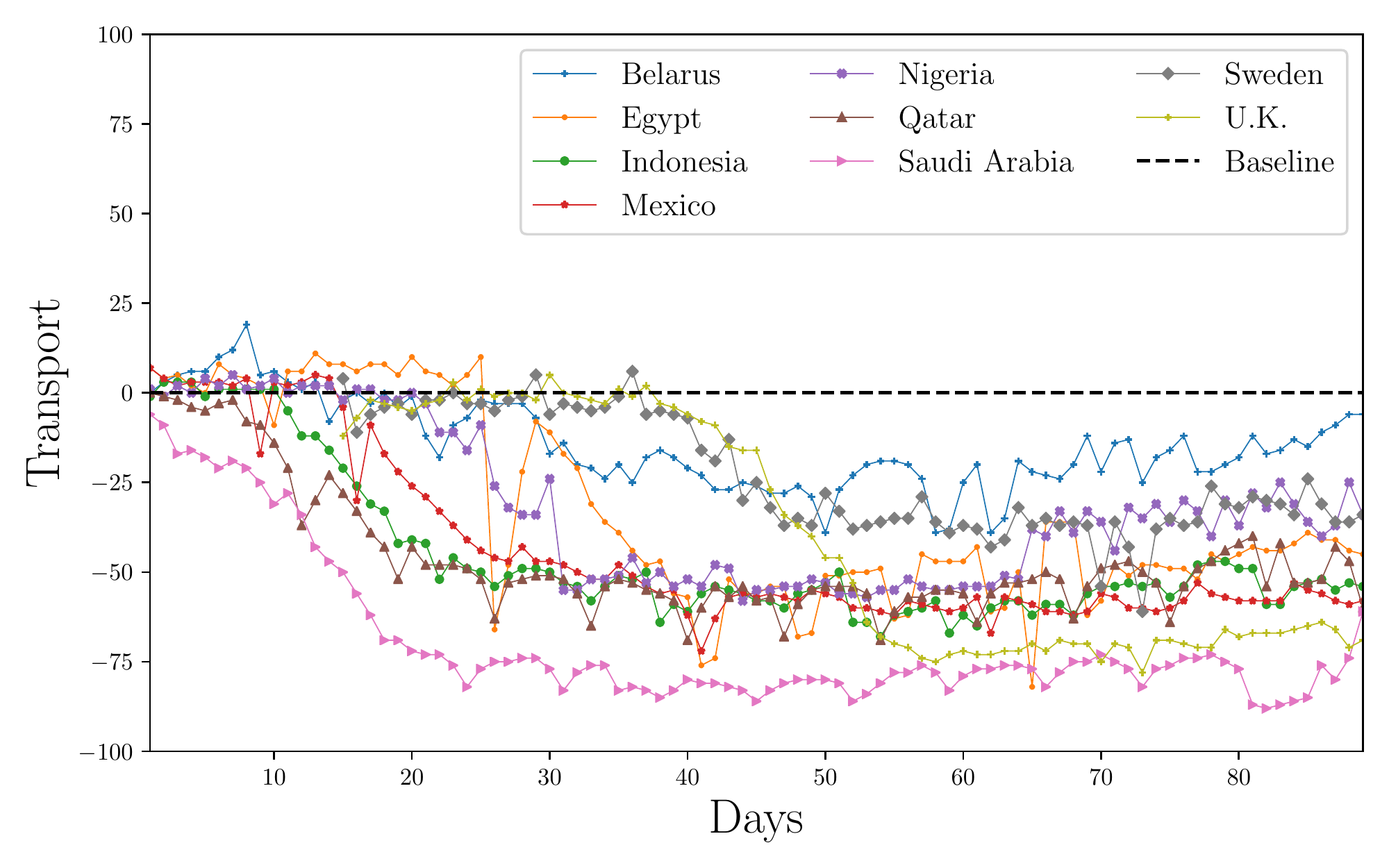}
		\subcaption{Mobility Pattern.}
		\label{fig:early_cluster_4_2_mobility}
	\end{subfigure}\hfill
	\caption{Early, middle and post stage spread patterns of Belarus, Egypt,  Indonesia,  Mexico, Nigeria, Qatar, Saudi Arabia, Sweden, U.K., and their mobility during the same period.}
\end{figure*}

The actual outbreak occurred in the middle stages in Germany, Italy, and Belgium.
Strict lockdowns or social distancing measures, contact tracing, and widespread testing helped to contain the spread in the post-stage.

According to Figure~\ref{fig:early_cluster_4_2}, the remaining nine countries, including Belarus, Egypt, Indonesia, Mexico, Nigeria, Qatar, Saudi Arabia, Sweden, and the U.K. also had very low spread patterns in the early stages.
However, the spread starts to increase gradually from the middle stages to the post stages.
Among these countries, Belarus did not impose any lockdown~\cite{belarus}.
Belarus opted for self-isolation of the infected patients~\cite{why_belarus}.
Lack of a lockdown or strict social distancing measures have not decreased the spread.
Like Belarus, Sweden also did not impose any lockdowns~\cite{Sweden_no_lock}.
Not maintaining strict social distancing measures, lack of testing, and high disease transmission compared to its neighboring countries contributed to the unsuccessful efforts of containing the spread in the post-stage of Sweden~\cite{Why_Sweden}.
Initially, the U.K. decided for a herd immunity approach to the COVID-19 pandemic but later they changed their decision and imposed a lockdown on March 23 (day 52)~\cite{UK_lock}.
The delay in implementing strict lockdowns or social distancing measures affected the spread in the country.
Egypt, Mexico, Qatar, and Saudi Arabia are still enforcing lockdowns~\cite{egypt_lock, Mexico_lock, Qatar_lock, algaissi2020preparedness} while Nigeria started to ease the restrictions from May 4 (day 66)~\cite{Nigeria_lock_lift}.
In Egypt, lack of sanitation facilities, poor hygiene, and weak healthcare system are the catalysts, as the lockdown failed to contain the spread~\cite{why_Egypt}.
Indonesia imposed strict restrictions on the mobility and social distancing instead of a lockdown~\cite{Indonesia_lock}.
Figure~\ref{fig:early_cluster_4_2_mobility} supports this fact as mobility is reduced by more than 50\%.
However, it is reported that the delayed testing and definite preventive measures along with lack of testing and rigorous lockdowns or social distancing measures in Indonesia caused a high spread~\cite{why_Indonesia_1, why_Indonesia_2, why_Indonesia_3}.
The spread in Mexico also suffered from the lack of testing like Indonesia along with the lack of contact tracing and negligence of the government to take any rigorous actions early in the pandemic~\cite{why_Mexico_1, why_Mexico_2}.
Qatar, on the other hand,  has large labor camps which have been often criticized for poor living conditions and overcrowding.
Almost 60\% of the infected patients are asymptotic.
Together, these factors played a pivotal role in increasing the spread in Qatar~\cite{why_Qatar_1, why_Qatar_2}.
Saudi Arabia ordered to follow precautionary measures even before detecting its first case~\cite{algaissi2020preparedness}.
Like Qatar, Saudi Arabia has labor camps with people living in close proximity~\cite{why_Saudi_1}.
Large social gatherings also played a significant role in the spread in Saudi Arabia~\cite{why_Saudi_2}.
In Nigeria, the lack of testing and social awareness contributed to the spread~\cite{why_Nigeria_1}.


The spread in these countries continues to rise in the post-stage despite implementing different preventive measures.
Lack of strict measures, not maintaining social distancing, lack of testing and contact tracing, delay in testing and implementing preventive measures, large labor camps with unhealthy living conditions are the most vital factors contributing the spread in these countries.

\subsection{Early-Stage: Clusters 5-8}

\begin{figure*}[t!]
	\centering
	\begin{subfigure}[b]{0.48\linewidth}
		\includegraphics[width=\linewidth]{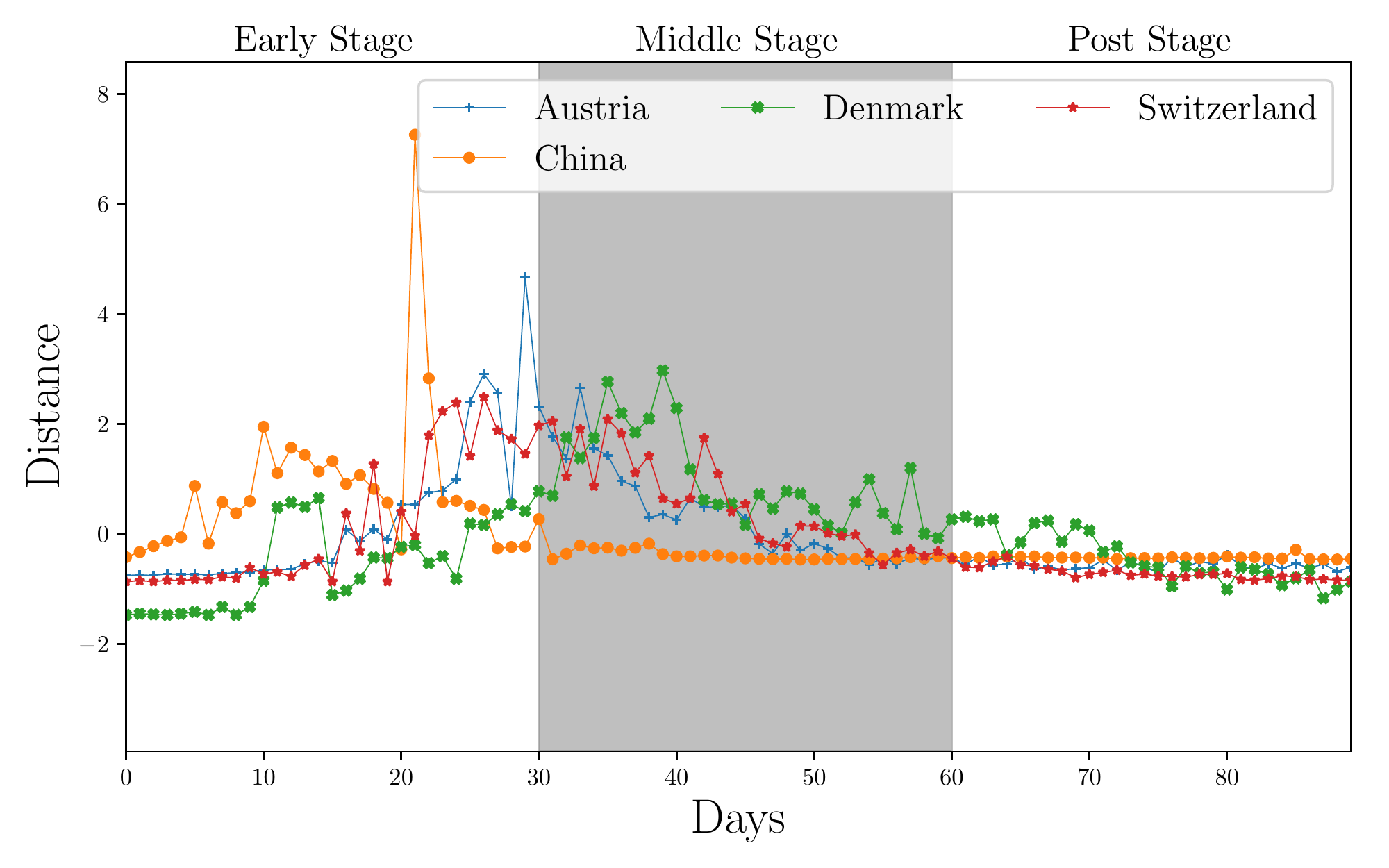}
		\subcaption{Spread Pattern}
		\label{fig:early_cluster_5_1}
	\end{subfigure}\hfill
	\begin{subfigure}[b]{0.48\linewidth}
		\includegraphics[width=\linewidth]{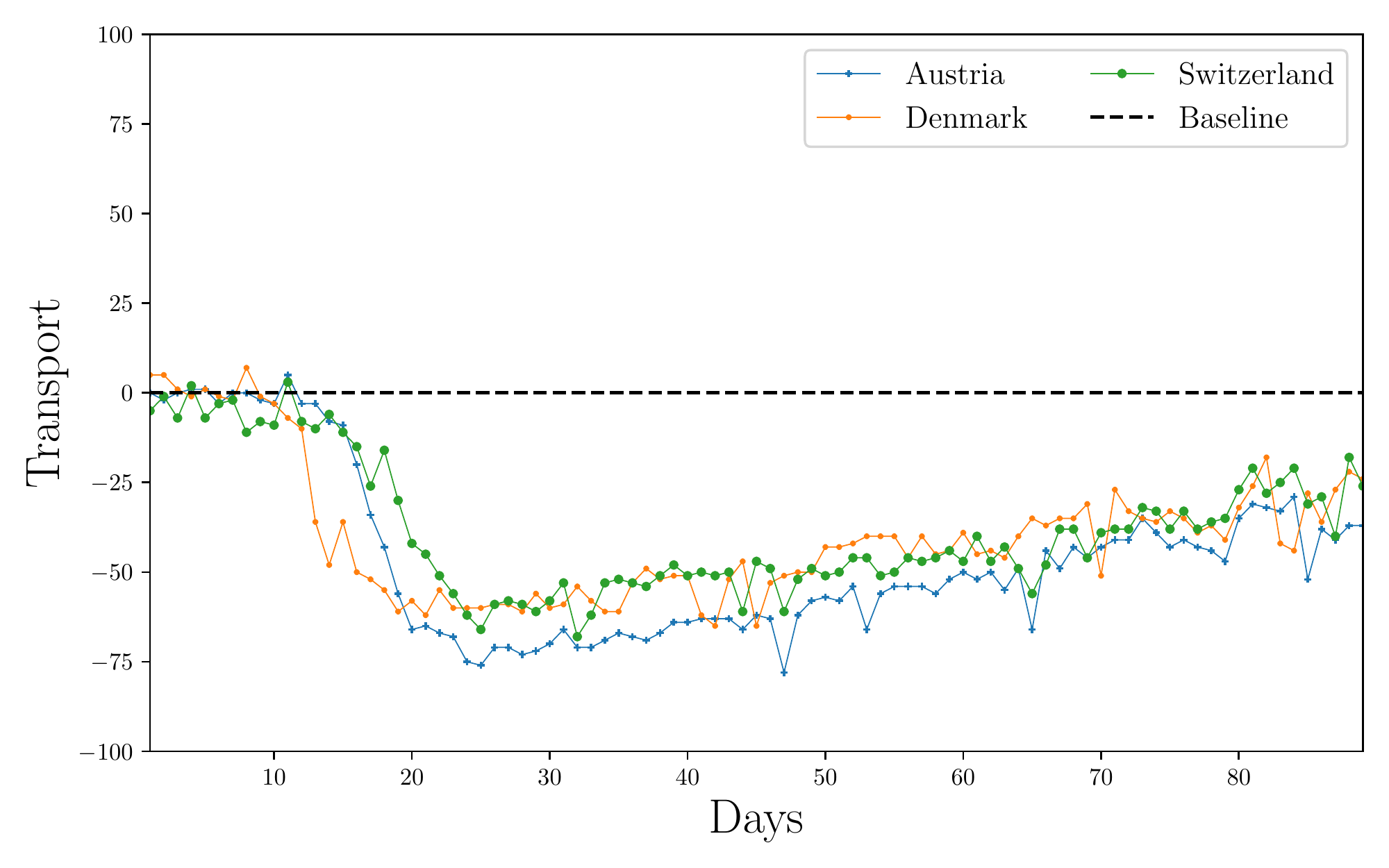}
		\subcaption{Mobility Pattern.}
		\label{fig:early_cluster_5_1_mobility}
	\end{subfigure}\hfill
	\caption{Early, middle and post stage spread patterns of Austria, Denmark, Switzerland and China, and their mobility during the same period.}
\end{figure*}

Early-stage clustering in~\cref{fig:Early_stage_cluster_5,fig:Early_stage_cluster_6,fig:Early_stage_cluster_7,fig:Early_stage_cluster_8} show that Austria, Denmark, Switzerland, and China reside in different singleton clusters.
These countries have dissimilar peaks in their respective timelines.
However, they reside in the same cluster in the post-stage (Cluster-5) due to the low spread patterns.
Figure~\ref{fig:early_cluster_5_1} shows the dissimilarities among these countries in the early-stage.
Switzerland imposed strict social distancing measures, while others went for complete lockdowns.
Austria imposed lockdown measures on March 16 (day 20)~\cite{austria_denmark_lock_lift} as the daily new cases started to increase.
According to Figure~\ref{fig:early_cluster_5_1}, Austria saw the peak around day 25-30.
Figure~\ref{fig:early_cluster_5_1_mobility} shows that their mobility data has reduced by 75\% on day 20. 
This figure indicates that the preventive measures were strictly implemented.
The Independent reported that the early lockdown, and mobile testing facilities are the core reasons for Austria to successfully contain the spread~\cite{why_austria}.
Thus, Austria has been able to flatten the curve as early as day 40.
The government lifted the restrictions on April 13 (day 48)~\cite{austria_denmark_lock_lift}.
Austria continues to exhibit a very low spread pattern in the post-stage.

Denmark exhibits different peaks in the early-stage compared to Austria.
According to Figure~\ref{fig:early_cluster_5_1}, Denmark had its first peak around day 10-13.
The government was quick to respond and imposed a national lockdown and closed its borders on March 11 (day 13)~\cite{austria_denmark_lock_lift}.
Denmark was among the first countries to impose strict preventive measures such as lockdowns, ban of large gatherings, restrictions on mobility, and air travel.
Figure~\ref{fig:early_cluster_5_1_mobility} shows that their mobility has reduced by more than 50\% since the lockdown was imposed.
A combination of factors such as the early lockdown, trust in government orders, and cultural attributes contribute to the containment of the spread~\cite{olagnier2020covid}.
As a result, the spread in Denmark starts to decrease from day 40 (see Figure~\ref{fig:early_cluster_5_1}).
Hence, the government lifted the restrictions on April 13 (day 46)~\cite{austria_denmark_lock_lift}.
The spread in Denmark continues to decrease in the latter part of the middle-stage and throughout the post-stage.

The spread in Switzerland starts to increase around day 15.
In response, the government imposed strict restrictions such as social distancing, and close down of borders on March 16 (day 20)~\cite{Switzerland_lock}.
The mobility data of Switzerland in Figure~\ref{fig:early_cluster_5_1_mobility} shows that such measures help the country to contain the spread around day 45 (see Figure~\ref{fig:early_cluster_5_1}) in the middle-stage.
Joseph~\emph{et al.}~\cite{lemaitre2020assessing} also verifies that the early implementation of the preventive measures sharply reduces the spread.
Similar to Austria and Denmark, Switzerland continues to exhibit a very low spread pattern in the post-stage.

China was the first country to hit by the pandemic.
The statewide complete lockdown in Wuhan and restrictions in air travel and local mobility was imposed on January 23 (accumulated day 1)~\cite{china_lock}.
By implementing these strict preventive measures, China was able to contain the spread in the last part of the early-stage.
Chinazzi~\emph{et al.}~\cite{chinazzi2020effect} estimated that the restrictions not only slowed down the spread in China but also helped to reduce the spread in other countries.
The government lifted the restrictions on April 8 (day 76)~\cite{china_lock_lift}.
China continued to exhibit a very low spread pattern in both middle and post stages.

Despite exhibiting different high spread patterns in the early stages, various preventive measures have helped these countries to control the spread in the latter part of the early stages or the early part of the middle stages.
As a result, these countries exhibit low spread patterns in the post stages.
China followed more strict lockdown, whereas early lockdown and compliance of people with social distancing measures helped other countries to reduce the spread as well.

\subsection{Early-Stage: Cluster-9}
\begin{figure*}[t!]
	\centering
	\begin{subfigure}[b]{0.48\linewidth}
		\includegraphics[width=\linewidth]{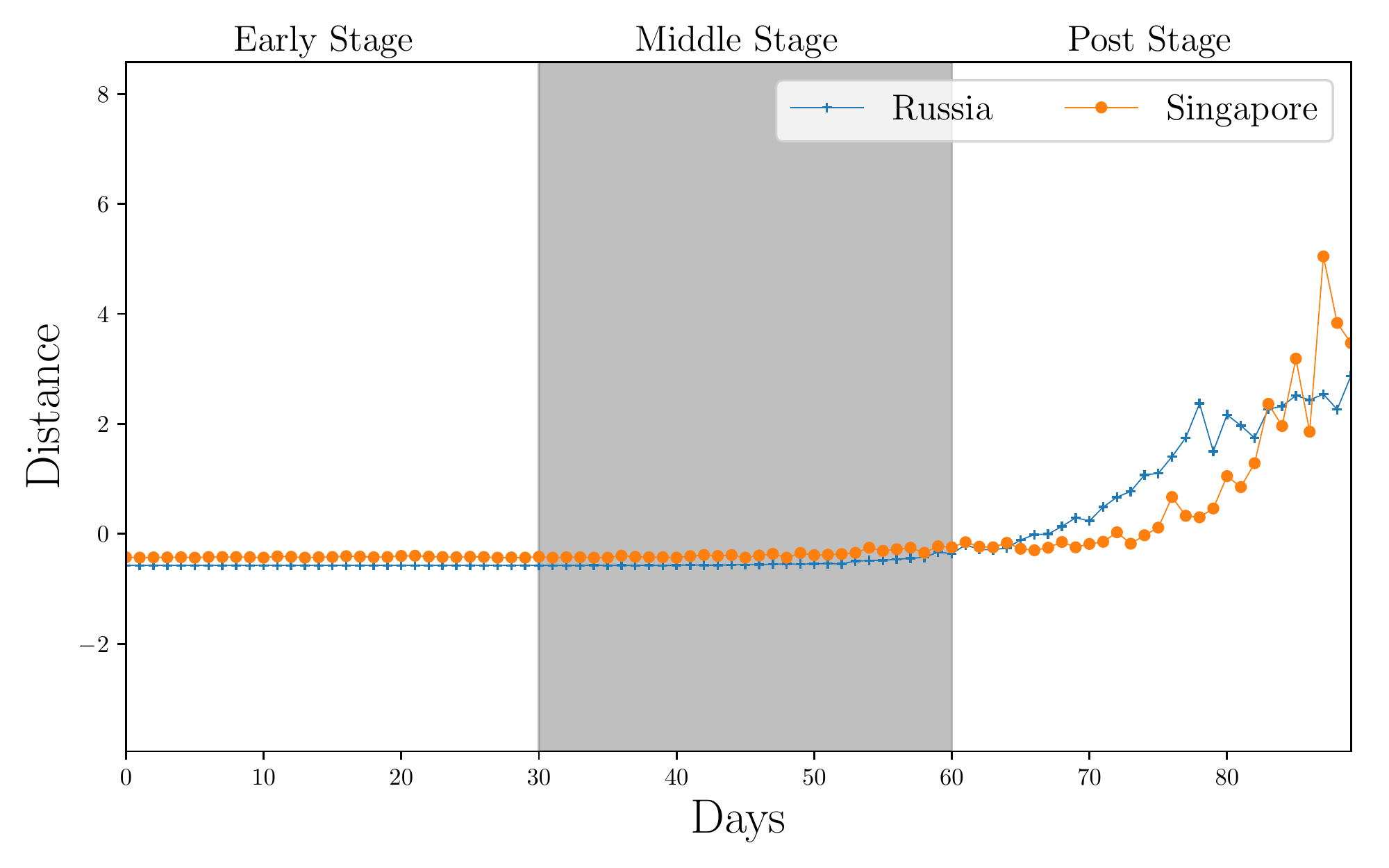}
		\subcaption{Spread Pattern}
		\label{fig:early_cluster_9_3}
	\end{subfigure}\hfill
	\begin{subfigure}[b]{0.48\linewidth}
		\includegraphics[width=\linewidth]{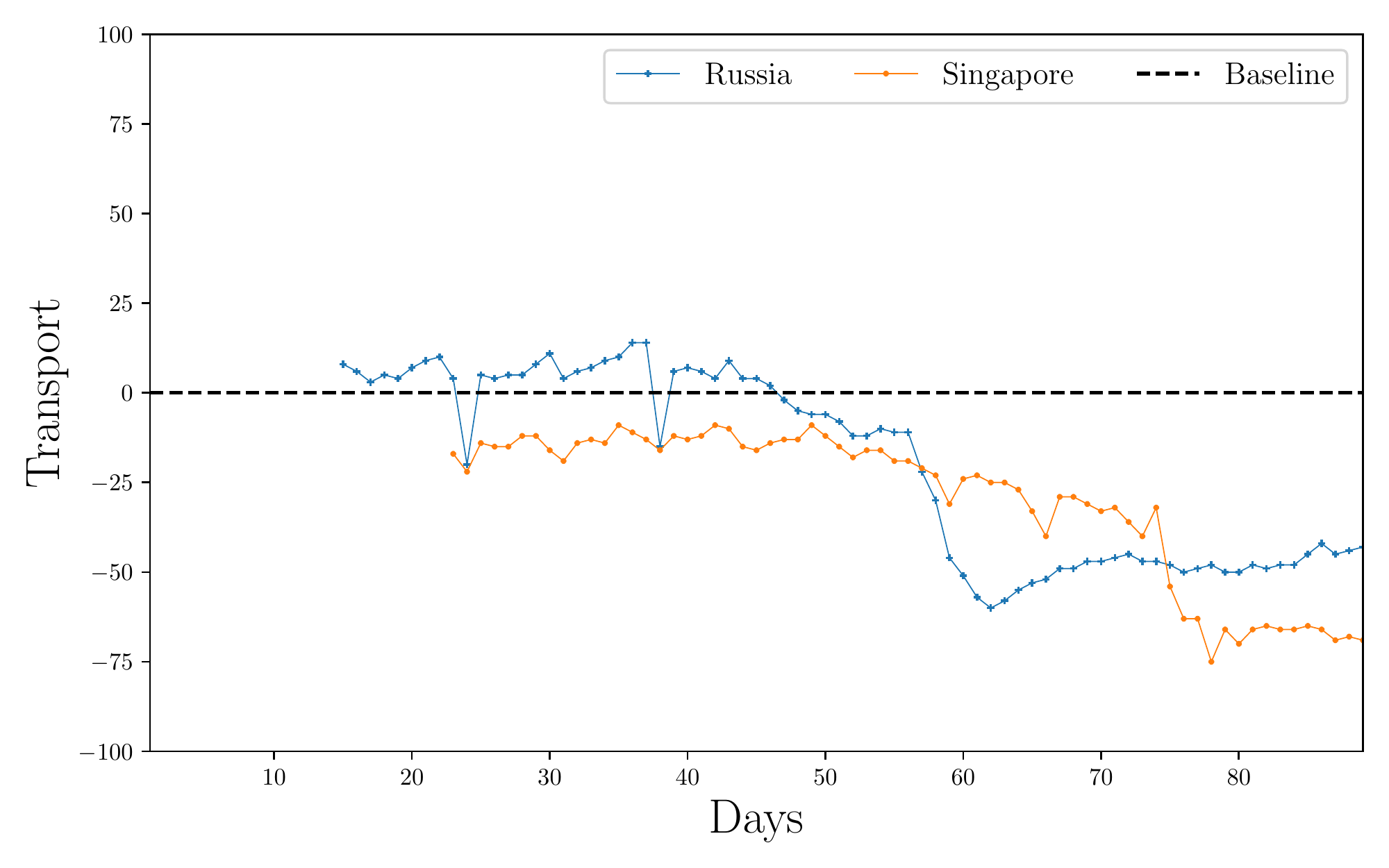}
		\subcaption{Mobility Pattern.}
		\label{fig:early_cluster_9_3_mobility}
	\end{subfigure}\hfill
	\caption{Early, middle and post stage spread patterns of Russia and Singapore, and their mobility during the same period.}
\end{figure*}

Early-stage clustering in Figure~\ref{fig:Early_stage_cluster_9}, Ecuador, France, Japan, Russia, and Singapore reside in cluster 9. 
In the post-stage, only Russia and Singapore remain clustered together in Cluster-2 as other countries move to different clusters.
These two countries have a very low spread in the early and middle stages according to Figure~\ref{fig:early_cluster_9_3}.
However, the daily new cases of these countries start to increase in their post stages.
As a result, the governments of these two countries imposed  lockdowns on 30 March (day 59) and 7 April (day 75), respectively.
Figure~\ref{fig:early_cluster_9_3_mobility} shows that mobility has initially been reduced by more than 60\% in Russia once the lockdown was imposed.
Even after imposing a lockdown, air travel ban, stay-at-home order, and social distancing measures, Russia suffered a spike in cases in the post-stage.
Most of the confirmed cases in the early and middle stages were due to travelers from European countries~\cite{Russia_peak}.
Several news articles reported that the lack of social awareness among people, delaying the lockdown order, unclear guidelines about the restrictions, and not complying with the government orders contributed to the rise of the spread in the post-stage in Russia~\cite{why_Russia_1, why_Russia_2}.
Singapore also shares a similar spread pattern as Russia according to Figure~\ref{fig:early_cluster_9_3}.
Singapore imposed the lockdown as soon as the spread started rising.
Most of the new cases were from the migrant workers as they continued to work during the pandemic~\cite{Why_Singapore}.
The lack of social awareness and failure to implement preventive measures earlier are the reasons for the increasing spread in Singapore.

\begin{figure*}[t!]
	\centering
	\begin{subfigure}[b]{0.48\linewidth}
		\includegraphics[width=\linewidth]{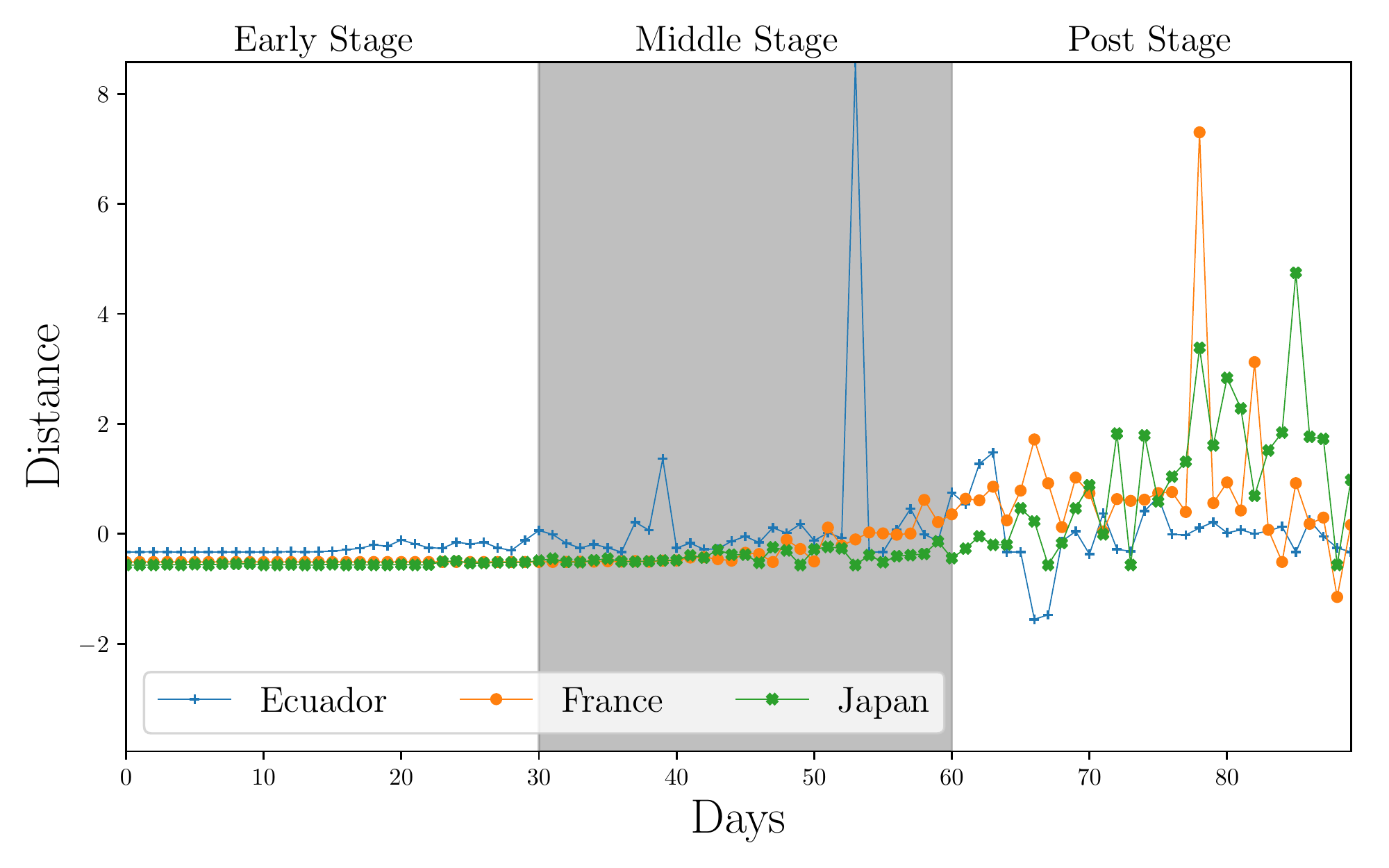}
		\subcaption{Spread Pattern}
		\label{fig:early_cluster_9_2}
	\end{subfigure}\hfill
	\begin{subfigure}[b]{0.48\linewidth}
		\includegraphics[width=\linewidth]{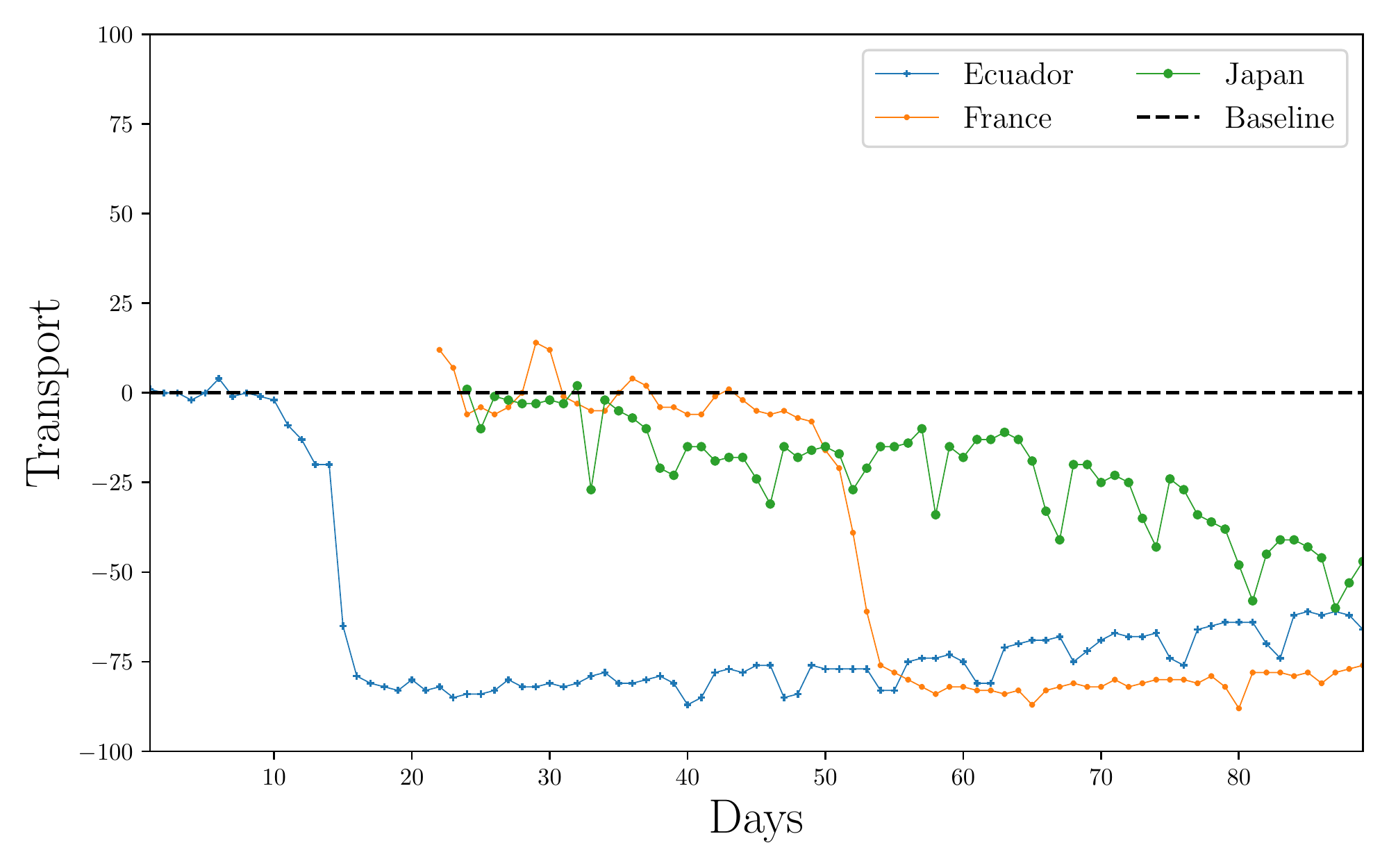}
		\subcaption{Mobility Pattern.}
		\label{fig:early_cluster_9_2_mobility}
	\end{subfigure}\hfill
	\caption{Early, middle and post stage spread patterns of Ecuador, France, and Japan, and their mobility during the same period.}
\end{figure*}

Ecuador, France, and Japan exhibit different spread patterns in their middle stages.
Hence, these countries move to different clusters in the post-stage (clusters 4, 8, and 6, respectively).
Ecuador has been in lockdown since 15 March (day 14)~\cite{ecu_lock}.
In the middle-stage, Ecuador exhibits inconsistent peaks compared to Japan and France.
It is reported that the increased capacity of COVID-19 tests caused the inconsistent peaks~\cite{why_ecu}.
Japan and France exhibit dissimilar peaks as both countries reside in singleton clusters in the post stages (clusters 6 and 8, respectively).
Both countries took different approaches to fight against the pandemic.
Japan did not impose any strict lockdown policies, rather they declared a state of emergency and social distancing measures.
However, these measures were lifted during the Hanami festival season on March 19 (day 57)~\cite{why_Japan}. 
The mobility data of Japan in Figure~\ref{fig:early_cluster_9_2_mobility} shows that mobility is reduced only after the festival season.
As a consequence, the spread is increased in the post-stage.
On the other hand, the daily new cases start to rise after day 45 in France. 
The French government imposed a total lockdown since March 17 (day 53)~\cite{france_lock}.
Figure~\ref{fig:early_cluster_9_2_mobility} shows that the mobility in France reduced significantly only after the lockdown was imposed.
Despite the lockdown, the spread kept increasing in France.
Vox and Politico reported that the lack of testing kits and large gatherings in a week-long festival in February contributed to the spread~\cite{why_France_2, why_France_3} in France.
The infrequent peaks in the post-stage are caused by adding previous data from nursing and retirement homes~\cite{why_France}.
The spread starts to decline in the last part of the post-stage.

Several factors affected the spread in Cluster-9. 
Lack of social awareness, lack of testing kits, large gatherings during festivals, or delaying lockdown orders, all contributed to the rise in the spread in the post stages.

\section{Conclusions}\label{conclusion}
In this study, we analyze the spread patterns of different countries from different geographic locations in the early period of the pandemic.
We divided the time-series data of each country into three stages~\emph{i.e.,} early, middle, and post stages.
We applied the agglomerative hierarchical clustering into the early and post stages, which divided 41 countries into nine clusters.
We used the middle-stage time-series data to explain the changes in clusters of early and post stages.
We further investigated the impact of different confinement policies or preventive measures adopted by the governments of each country to contain the virus spread.
We found that some countries were successful in containing the spread by implementing strict preventive measures (\emph{e.g.,} Romania, Portugal, Austria, Germany, France, Italy or China), while others were successful without any strict restrictions (\emph{e.g.,} Denmark, Netherlands, or South Korea).
The spread in some countries is still increasing despite implementing strict preventive measures (\emph{e.g.,} USA, Brazil, Canada, India, or Singapore).
These countries need to identify the point of transmission of spread and implement even more strict measures to contain the spread.
We also found that the effect of lockdowns or other strict preventive measures varies from country to country.
Lockdowns are less effective in densely populated countries with a lack of social awareness and government trust.
In such cases, bolstering the testing capacity, isolation of infected patients and regions, and strict social distancing measures can help to reduce the spread.
Implementing a single counter-measure may not be enough to contain the spread but when implemented together, these measures can certainly control the spread depending on the dynamics of the individual countries.

\appendix

\section{Cluster}\label{appendix_cluster}
\begin{figure}[H]
  \centering
  \begin{subfigure}[b]{\linewidth}
    \includegraphics[width=\linewidth]{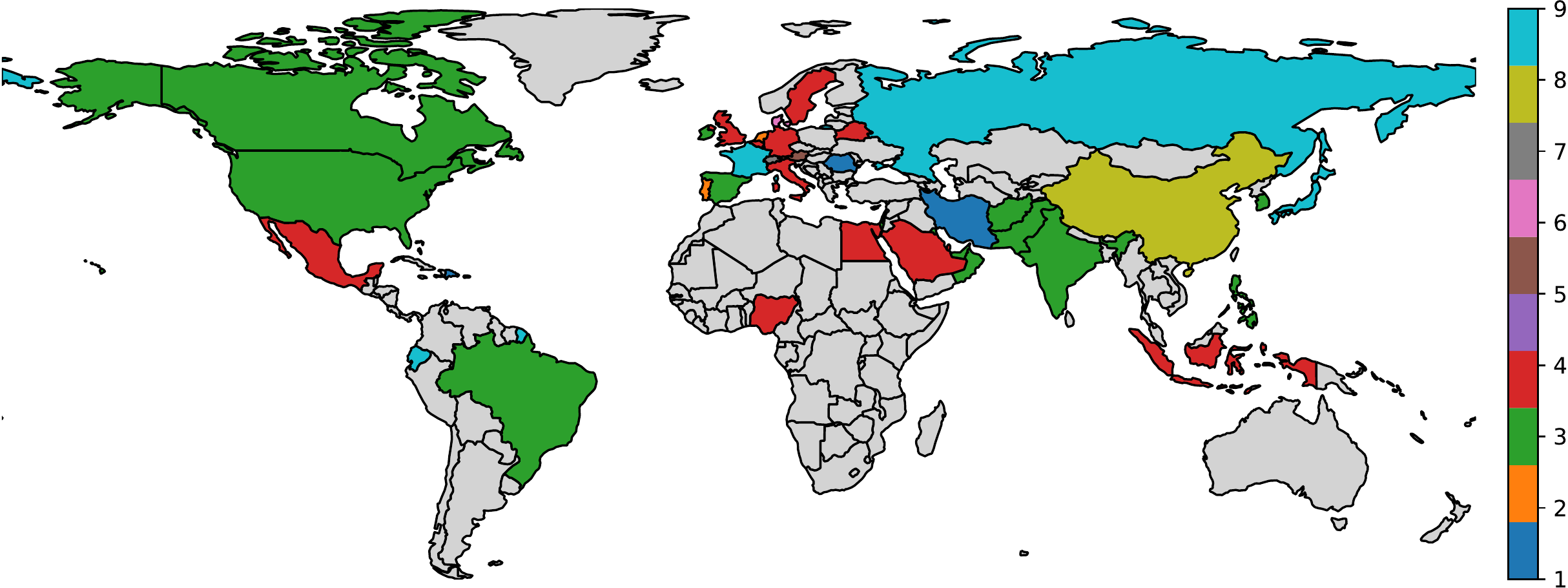}
     \subcaption{Early-stage }
  \end{subfigure}\hfill
  \begin{subfigure}[b]{\linewidth}
    \includegraphics[width=\linewidth]{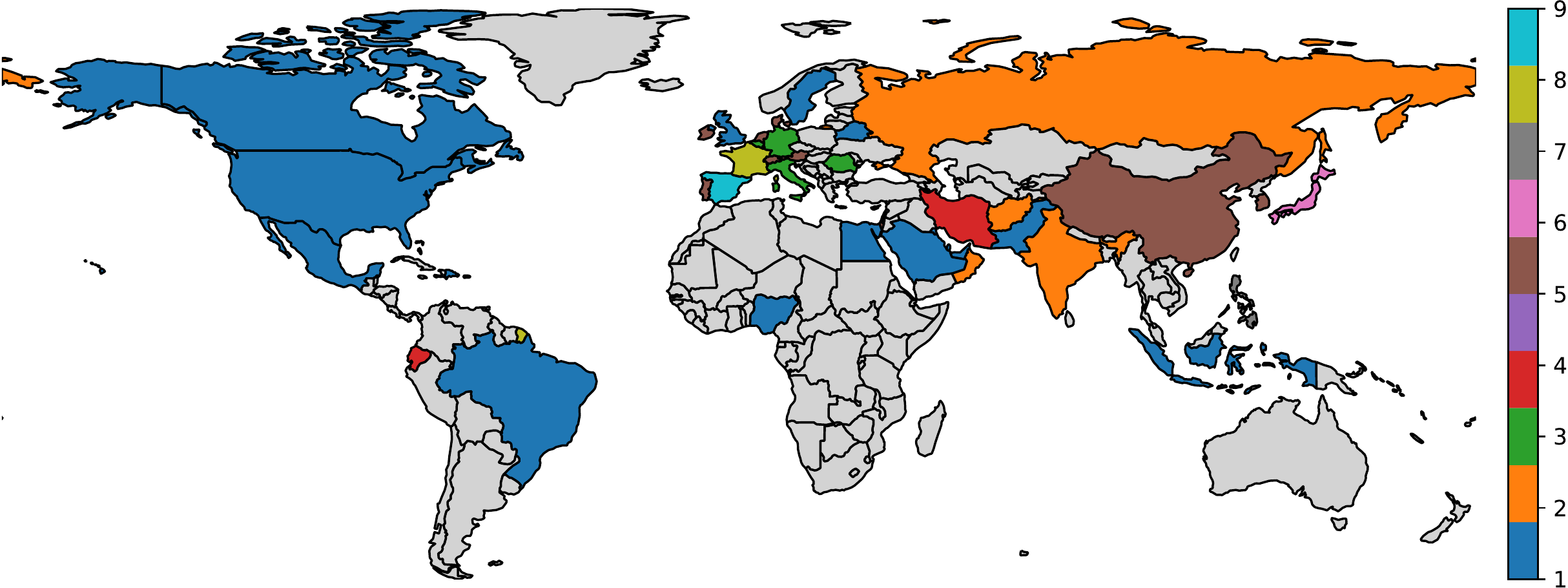}
    \subcaption{Post-stage }
  \end{subfigure}\hfill
  \caption{Early and post-stage clusters of countries on a Choropleth map.}
  \label{fig:world_map_clusters}
\end{figure}

\section{Cluster Labels}\label{appendix_cluster_labels}
\begin{table}[H]
\caption{Cluster labels after applying agglomerative hierarchical clustering on early and post-stage.}
\label{tab:cluster_labels}
\resizebox{0.43\textwidth}{!}{%
\begin{tabular}{|c|c|c|}
\hline
\textbf{Countries} & \textbf{\begin{tabular}[c]{@{}c@{}}Early-stage\\ Cluster Labels\end{tabular}} & \textbf{\begin{tabular}[c]{@{}c@{}}Post-stage\\ Cluster Labels\end{tabular}} \\ \hline \hline
D.R.               & 1                                                                             & 1                                                                            \\ \hline
Iran               & 1                                                                             & 4                                                                            \\ \hline
Romania            & 1                                                                             & 3                                                                            \\ \hline
Netherlands        & 2                                                                             & 5                                                                            \\ \hline
Portugal           & 2                                                                             & 5                                                                            \\ \hline
Afghanistan        & 3                                                                             & 2                                                                            \\ \hline
Bahrain            & 3                                                                             & 1                                                                            \\ \hline
Brazil             & 3                                                                             & 1                                                                            \\ \hline
Canada             & 3                                                                             & 1                                                                            \\ \hline
India              & 3                                                                             & 2                                                                            \\ \hline
Ireland            & 3                                                                             & 5                                                                            \\ \hline
Israel             & 3                                                                             & 5                                                                            \\ \hline
S.Korea            & 3                                                                             & 5                                                                            \\ \hline
Kuwait             & 3                                                                             & 1                                                                            \\ \hline
Oman               & 3                                                                             & 2                                                                            \\ \hline
Pakistan           & 3                                                                             & 1                                                                            \\ \hline
Philippines        & 3                                                                             & 7                                                                            \\ \hline
Spain              & 3                                                                             & 9                                                                            \\ \hline
US                 & 3                                                                             & 1                                                                            \\ \hline
U.A.E.             & 3                                                                             & 1                                                                            \\ \hline
Belarus            & 4                                                                             & 1                                                                            \\ \hline
Belgium            & 4                                                                             & 3                                                                            \\ \hline
Egypt              & 4                                                                             & 1                                                                            \\ \hline
Germany            & 4                                                                             & 3                                                                            \\ \hline
Indonesia          & 4                                                                             & 1                                                                            \\ \hline
Italy              & 4                                                                             & 3                                                                            \\ \hline
Mexico             & 4                                                                             & 1                                                                            \\ \hline
Nigeria            & 4                                                                             & 1                                                                            \\ \hline
Qatar              & 4                                                                             & 1                                                                            \\ \hline
Saudi Arabia       & 4                                                                             & 1                                                                            \\ \hline
Sweden             & 4                                                                             & 1                                                                            \\ \hline
U.K.               & 4                                                                             & 1                                                                            \\ \hline
Austria            & 5                                                                             & 5                                                                            \\ \hline
Denmark            & 6                                                                             & 5                                                                            \\ \hline
Switzerland        & 7                                                                             & 5                                                                            \\ \hline
China              & 8                                                                             & 5                                                                            \\ \hline
Ecuador            & 9                                                                             & 4                                                                            \\ \hline
France             & 9                                                                             & 8                                                                            \\ \hline
Japan              & 9                                                                             & 6                                                                            \\ \hline
Russia             & 9                                                                             & 2                                                                            \\ \hline
Singapore          & 9                                                                             & 2                                                                            \\ \hline
\end{tabular}
}
\end{table}


\bibliographystyle{ACM-Reference-Format}
\bibliography{library}

\end{document}